\documentclass[11pt,a4paper]{article}
\usepackage{amsfonts}
\usepackage{amsmath,amssymb}
\usepackage{epsfig,graphicx}

\input{tcilatex}

\begin{document}

\begin{center}
{\huge Gauge Fields as Goldstone Bosons Triggered by Spontaneously Broken
Supersymmetry}{\Huge \ }

{\Huge \bigskip \bigskip \bigskip }

\textbf{J.L.~Chkareuli}$^{1,2}$

$^{1}$\textit{Center for Elementary Particle Physics, Ilia State University,
0162 Tbilisi, Georgia\ \vspace{0pt}\\[0pt]
}

$^{2}$\textit{E. Andronikashvili} \textit{Institute of Physics, 0177
Tbilisi, Georgia\ }

\textit{\bigskip }

\bigskip

\bigskip

\bigskip

\bigskip

\textbf{Abstract}

\bigskip

\bigskip
\end{center}

The emergent gauge theories are reconsidered in light of supersymmetry and
an appropriate emergence conjecture is formulated. Accordingly, it might be
expected that only global symmetries are fundamental symmetries of Nature,
whereas local symmetries and associated massless gauge fields could solely
emerge due to spontaneous breaking of underlying spacetime symmetries
involved, such as relativistic invariance and supersymmetry. We further
argue that this breaking, taken in the form of the nonlinear $\sigma $-model
type pattern for vector fields or superfields, puts essential restrictions
on geometrical degrees of freedom of a physical field system that makes it
to adjust itself in such a way that its global internal symmetry $G$ turns
into the local symmetry $G_{loc}$. Otherwise, a given field system could
loose too many degrees of freedom thus getting unphysical that would make it
impossible to set the required initial conditions in an appropriate Cauchy
problem, or to choose self-consistent equal-time commutation relations in
quantum theory. Remarkably, this emergence process may naturally be
triggered by supersymmetry, as is illustrated in detail by an example of a
general supersymmetric QED model which is then extended to the Standard
Model and GUTs. The requirement of vacuum stability in such class of models
makes both Lorentz invariance and supersymmetry to become spontaneously
broken in the visible sector. As a consequence, massless photon and other
gauge bosons appear as the corresponding Goldstone and pseudo-Goldstone zero
modes and special local invariance is simultaneously generated. Due to this
invariance all possible Lorentz violations are turned out to be completely
cancelled out among themselves. However, broken supersymmetry effects
related to an existence of a light pseudo-goldstino (being essentially a
photino) are still left in the theory. It typically appears in the
low-energy particle spectrum as the $\mathrm{eV}$ scale stable LSP or the
electroweak scale long-lived NLSP, being in both cases accompanied by a very
light gravitino, that could be considered as some observational signature in
favor of emergent supersymmetric theories.

\bigskip

\bigskip

\thispagestyle{empty}\newpage

{\LARGE Contents}{\Huge \ \bigskip \bigskip\bigskip\bigskip }

\textbf{1 \ \ \ Introduction} \ \ \ \ \ \ \ \ \ \ \ \ \ \ \ \ \ \ \ \ \ \ \
\ \ \ \ \ \ \ \ \ \ \ \ \ \ \ \ \ \ \ \ \ \ \ \ \ \ \ \ \ \ \ \ \ \ \ \ \ \
\ \ \ \ \ \ \ \ \ \ \ \ \ \ \ \ \ \ \ \ \ \ \ \ \ \ \ \ \ \ \ \ \ \ \ \ \ 2\
\ \ \ \ \ \ \ \ \ \ \ \ \ \ \ \ \ \ \ \ \ \ \ \ \ \ \ \ \ \ \ \ \ \ \ \ \ \
\ \bigskip

\textbf{2\ \ \ Photons as Nambu-Goldstone zero modes: a brief sketch \ \ \ \
\ \ \ \ \ \ \ \ \ \ \ \ \ \ \ \ \ \ \ \ \ \ \ \ \ \ }3\textbf{\ \ \ } \ \ \
\ \ \ \ \ \ \ \ \ \ \ \ \ \ \ \ \ \ \ \ \ \ \ \ \ \ \ \ \ \ \ \ \ \ \ \ \ \
\ \ \ \ \ \ \ \ \ \ \ \ 

\ \ \ 2.1 \ Composite models\ \ \ \ \ \ \ \ \ \ \ \ \ \ \ \ \ \ \ \ \ \ \ \
\ \ \ \ 

\ \ \ 2.2 \ Potential-based models \ \ \ \ \ \ \ \ \ \ \ \ \ \ \ \ \ \ \ \ \
\ \ \ 

\ \ \ 2.3 \ Constraint-based models\ 

\ \ \ 2.4 \ Models with external vector backgrounds

\ \ \ 2.5 \ Supersymmetric models\ \ \ \ \ \ \ \ \ \ \ \ \ \ \ \ \ \ \ \ \ \
\ \ \ \ \ \ \ \ \ \ \ \ \ \ \bigskip

\textbf{3\ \ \ Gauge theories emerging from constraints \ \ \ \ \ \ \ \ \ \
\ \ \ \ \ \ \ \ \ \ \ \ \ \ \ \ \ \ \ \ \ \ \ \ \ \ \ \ \ \ \ \ \ \ \ \ \ \
\ \ \ \ \ \ }11

\ \ \ 3.1 \ An emergence conjecture revised \ \ \ \ \ \ \ \ \ \ \ \ \ \ \ \
\ \ \ \ \ \ \ \ \ \ \ \ \ \ \ \ \ \ \ \ \ \ \ \ \ \ \ \ \ \ \ \ \ \ \ \ \ \
\ \ \ 

\ \ \ 3.2 \ Emergent Abelian gauge invariance

\ \ \ 3.3 \ Non-Abelian gauge fields as pseudo-Goldstone modes

\ \ \ 3.4 \ Constraints inducing and uninducing gauge invariance

\ \ \ 3.5 \ Gauge invariance \textit{versus }spontaneous Lorentz violation

\bigskip

\textbf{4 \ \ Emergent SUSY theories: a QED primer} \ \ \ \ \ \ \ \ \ \ \ \
\ \ \ \ \ \ \ \ \ \ \ \ \ \ \ \ \ \ \ \ \ \ \ \ \ \ \ \ \ \ \ \ \ \ \ \ \ \
\ \ \ \ \ 23

\ \ \ 4.1 \ Spontaneous SUSY violation \ \ \ \ \ \ \ \ \ \ \ \ \ \ \ \ \ \ \
\ \ \ \ \ \ \ \ \ \ \ \ \ \ \ \ \ \ \ \ \ \ \ \ \ \ \ \ \ \ \ \ \ \ \ \ \ \
\ \ \ 

\ \ \ 4.2 \ Restoration of gauge invariance \ \ \ \ \ \ \ \ \ \ \ \ \ \ \ \
\ \ \ \ \ \ \ \ \ \ \ \ \ \ \ \ \ \ \ \ \ \ \ \ \ \ \ \ \ \ \ \ \ \ \ \ \ \
\ \ \ 

\ \ \ 4.3 \ Instability of superfield polynomial potential \ \ \ \ \ \ \ \ \
\ \ \ \ \ \ \ \ \ \ \ \ \ \ \ \ \ \ \ \ \ \ \ \ \ \ \ \ \ \ \ \ \ \ \ \ \ \
\ \ \ 

\ \ \ 4.4 \ Stabilization of vacuum by constraining vector superfield \ \ \ 

\ \ \ 4.5 \ Constrained vector superfield: a formal view

\bigskip

\textbf{5 \ \ On emergent supersymmetric Standard Models and GUTs}\ \ \ \ \
\ \ \ \ \ \ \ \ \ \ \ \ \ \ \ \ \ \ \ \ \ \ \ 34\ \ \ \ \ \ \ \ \ \ \ \ 

\ \ \ 5.1 \ Potential of Abelian and non-Abelian vector superfields

\ \ \ 5.2 \ Constrained vector supermultiplets

\ \ \ 5.3 \ Broken SUSY phase: an emergent $U(1)\times SU(N)$ theory

\ \ \ 5.4 \ Some immediate outcomes\ 

\bigskip

\textbf{6 \ \ Phenomenological implications: photino as pseudo-goldstino\ \
\ \ \ \ \ \ \ \ \ \ \ \ \ \ \ \ \ \ }43

\ \ \ 6.1 \ Two-sector SUSY breaking \ \ \ \ \ \ \ \ \ \ \ \ \ \ \ \ \ \ \ \
\ \ \ 

\ \ \ \ 6.2 \ Pure visible sector SUSY breaking scenario \ \ \ \ \ \ \ \ \ \
\ \ \ \ \ \ \ \ \ \ \ \ \ \ \ \ \ \ \ \ \ \ \ \ \ \ \ \ \ \ \ \ \ \ \ \ \ \
\ \ \ \ \ \ \ \ \ \ \ \ \ \ \ \ \ \ \ \ \ \ \ \ \ \ \ \ \ \ \ \ \ \ \ \ \ \
\ \ \ \ \ \ \ \ \ \ \ \ \ \ \ \ \ \ \ \ \ \ \ \ \ \ \ \ \ \ \ \ \ \ 

\bigskip

\textbf{7 \ \ Summary and conclusion \ \ \ \ \ \ \ \ \ \ \ \ \ \ \ \ \ \ \ \
\ \ \ \ \ \ \ \ \ \ \ \ \ \ \ \ \ \ \ \ \ \ \ \ \ \ \ \ \ \ \ \ \ \ \ \ \ \
\ \ \ \ \ \ \ \ \ \ \ \ \ \ \ \ \ \ \ \ \ \ }50

\bigskip \bigskip\bigskip

{\Large References}{\large \ \ }{\LARGE \ \ }\textbf{\ \ \ \ \ \ \ \ \ \ \ \
\ \ \ \ \ \ \ \ \ \ \ \ \ \ \ \ \ \ \ \ \ \ \ \ \ \ \ \ \ \ \ \ \ \ \ \ \ \
\ \ \ \ \ \ \ \ \ \ \ \ \ \ \ \ \ \ \ \ \ \ \ \ \ \ \ \ \ \ \ \ \ \ \ \ \ \
\ \ \ \ \ \ \ \ \ \ \ }52

%
\thispagestyle{empty}\newpage

\section{Introduction}

It is now conventional wisdom that internal gauge symmetries form the basis
of modern particle physics being most successfully realized within the
celebrated Standard Model (SM) of quarks and leptons and their fundamental
strong, weak and electromagnetic interactions. At the same time, local gauge
invariance, contrary to a global symmetry case, may look like a cumbersome
geometrical input rather than a "true" physical principle, especially in the
framework of an effective quantum field theory\ (QFT) becoming, presumably,
irrelevant at very high energies. In this connection, one could wonder
whether there is any basic dynamical reason that necessitates gauge
invariance and the associated masslessness of gauge fields as some emergent
phenomenon arising from a more profound level of dynamics. By analogy with a
dynamical origin of massless scalar particle excitations, which is very well
understood in terms of spontaneously broken global internal symmetries \cite%
{NJL}, one could think that the origin of massless gauge fields as vector
Nambu-Goldstone (NG) bosons are related to the spontaneous violation of
Lorentz invariance which is in fact the minimal spacetime global symmetry
underlying particle physics. This well-known approach providing a viable
alternative to quantum electrodynamics \cite{bjorken}, gravity \cite{ph} and
Yang-Mills theories \cite{eg} has a long history started over fifty years
ago.

However, the role of Lorentz invariance may change, and its spontaneous
violation may not be the only reason why massless photons could dynamically
appear, if spacetime symmetry is further enlarged. In this connection,
special interest is related to supersymmetry which has made a serious impact
on particle physics in the last decades (though has not been yet
discovered). Actually, as we will see, the situation is changed dramatically
in the SUSY inspired emergent gauge theories. In sharp contrast to non-SUSY
analogs, it appears that the spontaneous Lorentz invariance violation (SLIV)
caused by an arbitrary potential of vector superfield $V(x,\theta ,\overline{%
\theta })$ never goes any further than some nonlinear gauge constraint put
on its vector field component $A_{\mu }(x)$\ associated with a photon. This
allows to think that physical Lorentz invariance is somewhat protected by
SUSY, thus only requiring the "condensation" of the gauge degree of freedom
in the vector field $A_{\mu }$. The point is, however, that even in the case
when SLIV is not physical it inevitably leads to the generation of massless
photons as vector NG bosons provided that SUSY itself is spontaneously
broken. In this sense, a generic trigger for massless photons to dynamically
emerge happens to be spontaneously broken supersymmetry rather than
physically manifested Lorentz noninvariance.

The paper is organized in the following way. In the next section we give a
brief sketch of existing emergent gauge theories in light of supersymmetry.
This help us to see more clearly the significant changes which appears
necessary in a supersymmetric context, and properly formulate an emergence
conjecture in section 3. We give a detailed presentation of emergent gauge
invariant Abelian and non-Abelian theories and show somewhat fundamental
relationship between spontaneous Lorentz violation and emergent gauge
invariance due to which all SLIV contributions to physical processes
completely cancel out among themselves. In essence, the only way for SLIV to
manifest itself observationally may appear if gauge invariance in these
theories turns out to be broken in an explicit rather than spontaneous way.
As a result, the SLIV cancellation mechanism does not work longer and one
inevitably comes to physical Lorentz violation, as is explicitly
demonstrated in section 3.5. In the next section 4 we consider
supersymmetric QED model extended by an arbitrary polynomial potential of
massive vector superfield that breaks gauge invariance in the SUSY invariant
phase. However, the requirement of vacuum stability in such class of models
makes both supersymmetry and Lorentz invariance to become spontaneously
broken. As a consequence, the massless photino and photon appear as the
corresponding Nambu-Goldstone zero modes in an emergent SUSY QED, and also a
special gauge invariance is simultaneously generated. Due to this invariance
all observable relativistically noninvariant effects appear to be completely
cancelled out and physical Lorentz invariance is recovered. Further in
section 5, all basic arguments developed in SUSY QED are generalized
successively to the Standard Model and Grand Unified Theories (GUTs). For
definiteness, we focus on the $U(1)\times SU(N)$ symmetrical theories. Such
a split group form is dictated by the fact that in the pure non-Abelian
symmetry case one only has the SUSY invariant phase in the theory that makes
it inappropriate for an outgrowth of an emergence process. As possible
realistic realizations, the Standard Model case with the \ electroweak $%
U(1)\times SU(2)$ symmetry and flipped $SU(5)$ GUT are briefly discussed.
Phenomenological implications are largely given in section 6. The most
interesting part of them are related to the massless photino mentioned
above. This photino being mixed with another goldstino appearing from a
spontaneous SUSY violation in the hidden sector essentially turns into the
light pseudo-goldstino whose physics is considered in significant detail.
This physics is unambiguously related to the class of models where SUSY
breaks, at least partially, in the visible sector as well. This is the only
class of models where emergent supersymmetric QED or Standard Model can be
self-consistently realized. And finally in section 7, we summarize the main
results and conclude.

\section{Photons as Nambu-Goldstone zero modes:\textbf{\ a brief sketch}}

Below, we briefly comment on some known models where an idea of emergent
gauge theory according to which photons and other gauge fields may appear as
Nambu-Goldstone zero modes is realized in one way or the other. They include
the composite models, where this idea was considered for the first time \cite%
{bjorken, jb, kraus, jen}, and three other ones - the vector field
potential-based models \cite{ks, bluhm}, the vector field constraint-based
models \cite{nambu, az, wiki} and models with external vector backgrounds 
\cite{car, 2} together with their supersymmetric extensions \cite{bs, pos,
pos1}. Some quick summary on them may appear useful before we finally turn
to emergent SUSY models introduced recently \cite{c}, which we consider in
significant detail in subsequent sections.

\subsection{Composite models}

The first models \cite{bjorken} realizing the SLIV conjecture were based on
the four-fermi interaction where the photon appears as a fermion-antifermion
pair composite state in complete analogy with massless composite scalar
fields (identified with pions) in the original Nambu-Jona-Lazinio model \cite%
{NJL}. This old idea is better expressed nowadays in terms of effective
field theory where the standard \ QED Lagrangian is readily obtained through
the corresponding loop radiative effects due to $N$ fermion species involved 
\cite{jb, kraus, jen}. Also, instead of the old four-fermi model one can
start with the generalized effective action with all possible multi-fermi
interactions \cite{kraus} 
\begin{equation}
L(\psi ,\overline{\psi })=\text{ }\overline{\psi }_{i}(i\gamma \partial -%
\mathrm{m})\psi _{i}+N\sum_{n=1}^{\infty }\mathrm{G}_{2n}\left[ \frac{%
\overline{\psi }_{i}\gamma _{\mu }\psi _{i}}{N}\right] ^{2n}\text{ .}
\label{4f}
\end{equation}%
Here summation over flavor indices $i$ (and spacetime indices $\mu $) is
implied so that the Lagrangian $L(\psi ,\overline{\psi })$ possesses a $U(N)$
global flavor symmetry. This model is evidently non-renormalizable and can
be only considered as an effective theory valid at sufficiently low
energies. The dimensionful couplings $\mathrm{G}_{2n}$ are proportional to
appropriate powers of some UV cutoff $\Lambda $ being ultimately related to
some energy scale up to which this effective theory is valid, $\mathrm{G}%
_{2n}\sim \Lambda ^{4-6n}$. Factors of $N$ in (\ref{4f}) are chosen in such
a way to provide a well defined large $N$ limit so that the correlators for
the properly normalized fermion bilinears $(\overline{\psi }_{i}\gamma _{\mu
}\psi _{i})/N$ will scale as $N^{0}$.

The action (\ref{4f}) can be re-written using the standard trick of
introducing an auxiliary field $A_{\mu }$%
\begin{equation}
L(\psi ,\overline{\psi },A_{\mu })=\text{ }\overline{\psi }_{i}(i\gamma
\partial -\gamma A-\mathrm{m})\psi _{i}-N\mathrm{V}(A_{\mu }A^{\mu })
\label{4f1}
\end{equation}%
The potential $V$ is a power series in $A_{\mu }A^{\mu }$%
\begin{equation}
\mathrm{V}(A_{\mu }A^{\mu })=\frac{\mathrm{\mu }^{2}}{2}A_{\mu }A^{\mu }-%
\frac{\mathrm{\lambda }_{c}^{(4)}}{4}(A_{\mu }A^{\mu })^{2}+\cdot \cdot \cdot
\label{exp}
\end{equation}%
with coefficients chosen as%
\begin{equation}
\mathrm{\mu }^{2}=\frac{1}{2\mathrm{G}_{2}}\text{ , \ }\mathrm{\lambda }%
_{c}^{(4)}=\frac{1}{4}\frac{\mathrm{G}_{4}}{\mathrm{G}_{2}^{4}}\text{ , }%
\cdot \cdot \cdot  \label{par}
\end{equation}%
that by solving the algebraic equations of motion for $A_{\mu }$ and
substituting back into (\ref{4f1}) one recovers the starting Lagrangian (\ref%
{4f}). If instead one integrates out the fermions $\psi _{i}$, an effective
action emerges in terms of the composite $A^{\mu }$\ field alone, which
acquires its own dynamics%
\begin{equation}
S_{eff}=N\int d^{4}x\left[ \frac{1}{4e^{2}}F_{\mu \nu }F^{\mu \nu }+\mathrm{V%
}(A_{\mu }A^{\mu })+A_{\mu }J^{\mu }+\cdot \cdot \cdot \right]  \label{efff}
\end{equation}%
where the coupling constant $\mathrm{e}^{2}$ is given by 
\begin{equation}
\mathrm{e}^{2}=\frac{12\pi ^{2}}{\ln (\Lambda ^{2}/\mathrm{m}^{2})}\text{ \ }
\label{v}
\end{equation}%
with $\Lambda $ standing for an UV cutoff being mentioned above\footnote{%
This value (\ref{v}) simply follows from the usual vacuum polarization
integral. Although quadratic divergence does not appear in the loop diagrams
thanks to the global current conservation, logarithmic divergences do. Note
that all couplings and masses (see also (\ref{par})) appearing in the
emergent effective theory are evaluated at zero four-momenta.}. Since the
fermions $\psi _{i}$ are minimally coupled to the vector field $A_{\mu }$ in
(\ref{4f1}), its kinetic term generated in this way appears gauge invariant
provided that a gauge invariant cutoff is chosen. Furthermore, since there
are $N$ species of fermions the effective action (\ref{efff}) has an overall
factor of $N$. And the last, introducing in the basic Lagrangian the minimal
couplings of some extra matter fields $\Psi _{I}$ ($I=1,2,...$) to the basic
fermions $\psi _{i}$ via conserved currents, $J^{\mu }(\Psi )\overline{\psi }%
_{i}\gamma _{\mu }\psi _{i}$, one generates the minimal matter couplings
given in (\ref{efff}) which are also gauge invariant.

Let us turn now to the Lorentz violation in the model. As is readily seen
from equations (\ref{exp}, \ref{par}), the quartic term in the effective
action $S_{eff}$ may only appear when higher-order terms, beyond the
four-fermi interaction, are activated in the basic Lagrangian (\ref{4f}).
This quartic term is enough to generate the familiar Mexican hat structure
of the potential \textrm{V} (\ref{exp})\ and induce spontaneous Lorentz
violation through the non-vanishing vacuum expectation value (VEV) of the
vector field $A_{\mu }$ (for more detail, see the next section). Thereby,
three Lorentz generators becomes broken for both time-like and space-like
Lorentz violation. As a result, the three massless NG modes associated with
this symmetry breaking emerge. They might be interpreted as the photon
components. However, owing to the lack of gauge invariance in the starting
fermion Lagrangian (\ref{4f}) the effective theory for the composite vector
field (\ref{efff}) is not entirely gauge invariant either. Apart from the
vector field polynomial terms, it also contains many vector field-derivative
terms (being represented by ellipses in (\ref{efff})). These terms may badly
break gauge invariance unless they are properly suppressed by taking the
number of fermion species $N$ to be large enough. The absence of well
defined approximate gauge invariance could make it hard to explicitly
demonstrate that these NG\ modes emerging due to spontaneous Lorentz
violation really form together a massless photon as a gauge field candidate.
Rather, there would be in general three separate massless Goldstone modes,
two of which may mimic the transverse photon polarizations, while the third
one must be appropriately suppressed.

Nevertheless, as was argued in \cite{kraus}, it appears possible to arrange
an effective theory the way that gauge invariance is violated at leading
order in $N$\ only by potential terms in (\ref{efff}). At this order the
gauge invariant form of the kinetic terms in (\ref{efff})\ implies that only
the transverse NG bosons propagate, exactly as in the conventional Lorentz
invariant electrodynamics. As a consequence, interactions between conserved
matter currents $J^{\mu }(\Psi )$ give the standard QED results at leading
order plus Lorentz noninvariant corrections occurring at order $1/N$. The
third NG boson effects are also suppressed by $1/N$. Altogether, one comes
to the emergent effective QED where the spontaneously broken Lorentz
invariance may appear as a controllable approximate symmetry in low energy
physics.

\subsection{Potential-based models}

One could think that composite models contain too many prerequisites and
complications related to the large number of basic fermion species involved,
their proper arrangement, non-renormalizability of the fundamental
multi-fermi Lagrangian, stability under radiative corrections, and so on
indefinitely. This approach contains in fact a cumbersome invisible sector
which induces the effective emergent theory. A natural question arises
whether one could start from the effective vector field theory (\ref{efff})
instead thus having spontaneous Lorentz violation from the outset.

Actually, making a proper redefinition of the vector field $A_{\mu
}\rightarrow ieA_{\mu }$ in (\ref{efff}) one comes to a conventional QED
type Lagrangian extended by an arbitrary vector field potential energy terms
which explicitly break gauge invariance. For a minimal potential containing
bilinear and quartic vector field terms one comes to the Lagrangian%
\begin{equation}
L_{V}=L_{QED}-\frac{\mathrm{\lambda }_{c}}{4}\left( A_{\mu }A^{\mu }-n^{2}%
\mathrm{M}^{2}\right) ^{2}  \label{lag2}
\end{equation}%
where the coupling constant $\mathrm{\lambda }_{c}$ is determined as in (\ref%
{efff}), the mass term $n^{2}\mathrm{M}^{2}$ (properly expressed through
parameters of the effective theory) stands for the proposed SLIV scale,
while $n_{\mu }$ is a properly-oriented unit Lorentz vector, $n^{2}=n_{\mu
}n^{\mu }=\pm 1$. This partially gauge invariant model being sometimes
referred to as the \textquotedblleft bumblebee\textquotedblright\ model \cite%
{ks} (see also \cite{bluhm} and references therein) means in fact that the
vector field $A_{\mu }$ develops a constant background value%
\begin{equation}
<A_{\mu }>\text{ }=n_{\mu }\mathrm{M}  \label{vev1}
\end{equation}%
and Lorentz symmetry $SO(1,3)$ breaks down to $SO(3)$ or $SO(1,2)$ depending
on whether $n_{\mu }$ is time-like ($n^{2}=+1$) or space-like ($n^{2}=-1$)%
\footnote{%
Note that such freedom in choice of either of $n^{2}$ value exists in fact
for the minimal vector field potential in (\ref{lag2}). For the higher order
terms included, the potential may have minimum for only positive or only
negative $n^{2}$.}. Expanding the vector field around this vacuum
configuration, 
\begin{equation}
A_{\mu }(x)=a_{\mu }(x)+n_{\mu }(\mathrm{M}+\mathcal{A}),\text{ \ \ }n_{\mu
}a_{\mu }=0  \label{2}
\end{equation}%
one finds that the $a_{\mu }$ field components, which are orthogonal to the
Lorentz violating direction $n_{\mu }$, describe a massless vector
Nambu-Goldstone boson, while the $\mathcal{A}$ field corresponds to a
massive Higgs mode away from the potential minimum. Due to the presence of
this mode the model may lead to some physical Lorentz violation in terms of
the properly deformed dispersion relations for photon and matter fields
involved that appear from the corresponding radiative corrections to their
kinetic terms \cite{kraus}.

However, as was argued in \cite{vru}, a bumblebee-like model appears
generally unstable, its Hamiltonian is not bounded from below beyond the
constrained phase space determined by the nonlinear condition $A_{\mu
}A^{\mu }=n^{2}\mathrm{M}^{2}$. \ With this condition imposed, the massive
Higgs mode never appears, the Hamiltonian is positive, and the model is
physically equivalent to the nonlinear constraint-based QED, which we
consider in the next section. Apart from the instability, the
potential-based models were shown \cite{aki} to be obstructed from having a
consistent ultraviolet completion, whereas the most of viable effective
theories possess such a completion. The problems mentioned certainly appear
in the effective theories emerging from the composite models as well.
Nevertheless, since a\ natural mass scale associated with spontaneous
Lorentz violation is presumably of the Planck scale order, only
quantum-gravity theory might make the ultimate conclusion on physical
viability of such models at all energy scales.

\subsection{Constraint-based models}

This class of models starts directly with the nonlinearly realized Lorentz
symmetry for underlying vector field $A_{\mu }(x)$ through the
"length-fixing" constraint 
\begin{equation}
A_{\mu }A^{\mu }=n^{2}\mathrm{M}^{2}\text{ }  \label{const}
\end{equation}%
implemented into a conventional QED. The constraint-based models were first
studied by Nambu a long ago \cite{nambu} (see also \cite{jov}), and in more
detail in recent years \cite{az, wiki, kep, jej, urr, gra}. The constraint (%
\ref{const}) is in fact very similar to the constraint appearing in the
nonlinear $\sigma $-model for pions \cite{GL}, $\sigma ^{2}+\pi ^{2}=f_{\pi
}^{2}$, where $f_{\pi }$ is the pion decay constant. Rather than impose by
postulate, the constraint (\ref{const}) may be implemented into the standard
QED Lagrangian $L_{QED}$ through the invariant Lagrange multiplier term

\begin{equation}
L=L_{QED}-\frac{\mathrm{\lambda }}{2}\left( A_{\mu }A^{\mu }-n^{2}\mathrm{M}%
^{2}\right) \text{ , \ }n^{2}=n_{\mu }n^{\mu }=\pm 1  \label{lag}
\end{equation}%
provided that initial values for all fields (and their momenta) involved are
chosen so as to restrict the phase space to values with a vanishing
multiplier function \textrm{$\lambda $}$(x)$, \textrm{$\lambda $}$=0$.
Actually, due to an automatic conservation of the matter current in QED an
initial value $\mathrm{\lambda }=0$ will remain for all time\footnote{%
Interestingly, this solution with the Lagrange multiplier field $\mathrm{%
\lambda }(x)$ being vanished can technically be realized by introducing in
the Lagrangian (\ref{lag}) an additional Lagrange multiplier term of the
type $\mathrm{\xi \lambda }^{2}$, where $\mathrm{\xi }(x)$ is a new
multiplier field. One can now easily confirm that a variation of the
modified Lagrangia $L+$ $\mathrm{\xi \lambda }^{2}$ with respect to the $%
\mathrm{\xi }$ field leads to the condition $\mathrm{\lambda }=0$, whereas a
variation with respect to the basic multiplier field $\mathrm{\lambda }$
preserves the vector field constraint (\ref{const}).}. In a general case,
when nonzero values of $\mathrm{\lambda }$ are also allowed, it appears
problematic to have stable theory with a positive Hamiltonian (for a
detailed discussion see \cite{vru}). It is worth noting that, though the
Lagrange multiplier term formally breaks gauge invariance in the Lagrangian (%
\ref{lag}), this breaking is in fact reduced to the nonlinear gauge choice (%
\ref{const}). On the other hand, since gauge invariance is no longer
generically assumed, it seems that the vector field constraint (\ref{const})
might be implemented into the general vector field theory (\ref{lag2})
rather than the gauge invariant QED in (\ref{lag}). The point is, however,
that both theories are equivalent once the constraint (\ref{const}) holds.
Indeed, due to a simple structure of vector field polynomial in (\ref{lag2}%
), they lead to practically the same equations of motion in both cases.

One way or another, the constraint (\ref{const}) means in essence that the
vector field $A_{\mu }$ develops the VEV (\ref{vev1}) causing again an
appropriate (time-like or space-like) Lorentz violation at a scale $\mathrm{M%
}$. The point is, however, that in sharp contrast to the nonlinear $\sigma $
model for pions, the nonlinear QED theory, due to gauge invariance in the
starting Lagrangian $L_{QED}$ in (\ref{lag}) or in (\ref{lag2}), ensures
that all the physical Lorentz violating effects turn out to be
non-observable. Actually, as was shown in the tree \cite{nambu} and one-loop
approximations \cite{az}, the nonlinear constraint (\ref{const}) implemented
as a supplementary condition appears in essence as a possible gauge choice
for the vector field $A_{\mu }$, while the $S$-matrix remains unaltered
under such a gauge convention. So, as generally expected, the SLIV inspired
by the nonlinear constraint (\ref{const}), while producing an ordinary
photon as a true Goldstone vector boson ($a_{\mu }$) 
\begin{equation}
A_{\mu }=a_{\mu }+n_{\mu }\sqrt{\mathrm{M}^{2}-n^{2}a^{2}}\text{ , \ }n_{\mu
}a_{\mu }=0\text{ \ }(a^{2}\equiv a_{\mu }a^{\mu })  \label{gol}
\end{equation}%
leaves physical Lorentz invariance intact\footnote{%
Indeed, the nonlinear QED contains a plethora of Lorentz and $CPT$ violating
couplings when it is expressed in terms of the zero photon modes $a_{\mu }$.
However, the contributions of all these couplings to physical processes
completely cancel out among themselves.}. Similar result was also confirmed
for spontaneously broken massive QED \cite{kep}, non-Abelian theories \cite%
{jej} and tensor field gravity \cite{gra} that will be discussed from a
supersymmetry point of view in later sections.

To conclude, the constraint-based emergent gauge theories are in fact
indistinguishable from conventional gauge theories. Their emergent nature
could only be seen when taking the nonlinear gauge condition (\ref{const}).
Any other gauge, e.g. Coulomb gauge, is not in line with emergent picture,
since it breaks Lorentz invariance in an explicit rather than spontaneous
way. As to an observational evidence in favor of emergent theories the only
way for SLIV to cause physical Lorentz violation would appear only if gauge
invariance in these theories were really broken \cite{par} rather than
merely constrained by some gauge condition. This leads us to some general
observation that, in contrast to the spontaneous violation of internal
symmetries, SLIV seems not to necessarily imply a physical breakdown of
Lorentz invariance. Rather, when appearing in a gauge theory framework, this
may ultimately result in a nonlinear gauge choice in an otherwise gauge
invariant and Lorentz invariant theory. In substance, the SLIV ansatz, due
to which the vector field $A_{\mu }(x)$ develops the VEV (\ref{vev1}), may
itself be treated as a pure gauge transformation with a gauge function
linear in coordinates, $\omega (x)=$ $n_{\mu }x^{\mu }\mathrm{M}$. From this
viewpoint gauge invariance in QED leads to the conversion of SLIV into gauge
degrees of freedom of the massless photon emerged. This is what one could
refer to as the generic non-observability of SLIV in QED. Moreover, as was
shown some time ago \cite{cfn}, gauge theories, both Abelian and
non-Abelian, can be obtained by themselves from the requirement of the
physical non-observability of SLIV induced by vector fields rather than from
the standard gauge principle.

We will hereafter refer to this case of the constraint-based models as an
"inactive" SLIV, as opposed to an "active" SLIV leading to physical Lorentz
violation which appears if gauge invariance is explicitly broken, as is
presented later in section 3.5.

\subsection{Models with external vector backgrounds}

Although we are mainly focused here on spontaneous Lorentz violation, it
must not be ruled out that Lorentz invariance might be explicitly, rather
than spontaneously, broken at high energies. This has attracted considerable
attention in recent years as an interesting phenomenological possibility
appearing in direct Lorentz noninvariant extensions of QED and Standard
Model \cite{car, 2, refs, 3}. They are generically regarded as effective
theories being originated from a more fundamental theory at some large scale
probably related to the Planck scale $\mathrm{M}_{P}$. These extensions may
be in a certain measure motivated \cite{ks} by a string theory according to
which an explicit (from a QFT point of view) Lorentz violation might be in
essence a spontaneous Lorentz violation related to hypothetical
tensor-valued fields acquiring non-zero VEVs in some non-perturbative
vacuum. These VEVs appear effectively as a set of external background
constants so that interactions with these coefficients have preferred
spacetime directions in an effective QFT framework. The full SM extension
(SME) \cite{2} is then defined as an effective gauge invariant field theory
obtained when all such Lorentz violating vector and tensor field backgrounds
are contracted term by term with SM (and gravitational) fields. However,
without a completely viable string theory, it is not possible to assign
definite numerical values to these coefficients. Moreover, not to have
disastrous consequences (especially when these coefficients are contracted
with non-conserved currents) one also has to additionally propose that
observable Lorentz violating effects are properly suppressed \cite{car, 2,
refs, 3} that in many cases is a serious theoretical problem. Therefore, one
has in essence a pure phenomenological approach treating the above arbitrary
coefficients as quantities to be bounded in experiments as if they would
simply appear due to explicit Lorentz violation. Actually, in sharp contrast
to the above formulated SLIV\ in a pure QFT framework, there is nothing in
the SME itself that requires that these Lorentz-violation coefficients
emerge due to a process of a spontaneous Lorentz violation. Indeed, neither
the corresponding massless vector (tensor) NG bosons are required to be
generated, nor these bosons have to be associated with photons or any other
gauge fields of SM.

Apart from Lorentz violation in the Standard Model, one can generally think
that the vacuum in quantum gravity may also determine a preferred rest frame
at the microscopic level. If such a frame exists, it must be very much
hidden in low-energy physics since, as was mentioned above, numerous
observations severely limit the possibility of Lorentz violating effects for
the SM fields \cite{car, 2, refs, 3}. However, the constraints on Lorentz
violation in the gravitational sector are generally far weaker. This allows
to introduce a pure gravitational Lorentz violation having no significant
impact on the SM physics. An elegant way being close in spirit to our SLIV
model (\ref{lag}, \ref{gol}) seems to appear in the so called
Einstein-aether theory \cite{ted}. This is in essence a general covariant
theory in which local Lorentz invariance is broken by some vector
\textquotedblleft aether\textquotedblright\ field $u_{\mu }$ defining the
preferred frame. This field is similar to our constrained vector field $%
A_{\mu }$, apart from that this field is taken to be unit $u_{\mu }u^{\mu
}=1 $. It spontaneously breaks Lorentz symmetry down to a rotation subgroup,
just like as our constrained vector field $A_{\mu }$ does it for a timelike
Lorentz violation. So, they both give nonlinear realization of Lorentz
symmetry thus leading to its spontaneous violation and induce the
corresponding Goldstone-like modes. The crucial difference is that, while
modes related to the vector field $A_{\mu }$ are collected into the physical
photon, modes associated with the unit vector field $u_{\mu }$ (one
helicity-0 and two helicity-1 modes) exist by them own appearing in some
effective SM and gravitational couplings. Some of them might disappear being
absorbed by the corresponding spin-connection fields related to local
Lorentz symmetry in the Einstein-aether theory. In any case, while aether
field $u_{\mu }$ can significantly change dispersion relations of fields
involved, thus leading to many gravitational and cosmological consequences
of preferred frame effects, it certainly can not be a physical gauge field
candidate (say, the photon in QED).

\subsection{Supersymmetric models}

While there are many papers in the literature on Lorentz noninvariant
extensions of supersymmetric models (for some interesting ideas, see \cite%
{bs, pos, pos1, lor} and references therein), an emergent gauge theory in a
SUSY context has only recently been introduced \cite{c}. Actually, the
situation was shown to be seriously changed in a SUSY context which
certainly disfavors some emergent models considered above. It appears that,
while the constraint-based models of an inactive SLIV successfully matches
supersymmetry, the composite and potential-based models of an active SLIV
leading to physical Lorentz violation cannot be conceptually realized in the
SUSY context. The reason is that, in contrast to an ordinary vector field
theory where all kinds of polynomial terms $(A_{\mu }A^{\mu })^{n}$ ($%
n=1,2,...$) can be included into the Lagrangian in a Lorentz invariant way,
SUSY theories only admit the bilinear mass term $A_{\mu }A^{\mu }$ in the
vector field potential energy. As a result, without a stabilizing
high-linear (at least quartic, as in (\ref{lag2})) vector field terms, the
potential-based SLIV never occurs in SUSY theories. The same could be said
about composite models as well: the fundamental Lagrangian with multi-fermi
current-current interactions (\ref{4f}) can not be constructed from any
matter chiral superfields. So, all the models considered above, but the
constraint-based models, are ruled out in the SUSY framework and, therefore,
between the two basic SLIV versions, active and inactive, SUSY unambiguously
chooses the inactive SLIV case.

Meanwhile, some efforts have been made \cite{bs, pos, pos1} over the last
decade to construct Lorentz violating operators for matter and gauge fields
interacting with external vector field backgrounds in the supersymmetric QED
and Standard Model. These backgrounds, according to the SME approach \cite{2}
discussed above, are generated by some Lorentz violating dynamics at an
ultraviolet scale of order the Planck scale. As some advantages over the
ordinary SME, it was shown that in the supersymmetric Standard Model the
lowest possible dimension for such operators is five. Therefore, they are
suppressed by at least one power of an ultraviolet energy scale, providing a
possible explanation for the smallness of Lorentz violation and its
stability against radiative corrections. There were classified all possible
dimension five and six Lorentz violating operators in the SUSY QED \cite%
{pos1}, analyzed their properties at the quantum level and described their
observational consequences in this theory. These operators, as was
confirmed, do not induce destabilizing $D$-terms, nor gauge anomaly and the
Chern-Simons term for photons. Dimension-five Lorentz violating operators
were shown to be constrained by low-energy precision measurements at $%
10^{-10}-10^{-5}$ level in units of the inverse Planck scale, while the
Planck-scale suppressed dimension six operators are allowed by observational
data.

Also, it has been constructed the supersymmetric extension of the
Einstein-aether theory \cite{puj} discussed above. It has been found that
the dynamics of the super-aether is somewhat richer than of its non-SUSY
counterpart. In particular, the model possesses a family of inequivalent
vacua exhibiting different symmetry breaking patterns while remaining stable
and ghost free. Interestingly enough, as long as the aether VEV preserves
spatial supersymmetry (SUSY algebra without boosts), the Lorentz breaking
does not propagate into the SM sector at the renormalizable level. The
eventual breaking of SUSY, that must be incorporated in any realistic model,
is unrelated to the dynamics of the aether. It is assumed to come from a
different source characterized by a lower energy scale. However, in spite of
its own merits an important final step which would lead to natural
accommodation of this super-aether model into the supergravity framework has
not yet been done.

\section{ \textbf{Gauge theories emerging from constraints}}

\subsection{An emergence conjecture revised}

Summarizing the current status of models considered above it may seem that
an "emergence level" of an effective theory is decreased when going from the
original composite models to the vector field theoretical ones. At the first
glance, the latters look less fundamental if it is granted that emergent
degrees of freedom (gauge bosons) are necessarily built of more fundamental
degrees of freedom (fermions). However, the compositeness itself hardly is
important for emergent theories and, in essence, one can equally specify the
emergent gauge bosons simply as the NG modes associated with spontaneous
Lorentz violation, no matter they are composite or elementary.

Another seemingly depreciating point might be that the vector field
theoretical models are taken to possess gauge invariance from the outset
(being partial in the potential-based models and full in the
constraint-based ones), whereas in the composite models \cite{bjorken} one
tries to derive it, though this has not yet been really achieved. However,
the most important side of the nonlinear vector field constraint (\ref{const}%
) was shown \cite{cj} (see also \cite{cfn2, c3}) to be that one does not
need to specially postulate the starting gauge invariance. Normally, one can
start in the framework of an arbitrary relativistically invariant Lagrangian
which is proposed only to possess some global internal symmetry.
Nonetheless, looking for the theories which are compatible with the vector
field constraint (\ref{const}) one inevitably comes to gauge invariance.
Namely, gauge invariance in such theories has to appear in essence as a
response of an interacting field system\textit{\ }to putting the covariant
constraint (\ref{const}) on its dynamics, provided that we allow parameters
in the corresponding Lagrangian density to be adjusted so as to ensure
self-consistency without losing too many degrees of freedom. Otherwise, a
given field system could get unphysical in a sense that a superfluous
reduction in the number of degrees of freedom would make it impossible to
set the required initial conditions in the appropriate Cauchy problem.
Namely, it would be impossible to specify arbitrarily the initial values of
the vector and other field components involved, as well as the initial
values of the momenta conjugated to them. Furthermore, in quantum theory, to
choose self-consistent equal-time commutation relations would also become
impossible \cite{ogi3}.

Let us dwell upon this point in more detail. Conventionally, while a
standard variation principle requires the equations of motion to be
satisfied, for a general 4-vector field $A_{\mu }$ it is still possible (in
contrast to scalar and fermion fields) to eliminate one extra component in
order to describe a pure spin-1 particle by imposing a supplementary
condition. Typically, this is covariantly achieved by taking the divergence
from a general vector field equation of motion. In the massive vector field
case there are three physical spin-1 states to be described by the $A_{\mu }$
field. Similarly in the massless vector field case, although there are only
two physical (transverse) photon spin states, one cannot construct a
massless 4-vector field $A_{\mu }$ as a linear combination of creation and
annihilation operators for helicity $\pm 1$ states in a relativistically
covariant way, unless one fictitious state is added \cite{GLB}. So, in both
the massive and massless vector field cases, only one component of the $%
A_{\mu }$ field may be eliminated and still preserve physical Lorentz
invariance. Now, once the SLIV constraint (\ref{const}) is imposed, it is
therefore not possible to satisfy another supplementary condition since this
would superfluously restrict the number of degrees of freedom for the vector
field. To avoid this, its equation of motion should be automatically
divergenceless that is only possible in the gauge invariant theory. Thus,
due to spontaneous Lorentz violation determined by the constraint (\ref%
{const}), being the only possible covariant and holonomic vector field
constraint, the theory has to acquire on its own a gauge-type invariance,
which gauges the starting global symmetry of the interacting vector and
matter fields involved. In such a way, one comes to the\textbf{\ gauge
symmetry emergence (GSE)\ }conjecture:

\textit{Let there be given an interacting field system containing some
vector field (or vector field multiplet) }$A_{\mu }$\textit{\ together with
fermion (}$\psi $), \textit{scalar (}$\phi $\textit{) and other matter
fields in an arbitrary relativistically invariant Lagrangian\ }$L(A_{\mu
},\psi ,\phi ,...)$\textit{\ which possesses only global Abelian or
non-Abelian internal symmetry }$G$\textit{. Suppose that an underlying
relativistic invariance of this field system is spontaneously broken in
terms of t\textit{he} "length-fixing" covariant constraint put on vector
fields,} $A_{\mu }A^{\mu }=n^{2}\mathrm{M}^{2}$. \textit{If this constraint
is preserved under the time development given by the field equations of
motion, then in order to be\textit{\ protected }from further reduction in
degrees of freedom this system will modify its global symmetry }$G$\textit{\
into a local symmetry }$G_{loc},$ \textit{that will in turn convert the
vector field constraint itself into a gauge condition thus virtually
resulting in gauge invariant and Lorentz invariant theory.}

So, the nonlinear SLIV condition (\ref{const}), due to which true vacuum in
the theory is chosen and massless gauge fields are generated, may provide a
dynamical setting for all underlying internal symmetries involved through
the GSE conjecture. One might think that the length-fixing vector field
constraint (\ref{const}) itself seems not to be especially singled out in
the present context. Actually, it looks like that the GSE conjecture might
be equally formulated for any type of covariant constraint. However, as we
argue later in section 3.4, the SLIV constraint appears to be the only one
whose application leads to a full conversion of an internal global symmetry $%
G$\ into a local symmetry $G_{loc}$ that forces a given field system to
remain always physical.

\subsection{Emergent Abelian gauge invariance}

To see how technically a global internal symmetry may be converted into a
local one, let us consider in detail the question of consistency of a
possible constraint for a general 4-vector field $A_{\mu }$ with its
equation of motion in an Abelian symmetry case, $G=U(1)$. In the presence of
the SLIV constraint $C(A)=A_{\mu }A^{\mu }-n^{2}\mathrm{M}^{2}=0$ (\ref%
{const}), it follows that the equations of motion can no longer be
independent. The important point is that, in general, the time development
would not preserve the constraint. So the parameters in the Lagrangian have
to be chosen in such a way that effectively we have one less equation of
motion for the vector field. This means that there should be some
relationship between all the vector and matter field Eulerians ($E_{A}$, $%
E_{\psi }$, ...) involved\footnote{%
Hereafter, the notation $E_{A}$ stands for the vector field Eulerian $%
(E_{A})^{\mu }\equiv \partial L/\partial A_{\mu }-\partial _{\nu }[\partial
L/\partial (\partial _{\nu }A_{\mu })].$ We use similar notations for other
field Eulerians as well.}. Such a relationship can quite generally be
formulated as a functional - but by locality just a function - of the
Eulerians, $F(E_{A},E_{\psi },...)$, being put equal to zero at each
spacetime point with the configuration space restricted by the constraint $%
C(A)=0$: 
\begin{equation}
F(C=0;\text{ \ }E_{A},E_{\psi },...)=0\text{ .}  \label{FF}
\end{equation}%
This relationship must satisfy the same symmetry requirements of Lorentz and
translational invariance, as well as all the global internal symmetry
requirements, as the general starting Lagrangian does. This Lagrangian is
supposed to also include the standard Lagrange multiplier term with the
field $\mathrm{\lambda }(x)$ 
\begin{equation}
L^{tot}(A,\psi ,...,\mathrm{\lambda })=L(A,\psi ,...)-\frac{\mathrm{\lambda }%
}{2}\left( A_{\mu }A^{\mu }-n^{2}\mathrm{M}^{2}\right)  \label{44}
\end{equation}%
the variation under which results in the above constraint $C(A)=0$. In fact,
the relationship (\ref{FF}) is used as the basis for an emergence of gauge
symmetries in the SLIV constrained vector field theories \cite{cfn2, c3}.
Note that, while the Lagrange multiplier field is presented in the total
Lagrangian $L^{tot}$, it does not appear in the equation of motions of
vector field determined by the Eulerian $E_{A}$ in the expression (\ref{FF}%
). This is naturally occurred, as we explained in the previous section, if
initial values for all fields involved are chosen so as to restrict their
phase space to values with a vanishing multiplier function $\mathrm{\lambda }%
(x)$ (see also the footnote$^{\text{4}}$).

Let us now consider a \textquotedblleft Taylor expansion" of the function $F$
expressed as a linear combination of terms involving various field
combinations multiplying or derivatives acting on the Eulerians. We are
taking for simplicity only one matter (say, fermion) field $\psi $ in the
model. The constant term in this expansion is of course zero since the
relation (\ref{FF}) must be trivially satisfied when all the Eulerians
vanish, i.e. when the equations of motion are satisfied. We consider just
the terms containing field combinations (and derivatives) with the lowest
mass dimension 4, corresponding to the Lorentz invariant expressions 
\begin{equation}
\partial _{\mu }(E_{A})^{\mu },\text{ }A_{\mu }(E_{A})^{\mu },\text{ }%
E_{\psi }\psi ,\text{ }\overline{\psi }E_{\overline{\psi }}.  \label{fff}
\end{equation}%
to eventually have an emergent gauge theory at a renormalizible level. The
higher dimension terms we will discuss later in section 3.4. Now, under the
assumption that the SLIV constraint (\ref{const}) is preserved under the
time development given by the equations of motion, we show how gauge
invariance of the physical Lagrangian $L(A,\psi )$ in (\ref{44}) is
established. A conventional variation principle applied to the total
Lagrangian\ $L^{tot}(A,\psi ,\mathrm{\lambda })$ requires the following
equations of motion for the vector field $A_{\mu }$ and the auxiliary field $%
\mathrm{\lambda }$ to be satisfied%
\begin{equation}
(E_{A})^{\mu }=0\text{ , \ \ \ }C(A)=A_{\mu }A^{\mu }-n^{2}\mathrm{M}^{2}=0%
\text{\ }.  \label{em}
\end{equation}%
However, in accordance with general arguments given above, the existence of
five equations for the 4-component vector field $A^{\mu }$ (one of which is
the constraint) means that not all of the vector field Eulerian components
can be independent. Therefore, there must be a relationship of the form
given in equation (\ref{FF}). When being expressed as a linear combination
of the Lorentz invariant terms (\ref{fff}), this equation leads to the
identity between the vector and matter field Eulerians of the following type 
\begin{equation}
\partial _{\mu }(E_{A})^{\mu }=itE_{\psi }\psi -it\overline{\psi }E_{%
\overline{\psi }}.  \label{div}
\end{equation}%
where $t$ is some constant\footnote{%
Note the term proportional to the vector field itself, $A_{\mu }(E_{A})^{\mu
}$, which would correspond to the selfinteraction of vector field, is absent
in the identity (\ref{div}). In presence of this term the transformations of
the vector field given below in (\ref{transss}) would be changed to $\delta
A_{\mu }=\partial _{\mu }\omega +c\omega A_{\mu }$. The point is, however,
that these transformations cannot in general form a group unless the
constant $c$ vanishes, as can be readily confirmed by constructing the
corresponding Lie bracket operation for two successive vector field
variations. We shall see later that non-zero $c$-type coefficients
necessarilly appear in the non-Abelian internal symmetry case, resulting
eventually in a emergent gauge invariant Yang-Mills theory.}. This identity
immediately signals about invariance of the basic Lagrangian $L(A,\psi )$ in
(\ref{44}) under vector and fermion field local $U(1)$ transformations whose
infinitesimal form is given by 
\begin{equation}
\delta A_{\mu }=\partial _{\mu }\omega ,\text{ \ \ }\delta \psi \text{\ }%
=it\omega \psi \text{ .}  \label{transss}
\end{equation}%
Here $\omega (x)$ is an arbitrary function, only being restricted by the
requirement to conform with the nonlinear constraint (\ref{const})%
\begin{equation}
(A_{\mu }+\partial _{\mu }\omega )(A^{\mu }+\partial ^{\mu }\omega )=n^{2}%
\mathrm{M}^{2}\text{ .}  \label{transs}
\end{equation}%
Conversely, the identity (\ref{div}) follows from the invariance of the
physical Lagrangian $L(A,\psi )$ under the transformations (\ref{trans}).
Indeed, both direct and converse assertions are particular cases of
Noether's second theorem \cite{n}.

So, we have shown how the choice of a vacuum conditioned by the SLIV
constraint (\ref{const}) enforces the choice of the parameters in the
starting Lagrangian $L^{tot}(A,\psi ,\mathrm{\lambda })$, so as to convert
the starting global $U(1)$ charge symmetry into a local one, thus
demonstrating an emergence of gauge symmetry (\ref{transss}) that allows the
emerged Lagrangian to be determined in full. For a theory with
renormalizable couplings, it is in fact the conventional QED Lagrangian (\ref%
{lag}) extended by the Lagrange multiplier term%
\begin{equation}
L^{\mathfrak{em}}(A,\psi ,\mathrm{\lambda })=L_{QED}(A,\psi )-\frac{\mathrm{%
\lambda }}{2}\left( A_{\mu }A^{\mu }-n^{2}\mathrm{M}^{2}\right)  \label{333}
\end{equation}%
which provides the SLIV constraint (\ref{const}) imposed on the vector field 
$A_{\mu }$.

\subsection{Non-Abelian gauge fields as pseudo-Goldstone modes}

We still have only considered a single vector field case with an underlying
global $U(1)$ symmetry. However, an extension to a theory possessing from
the outset some global non-Abelian symmetry $G$ is quite straightforward 
\cite{cfn2, c3}. Suppose that this theory contains an adjoint vector field
multiplet \textbf{\ }$\boldsymbol{A}_{\mu }^{p}$ and some fermion matter
field multiplet $\boldsymbol{\psi }$ \ belonging to one of irreducible
representations of $G$ given by matrices $t^{p}$%
\begin{equation}
\lbrack t^{p},t^{q}]=if^{pqr}t^{r}\text{ , \ }Tr(t^{p}t^{q})=\delta ^{pq}%
\text{ }\ \text{(}p,q,r=0,1,...,\Upsilon -1\text{)\ }  \label{22}
\end{equation}%
where $f^{pqr}$ stand structure constants, while $\Upsilon $ is a dimension
of the $G$ group. The corresponding Lagrangian $\mathbb{L}^{tot}(\boldsymbol{%
A}_{\mu },\boldsymbol{\psi },\mathbf{\lambda })$ is supposed to also include
the standard Lagrange multiplier term with the field function $\mathbf{%
\lambda }(x)$ 
\begin{equation}
\mathbb{L}^{tot}(\boldsymbol{A}_{\mu },\boldsymbol{\psi },\mathbf{\lambda })=%
\mathbb{L}(\boldsymbol{A}_{\mu },\boldsymbol{\psi })-\text{ }\frac{\mathbf{%
\lambda }}{2}(\boldsymbol{A}_{\mu }^{p}\boldsymbol{A}^{p\mu }-\boldsymbol{n}%
^{2}\mathbf{M}^{2})\text{ , \ }\boldsymbol{n}^{2}\equiv \boldsymbol{n}_{\mu
}^{p}\boldsymbol{n}^{p\mu }=\pm 1  \label{33}
\end{equation}%
the variation under which results in the vector field length-fixing
constraint%
\begin{equation}
C(\boldsymbol{A})=\boldsymbol{A}_{\mu }^{p}\boldsymbol{A}^{p\mu }-%
\boldsymbol{n}^{2}\mathbf{M}^{2}=0  \label{constt}
\end{equation}%
(where $\boldsymbol{n}_{\mu }^{p}$ stands now for some properly-oriented
`unit' rectangular matrix, see below). The need to preserve the constraint $%
C(\boldsymbol{A})=0$ with time implies that the equations of motion for the
vector fields $\boldsymbol{A}_{\mu }^{p}$ cannot be all independent. As a
result, the so-called "emergence identity" analogous to the identity (\ref%
{div}) inevitably occurs 
\begin{equation}
\partial _{\mu }(\mathbb{E}_{\boldsymbol{A}}^{p})^{\mu }=f^{pqr}\boldsymbol{A%
}_{\mu }^{q}(\mathbb{E}_{\boldsymbol{A}}^{r})^{\mu }+\mathbb{E}_{\boldsymbol{%
\psi }}(it^{p})\boldsymbol{\psi }+\overline{\boldsymbol{\psi }}(-it^{p})%
\mathbb{E}_{\overline{\boldsymbol{\psi }}}\text{ .}  \label{id11}
\end{equation}%
An identification of the coefficients of the Eulerians on the right-hand
side of the identity (\ref{id11}) with the structure constants $f^{pqr}$ and
generators $t^{p}$(\ref{22}) of the group $G$ is quite transparent. This
simply follows from the fact the right-hand side of this identity must
transform in the same way as the left-hand side, which transforms as the
adjoint representation of $G$. Note that, in contrast to the Abelian case,
the term proportional to the vector field multiplet $\boldsymbol{A}_{\mu
}^{p}$ itself which corresponds to a self-interaction of non-Abelian vector
fields, also appears in the identity (\ref{id11}). Again, Noether's second
theorem \cite{n} can be applied directly to this identity in order to derive
the gauge invariance of the Lagrangian $\mathbb{L}(\boldsymbol{A}_{\mu },%
\boldsymbol{\psi })$ in (\ref{33}). Indeed, with the constraint (\ref{constt}%
) implied, the $\mathbb{L}(\boldsymbol{A}_{\mu },\boldsymbol{\psi })$ tends
to be invariant under vector and fermion field local transformations having
the infinitesimal form 
\begin{equation}
\delta \boldsymbol{A}_{\mu }^{p}=\partial _{\mu }\omega ^{p}+f^{pqr}%
\boldsymbol{A}_{\mu }^{q}\omega ^{r},\text{ \ \ }\delta \boldsymbol{\psi }%
\text{\ }=(it^{p})\omega ^{p}\boldsymbol{\psi },\text{ \ \ }\delta \overline{%
\boldsymbol{\psi }}\text{\ }=\overline{\boldsymbol{\psi }}(-it^{p})\omega
^{p}.  \label{trans1}
\end{equation}%
For a theory with renormalizable coupling constants, this emergent gauge
symmetry leads to the conventional Yang-Mills type Lagrangian 
\begin{equation}
\mathbb{L}^{\mathfrak{em}}(\boldsymbol{A}_{\mu },\boldsymbol{\psi },\mathbf{%
\lambda })=\mathbb{L}_{YM}(\boldsymbol{A}_{\mu },\boldsymbol{\psi })-\text{ }%
\frac{\mathbf{\lambda }}{2}(\boldsymbol{A}_{\mu }^{p}\boldsymbol{A}^{p\mu }-%
\boldsymbol{n}^{2}\mathbf{M}^{2})  \label{nab1}
\end{equation}%
where we also include the corresponding Lagrange multiplier term. As in the
above Abelian case, this term does not contribute to the vector field
equation of motion in the identity (\ref{id11}).

Now let us turn to the spontaneous Lorentz violation which is caused by the
nonlinear vector field constraint (\ref{constt}) determined by the Lagrange
multiplier term in (\ref{nab1}). Although the Lagrangian $\mathbb{L}^{%
\mathfrak{em}}(\boldsymbol{A}_{\mu },\boldsymbol{\psi },\mathbf{\lambda })$
only has an $SO(1,3)\times G$ invariance, the last term in it possesses a
much higher accidental symmetry $SO(\Upsilon ,3\Upsilon )$ according to the
dimension $\Upsilon $ of the adjoint representation of $G$ to which the
vector fields $\boldsymbol{A}_{\mu }^{p}$ belong. This symmetry is indeed
spontaneously broken at a scale $\mathbf{M}$ 
\begin{equation}
<\boldsymbol{A}_{\mu }^{p}(x)>\text{ }=\boldsymbol{n}_{\mu }^{p}\mathbf{M}%
\text{ }  \label{5'}
\end{equation}%
with the vacuum direction determined now by the `unit' rectangular matrix $%
\boldsymbol{n}_{\mu }^{p}$ which describes simultaneously both of the
non-Abelian SLIV cases, time-like 
\begin{equation}
SO(\Upsilon ,3\Upsilon )\rightarrow SO(\Upsilon -1,3\Upsilon \mathbb{)}
\label{ss}
\end{equation}%
or space-like 
\begin{equation}
SO(\Upsilon ,3\Upsilon )\rightarrow SO(\Upsilon ,3\Upsilon -1)  \label{sss}
\end{equation}%
depending on the sign of $\boldsymbol{n}^{2}\equiv \boldsymbol{n}_{\mu }^{p}%
\boldsymbol{n}^{p\mu }=\pm 1$. In both cases this matrix has only one
non-zero element, subject to the appropriate $SO(1,3)$ and (independently) $%
G $ rotations. They are, specifically, $\boldsymbol{n}_{0}^{0}$ or $%
\boldsymbol{n}_{3}^{0}$ provided that the vacuum expectation value (\ref{5'}%
) is developed along the $p=0$ direction in the internal space and along the 
$\mu =0$ or $\mu =3$ direction respectively in the ordinary four-dimensional
spacetime.

As was argued in \cite{jej, cj}, side by side with one true vector Goldstone
boson corresponding to the spontaneous violation of the actual $%
SO(1,3)\otimes G$ symmetry of the Lagrangian $\mathbb{L}^{\mathfrak{em}}(%
\boldsymbol{A}_{\mu },\boldsymbol{\psi },\mathbf{\lambda })$, the $\Upsilon
-1$ pseudo-Goldstone vector bosons (PGB) related to the breakings (\ref{ss}, %
\ref{sss}) of the accidental symmetry $SO(\Upsilon ,3\Upsilon )$ of the
constraint (\ref{constt}) per se are also produced\footnote{%
Note that in total there appear $4\Upsilon -1$ pseudo-Goldstone modes,
complying with the number of broken generators of $SO(\Upsilon ,3\Upsilon )$%
. From these $4\Upsilon -1$ pseudo-Goldstone modes, $3\Upsilon $ modes
correspond to the $\Upsilon $ three-component vector states as will be shown
below, while the remaining $\Upsilon -1$ modes are scalar states which will
be excluded from the theory.}. Remarkably, in contrast to the familiar
scalar PGB case \cite{GL}, the vector PGBs remain strictly massless being
protected by the simultaneously generated non-Abelian gauge invariance.
Together with the above true vector Goldstone boson, they also come into
play properly completing the whole gauge multiplet of the internal symmetry
group $G$ taken.

After the explicit use of this constraint (\ref{constt}), which virtually
appears as a single condition put on the vector field multiplet $\boldsymbol{%
A}_{\mu }^{p}$, one can identify the pure Goldstone field modes $\boldsymbol{%
a}_{\mu }^{p}$ as follows 
\begin{equation}
\text{\ \ }\boldsymbol{A}_{\mu }^{p}=\boldsymbol{a}_{\mu }^{p}+\boldsymbol{n}%
_{\mu }^{p}\sqrt{\mathbf{M}^{2}-\boldsymbol{n}^{2}\boldsymbol{a}^{2}}\text{ }%
,\text{ \ }\boldsymbol{n}_{\mu }^{p}\boldsymbol{a}^{p\mu }\text{\ }=0\text{
\ \ \ }(\boldsymbol{a}^{2}\equiv \boldsymbol{a}_{\mu }^{p}\boldsymbol{a}%
^{p\mu }).  \label{supp}
\end{equation}%
There is also an effective \textquotedblleft Higgs" mode $\boldsymbol{n}%
_{\mu }^{p}(\mathbf{M}^{2}-\boldsymbol{n}^{2}\boldsymbol{a}^{2})^{1/2}$
determined by the SLIV constraint. Note that, apart from the pure vector
fields, the general zero modes $\boldsymbol{a}_{\mu }^{p}$ contain $\Upsilon
-1$ scalar modes, $\boldsymbol{a}_{0}^{p^{\prime }}$ or $\boldsymbol{a}%
_{3}^{p^{\prime }}$ ($p^{\prime }=1,...,\Upsilon -1$), for the time-like ($%
\boldsymbol{n}_{\mu }^{p}=n_{0}^{0}g_{\mu 0}\delta ^{p0}$) or space-like ($%
\boldsymbol{n}_{\mu }^{p}=n_{3}^{0}g_{\mu 3}\delta ^{p0}$) SLIV,
respectively. They can be eliminated from the theory, if one imposes
appropriate supplementary conditions on the $\Upsilon -1$ fields $%
\boldsymbol{a}_{\mu }^{p}$ which are still free of constraints. Using their
overall orthogonality (\ref{supp}) to the physical vacuum direction $%
\boldsymbol{n}_{\mu }^{p}$, one can formulate these supplementary conditions
in terms of a general axial gauge for the entire $\boldsymbol{a}_{\mu }^{p}$
multiplet 
\begin{equation}
n\cdot \boldsymbol{a}^{p}\equiv n_{\mu }\boldsymbol{a}^{p\mu }=0,\text{ \ }%
p=0,1,...,\Upsilon -1.  \label{supp'}
\end{equation}%
Here $n_{\mu }$ is the unit Lorentz vector being analogous to the vector
introduced in the Abelian case, which is now oriented in Minkowskian
spacetime so as to be "parallel" to the vacuum unit $\boldsymbol{n}_{\mu
}^{p}$ matrix. This matrix can be taken hereafter in the "two-vector" form 
\begin{equation}
\boldsymbol{n}_{\mu }^{p}=n_{\mu }\boldsymbol{\epsilon }^{p}\text{ },\text{ }%
\boldsymbol{\epsilon }^{p}\boldsymbol{\epsilon }^{p}=1  \label{vec}
\end{equation}%
where $\boldsymbol{\epsilon }^{p}$ is unit $G$ group vector belonging to its
adjoin representation. As a result, in addition to the \textquotedblleft
Higgs" mode excluded earlier by the above orthogonality condition (\ref{supp}%
), all the other scalar fields are eliminated. Consequently only the pure
vector fields, $\boldsymbol{a}_{i}^{p}$ ($i=1,2,3$ ) or $\boldsymbol{a}_{\mu
^{\prime }}^{p}$ ($\mu ^{\prime }=0,1,2$), for time-like or space-like SLIV
respectively, are left in the theory. Clearly, the components $\boldsymbol{a}%
_{i}^{p=0}$ and $\boldsymbol{a}_{\mu ^{\prime }}^{p=0}$ correspond to the
true Goldstone boson, for each type of SLIV respectively, while all the
other (for $p=1,...,\Upsilon -1$) are vector PGBs. Substituting the
parameterization (\ref{supp}) into the emergent Lagrangian (\ref{nab1}) and
expanding the square root in powers of $\boldsymbol{a}^{2}/\mathbf{M}^{2}$,
one is led to a highly nonlinear theory in terms of the zero vector modes $%
\boldsymbol{a}_{\mu }^{p}$ which contains a variety of Lorentz and CPT
violating couplings. However, as in Abelian symmetry case, they do not lead
to physical Lorentz violation effects which turn out to be strictly
cancelled among themselves \cite{jej}, thus giving one more example of an
inactive SLIV.

\subsection{Constraints inducing and uninducing gauge invariance}

We now turn to a question that naturally arises : whether the length-fixing
vector field constraints (\ref{const}, \ref{constt}), both for Abelian and
non-Abelian symmetry case, are of fundamental importance for an emergence
conjecture. It seems that the basic relations between all fields Eulerians,
called above the `emergence identities' (\ref{div}, \ref{id11}), might occur
for any type of covariant constraints introduced through the corresponding
Lagrange multiplier terms. On the other hand, if one keeps in mind the
minimal single-field constraints there are only two possible covariant
constraints for vector fields in a relativistically invariant theory: the
holonomic SLIV constraints (\ref{const}, \ref{constt}) and the non-holonomic
one, known as the Lorentz condition%
\begin{equation}
C^{\prime }(A)=\partial _{\mu }A^{\mu }=0\text{, \ \ \ }C^{\prime }(%
\boldsymbol{A})=\partial _{\mu }\boldsymbol{A}^{p\mu }=0\text{ \ \ \ \ }
\label{nh}
\end{equation}%
for Abelian and non-Abelian vector fields, respectively (the index $p$
enumerates the $G$ group generators). In general, of course, many
non-minimal covariant constraints are also possible. However, as we argue
below, just the SLIV constraints (\ref{const}, \ref{constt}) seem to push
the origin of gauge invariance in a theory so as to provide a sufficient
number of degrees of freedom\ for a physical field system evolved over time.
Other covariant constraints, when being put on the fields, will lead, at
best, to partial gauge invariance.

We consider a general quantum field theory where the vector fields, by them
own or together with matter fields, are subject some covariant constraint(s)
whose precise form is yet unknown. Rather than postulate this form in terms
of the SLIV constraints (\ref{const}, \ref{constt}), as we have done in
previous sections, let us try to derive them. We suppose that such
constraints could be determined in general by the underlying Lagrangian
itself rather than introduced from outside through some Lagrange multiplier
terms. Let there be given an interacting field system containing vector
field(s) $A_{\mu }$\ together with fermion ($\psi $), scalar ($\phi $) and
other matter fields in a relativistically invariant Lagrangian $L(A_{\mu
},\psi ,\phi ,...)$ which only possesses global Abelian or non-Abelian
symmetry $G$.\textit{\ }Suppose that the Lagrangian $L$\ is separated into
two parts, $L=L_{g}+\widetilde{L}$, which we call the generic and
constraint-bearing ones, respectively. We assume that all possible
constraint(s) which can be put on the given field system are completely
determined by the variation of the Lagrangian $\widetilde{L}$ that specifies
some extra source current $J_{\mu }=(\widetilde{E}_{A})_{\mu }$ for vector
field $A_{\mu }$. We show below that in order to remain the given field
system physical this current has to be vanished or conserved that, in turn,
makes the generic Lagrangian $L_{g}$\ to become gauge invariant.

For the sake of generality, we consider the non-Abelian symmetry case (\ref%
{22}) writing the total Lagrangian in an appropriate notation taken above, $%
\mathbb{L=L}_{g}+\widetilde{\mathbb{L}}$. We suppose that vector field
multiplet $\boldsymbol{A}_{\mu }^{p}$\ belongs to an adjoint representation
of a group $G$ with structure constants $f^{pqr}$, while matter fields (we
leave only fermion fields, for simplicity) transform according to some
representation given by matrices $(t^{p})_{j}^{i}$. Consider first the case
when the extra source current for vector fields $\boldsymbol{A}_{\mu }^{p}$
vanishes, 
\begin{equation}
\mathbb{J}_{\mu }^{p}=(\widetilde{\mathbb{E}}_{\boldsymbol{A}}^{p})_{\mu }=0%
\text{ , }p=0,1,...,\Upsilon -1\text{ .}  \label{ai}
\end{equation}%
This allegedly happens due to the appropriately restricted vector field
configurations rather than vanishing coupling constants in the Lagrangian $%
\widetilde{\mathbb{L}}$. One can see, however, that such conditions
eliminate too many vector field degrees of freedom. Namely, $4\Upsilon $
degrees appear to be eliminated, whereas only $\Upsilon $ degrees - one for
each vector field specie - may be excluded. Additional constraints could
also appear for matter fields, if they are contained in the Lagrangian $%
\widetilde{\mathbb{L}}$. This means that for these constraints to be
admissible, only a special class of the constraint-bearing Lagrangians $%
\widetilde{\mathbb{L}}$ has to be taken. Actually, the only way to proceed,
as one may readily confirm, could be the case if the Lagrangian $\widetilde{%
\mathbb{L}}$ would depend on all the fields involved only through the
`length squared' invariants $\boldsymbol{A}_{\mu }^{p}\boldsymbol{A}^{p\mu }$%
, $\overline{\boldsymbol{\psi }}^{i}\boldsymbol{\psi }_{i}$, and so on. This
would mean that in the minimal case with the lowest mass dimension couplings
the Lagrangian $\widetilde{\mathbb{L}}$ only contains a conventional
four-order polynomial in vector field $\boldsymbol{A}_{\mu }^{p}$ 
\begin{equation}
\widetilde{\mathbb{L}}_{\min }=-\text{ }\frac{\mathbf{\lambda }_{c}}{4}(%
\boldsymbol{A}_{\mu }^{p}\boldsymbol{A}^{p\mu }-\boldsymbol{n}^{2}\mathbf{M}%
^{2})^{2}  \label{l}
\end{equation}%
where $\mathbf{\lambda }_{c}$ and $\boldsymbol{n}^{2}\mathbf{M}^{2}$ are the
corresponding vector field parameters ($\boldsymbol{n}^{2}=\pm 1$). In
general, there could be, of course, a variety of high-dimensional
vector-vector and vector-fermion couplings of type 
\begin{equation}
(\boldsymbol{A}_{\mu }^{p}\boldsymbol{A}^{p\mu })^{k},\text{ }k\geq 3;\text{ 
}(\boldsymbol{A}_{\mu }^{p}\boldsymbol{A}^{p\mu })^{l}(\overline{\boldsymbol{%
\psi }}^{i}\boldsymbol{\psi }_{i})^{m},\text{ }l\geq 0,\text{ }m\geq 1
\label{ll}
\end{equation}%
and so forth, being properly suppressed by some high scale mass(es). This
structure of the Lagrangian $\widetilde{\mathbb{L}}$ provides the following
expressions for vector and fermion field Eulerians \ 
\begin{equation}
\mathbb{J}_{\mu }^{p}=(\widetilde{\mathbb{E}}_{\boldsymbol{A}}^{p})^{\mu }=2%
\boldsymbol{A}^{p\mu }\frac{\partial \widetilde{\mathbb{L}}}{\partial (%
\boldsymbol{A}_{\mu }^{q}\boldsymbol{A}^{q\mu })}\text{ , \ }\widetilde{%
\mathbb{E}}_{\boldsymbol{\psi }}(it^{p})\boldsymbol{\psi }=\overline{%
\boldsymbol{\psi }}(it^{p})\widetilde{\mathbb{E}}_{\overline{\boldsymbol{%
\psi }}}\text{ .}  \label{r}
\end{equation}%
the first of which actually reduces all constraints (\ref{ai}) to the single
one%
\begin{equation}
\partial \widetilde{\mathbb{L}}/\partial (\boldsymbol{A}_{\mu }^{q}%
\boldsymbol{A}^{\mu q})=0\text{ ,}  \label{qqq}
\end{equation}%
while the second one is satisfied automatically. As a result, for the
minimal Lagrangian $\widetilde{\mathbb{L}}_{\min }$ (\ref{l}) the condition (%
\ref{qqq}) immediately leads to the SLIV constraint (\ref{constt}). Now,
just as in the previous section, assuming that this constraint is preserved
under the time development given by the equations of motion, the so-called
"emergence identity", analogous to identity (\ref{id11}), inevitably occurs%
\begin{equation}
\partial _{\mu }(\mathbb{E}_{\boldsymbol{A}}^{p}+\widetilde{\mathbb{E}}_{%
\boldsymbol{A}}^{p})^{\mu }=f^{pqr}\boldsymbol{A}_{\mu }^{q}(\mathbb{E}_{%
\boldsymbol{A}}^{r}+\widetilde{\mathbb{E}}_{\boldsymbol{A}}^{r})^{\mu }+(%
\mathbb{E}_{\boldsymbol{\psi }}+\widetilde{\mathbb{E}}_{\boldsymbol{\psi }%
})(it^{p})\boldsymbol{\psi }+\overline{\boldsymbol{\psi }}(-it^{p})(\mathbb{E%
}_{\overline{\boldsymbol{\psi }}}+\widetilde{\mathbb{E}}_{\overline{%
\boldsymbol{\psi }}})  \label{oii}
\end{equation}%
where the Eulerians for vector and fermion fields are generated by both
Lagrangians $\mathbb{L}_{g}$ and $\widetilde{\mathbb{L}}$, respectively. Due
to constraints taken (\ref{ai}) and equation for fermion Eulerians in (\ref%
{r}), all the Eulerians generated by the constraint-bearing Lagrangian $%
\widetilde{\mathbb{L}}$ disappear, so that only the generic Lagrangian $%
\mathbb{L}_{g}$ contributes to both sides of this identity. This implies
according to Noether's second theorem \cite{n} that the generic Lagrangian $%
\mathbb{L}_{g}$ is in fact gauge invariant. As to the constraint-bearing
Lagrangian $\widetilde{\mathbb{L}}$, it may only contain some constant term,
and also 2-fermi and multi-fermi interaction terms. They appear as soon as
the constraint equation (\ref{qqq}) is solved with respect to $\boldsymbol{A}%
_{\mu }^{q}\boldsymbol{A}^{\mu q}$ which then substituted back into the $%
\widetilde{\mathbb{L}}$ (\ref{l}, \ref{ll})\footnote{%
Such substitution is in principle an allowed procedure since virtually it
does not change the equations of motion of the fields involved.}. Actually,
the Lagrangian $\widetilde{\mathbb{L}}$ also appears to be gauge invariant
likewise the generic Lagrangian $\mathbb{L}_{g}$ (though the constraint (\ref%
{qqq}) itself breaks gauge invariance). For a minimal Lagrangian $\widetilde{%
\mathbb{L}}_{\min }$ (\ref{l}) the theory completely coincides with the
above SLIV constraint case given by the Lagrangian (\ref{nab1}) provided
that the constraint (\ref{qqq}) in its final form (\ref{constt}) is also
included through an appropriate Lagrange multiplier term. Remarkably,
symmetry of the constraint (\ref{qqq}) uniquely established above from the
requirement not to have too many degrees of freedom eliminated is much
higher than symmetry of the whole Lagrangian (\ref{nab1}). This, as we could
see in the previous section, allows to treat non-Abelian gauge fields as
pseudo-Goldstone bosons.

Let us now turn to the nonzero extra vector field source current $\mathbb{J}%
^{p\mu }$\ which is only required to be conserved%
\begin{equation}
\partial _{\mu }\mathbb{J}^{p\mu }=\partial _{\mu }(\widetilde{\mathbb{E}}%
_{A}^{p})^{\mu }=0\text{ .}  \label{66}
\end{equation}%
that gives in principle a sufficient number of constraints (one for each
vector field specie, $p=0,1,...,\Upsilon -1$). We start deriving the
divergenceless conditions for the equations of motion of the vector fields $%
\boldsymbol{A}_{\mu }^{p}$. Indeed, varying the total Lagrangian $\mathbb{L=L%
}_{g}+\widetilde{\mathbb{L}}$ and taking 4-divergence from the corresponding
vector field Eulerians one has 
\begin{equation}
\partial _{\mu }(\mathbb{E}_{\boldsymbol{A}}^{p})^{\mu }+\partial _{\mu }(%
\widetilde{\mathbb{E}}_{\boldsymbol{A}}^{p})^{\mu }=0\text{ .}  \label{7}
\end{equation}%
Next, since no other constraints than the proposed current conservation (\ref%
{66}) are admissible the first 4-diveregnce term in equation (\ref{7}) has
to vanish either identically or as a result of the equations of motion for
vector and fermion fields. This implies that in the absence of these
equations of motion there must hold a general identity given in (\ref{oii}).
However, in contrast to the previous case, the Eulerians generated by the
constraint-bearing Lagrangian $\widetilde{\mathbb{L}}$, while vanish on the
left-handed side of this identity, will give nonzero contributions to its
the right-hand side. Thus, having different vector field Eulerians in the
identity (\ref{oii}) one has to conclude that Noether's second theorem \cite%
{n} does not hold for this case. This means that gauge invariance fails in
general to emerge when the constraint in terms of source current
conservation (\ref{66}) for the vector field multiplet $\boldsymbol{A}_{\mu
}^{p}$ is required. In contrast to the previous case with the vanishing
source current $\mathbb{J}^{p\mu }$, where structure of the
constraint-bearing Lagrangian $\widetilde{\mathbb{L}}$ was virtually
established (\ref{l}, \ref{ll}) by the constraints (\ref{ai}) required, now
this Lagrangian, despite the constraints (\ref{66}) imposed, is still left
quite arbitrary. However, if we additionally propose that, as in the
previous case, the Lagrangian $\widetilde{\mathbb{L}}$ depends on all the
fields involved only through their `length squared' invariants, then all
goes well and gauge invariance arises. Indeed, using the Lagrangian form
determined above (\ref{l}, \ref{ll}) and corresponding expressions for
vector and fermion field Eulerians (\ref{r}), one can easily confirm that
all `tilded' terms induced by the constraint-bearing Lagrangian $\widetilde{%
\mathbb{L}}$ in the identity (\ref{oii}) are strictly canceled for obvious
symmetry reasons. So, this identity acquires a form to which Noether's
second theorem \cite{n} can be directly applied in order to finally
establish gauge invariance of the generic Lagrangian $\mathbb{L}_{g}$.

Eventually, for a minimal case with the mass squared dimension and
dimensionless coupling constants the whole emergent theory acquires a form 
\begin{equation}
\mathbb{L}^{\mathfrak{em}}(\boldsymbol{A}_{\mu },\boldsymbol{\psi }\text{ };%
\text{ }\mathbf{\lambda }_{c},\mathbf{M}^{2})=\mathbb{L}_{YM}(\boldsymbol{A}%
_{\mu },\boldsymbol{\psi })-\text{ }\frac{\mathbf{\lambda }_{c}}{4}(%
\boldsymbol{A}_{\mu }^{p}\boldsymbol{A}^{p\mu }-\boldsymbol{n}^{2}\mathbf{M}%
^{2})^{2}  \label{em1}
\end{equation}%
where the first term is a conventional Yang-Mills Lagrangian arising from a
generic Lagrangian $\mathbb{L}_{g}$, while the second term is a minimal
constraint-bearing Lagrangian $\widetilde{\mathbb{L}}_{\min }$ (\ref{l}). In
contrast to the previous case, we have obtained some gauge non-invariant
extension to Yang-Mills theory in the form of the potential with the mass
and self-interaction terms for vector fields. Note that for the Abelian
symmetry case the emergent Lagrangian (\ref{em1}) turns to the "bumblebee"
model (\ref{lag2}) considered in section 2.2. Interestingly, while the
Lagrangian $\widetilde{\mathbb{L}}$ taken above (\ref{l}, \ref{ll}) provides
an emergence of gauge invariance in the generic Lagrangian $\mathbb{L}_{g}$ (%
$\mathbb{L}_{g}\rightarrow $ $\mathbb{L}_{YM}$), it breaks this gauge
invariance by itself. In the simplest case ($\mathbf{\lambda }%
_{c}\rightarrow 0\mathrm{,}$ $\mathbf{\lambda }_{c}\mathbf{M}^{2}\rightarrow 
\mathbf{M}_{A}^{2}$) one has the massive Yang-Mills theory where the
constraint (\ref{66}) is reduced to the spin-1 condition (\ref{nh}) for
massive vector fields (having the mass $\mathbf{M}_{A}$). This particular
case was thoroughly studied in its own right quite a long ago \cite{ogi3}.

One can conclude that the length-fixing vector field constraints (\ref{const}%
, \ref{constt}) seems really to be of fundamental significance for emergent
gauge invariance. Actually, when constraints being put on the field system
are determined by the underlying Lagrangian itself, rather than taken ad hoc
through some Lagrange multiplier terms, the SLIV constraints (\ref{const}, %
\ref{constt}) appear strongly preferred over other ones. Indeed, as was
shown, only the strictly vanishing vector field source current, $\mathbb{J}%
_{\mu }^{p}=0$, that corresponds to the SLIV constraints (\ref{const}, \ref%
{constt}), leads to the full conversion of a starting global symmetry $G$ of
the total Lagrangian $\mathbb{L}=\mathbb{L}_{g}+\widetilde{\mathbb{L}}$ into
a local one $G_{loc}$. For nonzero current $\mathbb{J}_{\mu }^{p}$, on the
other hand, when the vector field constraint is solely determined by the
current conservation, $\partial ^{\mu }\mathbb{J}_{\mu }^{a}=0$, gauge
symmetry does not emerge or, at best, may only be partial.

\subsection{Gauge invariance \textit{versus} spontaneous Lorentz violation}

One can see that the gauge theory framework, be it taken from the outset or
emerged, makes in turn spontaneous Lorentz violation to be physically
unobservable both in Abelian and non-Abelian symmetry case. We referred to
it above as the inactive SLIV in contrast to the active SLIV case where
physical Lorentz invariance could effectively occur. From the present
standpoint, the only way for an active SLIV to occur would be if emergent
gauge symmetries presented above were slightly broken at small distances.
This could naturally happen, for example, in a partially gauge invariant
theory (\ref{em1}) which emerges due to properly chosen constraint (\ref{66}%
) being put on the physical field system, as was illustrated above. A more
radical point of view would be that the considered field system could become
unphysical at distances being presumably controlled by quantum gravity. One
could think that quantum gravity could in principle hinder the setting of
the required initial conditions in the appropriate Cauchy problem thus
admitting a superfluous restriction of vector fields in terms of some
high-order operators which occur at the Planck scale.

Recall in this connection that in the emergence equations (\ref{div}) and (%
\ref{id11}) we have only considered the lowest dimension terms which
eventually provide an emergent gauge theory at a renormalizible level. All
other terms (following from the expansion in (\ref{FF})) contain field
combinations and derivatives with higher mass dimension and must therefore
have coefficients with an inverse mass dimension. We expect the mass scale
associated with these coefficients should correspond to a large fundamental
mass (e.g. the Planck mass $\mathrm{M}_{P}$). Hence we may conclude that
such higher dimensional terms must be highly suppressed and can be neglected
for the effective low-energy gauge invariant theory. However, these terms
could lead to the breaking of an emergent gauge symmetry at high energies
just what is actually needed for SLIV to become active. This may be a place
where the emergent vector field theories may significantly differ from
conventional gauge theories that could have some observational evidence at
low energies. Below we present some particular model to see more clearly how
it may happen.

Looking for some appropriate examples of physical Lorentz violation in a QFT
framework one necessarily come across a problem of proper suppression of
gauge noninvariant high-dimension couplings where such violation can in
principle occur. Remarkably enough, for QED type theories with the
supplementary vector field constraint (\ref{const}) gauge symmetry breaking
naturally appears only for five- and higher-dimensional couplings. Indeed,
all dimension-four couplings are generically gauge invariant, if the vector
field kinetic term has a standard $F_{\mu \nu }F^{\mu \nu }$ and, apart from
relativistic invariance, the restrictions related to the conservation of
parity, charge-conjugation symmetry and fermion number conservation are
generally imposed on a theory \cite{par}. With these restrictions taken, one
can easily confirm that all possible dimension-five couplings are also
combined by themselves in some would-be gauge invariant form provided that
vector field is constrained by the SLIV condition (\ref{const}). Indeed, for
charged matter fermions interacting with vector field such couplings are
generally amounted to 
\begin{equation}
L_{\dim 5}=\frac{1}{\mathcal{M}}\check{D}_{\mu }^{\ast }\overline{\psi }%
\cdot \check{D}^{\mu }\psi +\frac{\mathrm{G}}{\mathcal{M}}A_{\mu }A^{\mu }%
\overline{\psi }\psi \text{ , \ }A_{\mu }A^{\mu }=n^{2}\mathrm{M}^{2}\text{ .%
}  \label{dim5}
\end{equation}%
Such couplings could presumably become significant at an ultraviolet scale $%
\mathcal{M}$ probably being close to the Planck scale $\mathrm{M}_{P}$.
They, besides covariant derivative terms, also include an independent
"sea-gull" fermion-vector field term with the coupling constant $\mathrm{G}$
being in general of the order $1$. The main point regarding the Lagrangian (%
\ref{dim5}) is that, while it is gauge invariant in itself, the coupling
constant $\mathrm{\check{e}}$ in the covariant derivative $\check{D}^{\mu
}=\partial ^{\mu }+i\mathrm{\check{e}}A^{\mu }$ differs in general from the
coupling $e$ in the covariant derivative $D^{\mu }=\partial ^{\mu }+i\mathrm{%
e}A^{\mu }$ in the standard Dirac Lagrangian (\ref{lag})%
\begin{equation}
L_{QED}=-\frac{1}{4}F_{\mu \nu }F^{\mu \nu }+\overline{\psi }(i\gamma _{\mu
}D^{\mu }-\mathrm{m})\psi \text{ .}  \label{qed}
\end{equation}%
Therefore, gauge invariance is no longer preserved in the total Lagrangian $%
L_{QED}+$ $L_{\dim 5}$. It is worth noting that, though the high-dimension
Lagrangian part $L_{\dim 5}$ (\ref{dim5}) usually only gives some small
corrections to a conventional QED Lagrangian (\ref{qed}), the situation may
drastically change when the vector field $A_{\mu }$ develops a VEV and SLIV
occurs.

Actually, putting the SLIV parameterization (\ref{gol}) into the basic QED
Lagrangian (\ref{qed}) one comes to the truly emergent model for QED being
essentially nonlinear in the vector Goldstone modes $a_{\mu }$ associated
with photons. This model contains, among other terms, the inappropriately
large (while false, see below) Lorentz violating fermion bilinear $-\mathrm{%
eM}\overline{\psi }\gamma _{\mu }n^{\mu }\psi $. This term appears when the
effective Higgs mode expansion in Goldstone modes $a_{\mu }$ (as is given in
the parametrization (\ref{gol})) is applied to the fermion current
interaction term $-\mathrm{e}\overline{\psi }\gamma _{\mu }A^{\mu }\psi $ in
the QED Lagrangian (\ref{qed}). However, due to local invariance this
bilinear term can be gauged away by making an appropriate redefinition of
the fermion field $\psi \rightarrow e^{-i\mathrm{e}\omega (x)}\psi $ with a
gauge function $\omega (x)$ linear in coordinates, $\omega (x)=$ $(x_{\mu
}n^{\mu })\mathrm{M}$. Meanwhile, the dimension-five Lagrangian $L_{\dim 5}$
(\ref{dim5}) is substantially changed under this redefinition that
significantly modifies fermion bilinear terms \ \ \ \ 
\begin{equation}
L_{\overline{\psi }\psi }=i\overline{\psi }\gamma _{\mu }\partial ^{\mu
}\psi +\frac{1}{\mathcal{M}}\partial _{\mu }\overline{\psi }\cdot \partial
^{\mu }\psi -i\Delta \mathrm{e}\frac{\mathrm{M}}{\mathcal{M}}n_{\mu }%
\overline{\psi }\overleftrightarrow{\partial ^{\mu }}\psi -\mathrm{m}_{f}%
\overline{\psi }\psi  \label{nl}
\end{equation}%
where we retained the notation $\psi $ for the redefined fermion field and
denoted, as usually, $\overline{\psi }\overleftrightarrow{\partial ^{\mu }}%
\psi =\overline{\psi }(\partial ^{\mu }\psi )-(\partial ^{\mu }\overline{%
\psi })\psi $. Note that the extra fermion derivative terms given in (\ref%
{nl}) is produced just due to the gauge invariance breaking that is
determined by the electromagnetic charge difference $\Delta \mathrm{e}=%
\mathrm{\check{e}}-\mathrm{e}$ in the total Lagrangian $L_{QED}+$ $L_{\dim
5} $. As a result, there appears the entirely new, SLIV\ inspired,
dispersion relation for a charged fermion (taken with 4-momentum $p_{\mu }$)
of type%
\begin{equation}
p_{\mu }^{2}\cong \lbrack \mathrm{m}_{f}+2\delta (p_{\mu }n^{\mu })]^{2},%
\text{ \ \ }\mathrm{m}_{f}=\mathrm{m}-\mathrm{GM}^{2}/\mathcal{M}-\delta
^{2}n^{2}\mathcal{M}  \label{m}
\end{equation}%
given to an accuracy of $O(\mathrm{m}_{f}^{2}/\mathcal{M}^{2})$ with a
properly modified total fermion mass $\mathrm{m}_{f}$. Here $\delta $ stands
for the small characteristic, positive or negative, parameter $\delta
=(\Delta e)\mathrm{M}/\mathcal{M}$ of physical Lorentz violation that
reflects the joint effect as is given, from the one hand, by the SLIV scale $%
\mathrm{M}$ and, from the other, by the charge difference $\Delta \mathrm{e}$
being a measure of an internal gauge noninvariance. Notably, the spacetime
in itself still possesses Lorentz invariance, however, fermions with SLIV
contributing into their total mass $\mathrm{m}_{f}$ (\ref{m}) propagate and
interact in it in the Lorentz non-covariant way. At the same time, the
photon dispersion relation\ is still retained undeformed in the order $1/%
\mathcal{M}$ considered.

So, it was shown that SLIV caused by the vector field VEV (\ref{vev1}),
while\ being\ superficial in a strictly gauge invariant theory, may become
physically significant when this gauge invariance is broken at the SLIV\
scale $\mathrm{M}$ being close to the scale $\mathcal{M}$, which is proposed
to be located near the Planck mass scale $\mathrm{M}_{P}$.\ This may happen
even at relatively low energies provided the gauge noninvariance caused by
high-dimension couplings of matter and vector fields is not vanishingly
small. As a consequence, through special dispersion relations appearing for
matter charged fermions, one is led a new class of phenomena which could be
of distinctive observational interest in particle physics and astrophysics 
\cite{par}. They include a significant change in the
Greizen-Zatsepin-Kouzmin cutoff for ultra-high energy cosmic-ray nucleons,
stability of high-energy pions and $W$ bosons, modification of nucleon beta
decays, and some others just in the presently accessible energy area in
cosmic ray physics.

However, though one could speculate about some generically broken or partial
gauge symmetry in a QFT framework \cite{par}, this appears to be too high a
price for an actual Lorentz violation which \ may stem from SLIV. And, what
is more, should one insist on physical Lorentz violation, if emergent gauge
fields are anyway generated through the \textquotedblleft
safe\textquotedblright\ inactive SLIV models which recover a conventional
Lorentz invariance? As will be seen in later sections, emergent SUSY
theories are most likely to give a negative answer to this question, thus
favouring just an inactive SLIV version.

\section{Emergent SUSY theories: a QED primer}

In contrast to theories probing physical Lorentz noninvariance, be it caused
by generically broken gauge symmetry or external tensor-valued backgrounds,
we are primarily focused here on a spontaneous Lorentz violation in an
actual gauge invariant QFT framework related to the Standard Model rather
than its hypothetical effective SME\ counterpart originated somewhere around
the Planck scale. In essence, we try to extend emergent gauge theories with
SLIV and an associated emergence of the SM\ gauge bosons as massless vector
Nambu-Goldstone modes studied earlier \cite{nambu, cfn, jb, kraus, bluhm} to
their supersymmetric analogs. Generally speaking, it may turn out that SLIV
is not the only reason why massless photons could dynamically appear, if
spacetime symmetry is further enlarged. In this connection, special interest
may be related to supersymmetry, as was recently argued in \cite{c}.
Actually, the situation is changed remarkably in the SUSY inspired emergent
models which, in contrast to non-SUSY theories, could naturally have some
clear observational evidence. Indeed, as we discussed in section 3.5,
ordinary emergent theories admit some experimental verification only if
gauge invariance is properly broken being caused by some high-dimension
couplings. Their SUSY counterparts, and primarily emergent SUSY QED, are
generically appear with supersymmetry being spontaneously broken in a
visible sector to ensure stability of the theory. Therefore, the
verification is now related to an inevitable emergence of a goldstino-like
photino state in the SUSY particle spectrum at low energies, while physical
Lorentz invariant is still left intact\footnote{%
Of course, physical Lorentz violation will also appear if one admits some
gauge noninvariance in the emergent SUSY theory as well. This may happen,
for example, through high-dimension couplings being supersymmetric analogs
of the couplings (\ref{dim5}).}. In this sense, a generic trigger for
massless photon to appear may be spontaneously broken supersymmetry rather
than physically manifested spontaneous Lorentz violation.

In this and subsequent sections the supersymmetric emergent gauge theories,
including their possible observational consequences, are considered in
significant detail.

\subsection{Spontaneous SUSY violation}

Precisely speaking, since gauge invariance is not generically assumed in an
emergent approach, some essential gauge-noninvariant couplings are
inevitably occurred in the theory in a pre-emergent phase. They, as seen
above, are basically related to the vector field self-interaction terms
triggering an emergence process in non-SUSY theories. Starting from this
standpoint, we consider a conventional supersymmetric QED being similarly
extended by an arbitrary polynomial potential of \ a general vector
superfield $V(x,\theta ,\overline{\theta })$ which in the standard
parametrization \cite{wess} has a form 
\begin{eqnarray}
V(x,\theta ,\overline{\theta }) &=&C+i\theta \chi -i\overline{\theta }%
\overline{\chi }+\frac{i}{2}\theta \theta S-\frac{i}{2}\overline{\theta }%
\overline{\theta }S^{\ast }  \notag \\
&&-\theta \sigma ^{\mu }\overline{\theta }A_{\mu }+i\theta \theta \overline{%
\theta }\overline{\lambda ^{\prime }}-i\overline{\theta }\overline{\theta }%
\theta \lambda ^{\prime }+\frac{1}{2}\theta \theta \overline{\theta }%
\overline{\theta }D^{\prime },  \label{par1}
\end{eqnarray}%
where its vector field component $A_{\mu }$ is usually associated with a
photon. Note that, apart from an ordinary photino field $\lambda $ and an
auxiliary $D$ field , the superfield (\ref{par1}) contains in general some
additional degrees of freedom in terms of the dynamical $C$ and $\chi $
fields and nondynamical complex scalar field $S$ (we have used the brief
notations, $\lambda ^{\prime }=\lambda +\frac{i}{2}\sigma ^{\mu }\partial
_{\mu }\overline{\chi }$ \ and $D^{\prime }=D+\frac{1}{2}\partial ^{2}C$
with $\sigma ^{\mu }=(1,\overrightarrow{\sigma })$ and $\overline{\sigma }%
^{\mu }=(1,-\overrightarrow{\sigma })$). The corresponding Lagrangian can be
written as%
\begin{equation}
\mathcal{L}=L_{SQED}+\frac{1}{2}D^{2}+\sum_{k=1}b_{k}V^{k}|_{D}  \label{slag}
\end{equation}%
where, besides a standard SUSY QED part, new potential terms are presented
in the sum by corresponding $D$-term expansions $V^{k}|_{D}$ of the vector
superfield (\ref{par1}) into the component fields ($b_{k}$ are some
constants). It can readily be checked that the first term in this expansion
is the known Fayet-Iliopoulos $D$-term, while other terms only contain
bilinear, trilinear and quartic combination of the superfield components $%
A_{\mu }$, $S$, $\lambda $ and $\chi $, respectively\footnote{%
Without loss of generality, we may restrict ourselves to the third degree
superfield polynomial in the Lagrangian $\mathcal{L}$ (\ref{slag}) to
eventually have a theory with dimesionless coupling constants for component
fields. However, for completeness sake, it seems better to proceed with a
general case. As we have recently learned, a similar self-interaction
polynomial for vector superfield (see also below the Lagrangian (\ref{lag3})
had been first considered quite a long ago \cite{fayet} to get some kind of
an economic Higgs model in a massive SUSY QED.}. Actually, the higher-degree
terms only appear for the scalar field component $C(x)$. Expressing them all
in terms of the $C$ field polynomial%
\begin{equation}
P(C)=\sum_{k=1}\frac{k}{2}b_{k}C^{k-1}(x)  \label{pot}
\end{equation}%
and its first three derivatives 
\begin{equation}
P_{C}^{\prime }\equiv \frac{\partial P}{\partial C}\text{ , \ \ }%
P_{C}^{\prime \prime }\equiv \frac{\partial ^{2}P}{\partial C^{2}}\text{ , \
\ }P_{C}^{\prime \prime \prime }\equiv \frac{\partial ^{3}P}{\partial C^{3}}%
\text{ }  \label{dd}
\end{equation}%
one has for the whole Lagrangian $\mathcal{L}$ 
\begin{eqnarray}
\mathcal{L} &=&L_{SQED}+\frac{1}{2}D^{2}+\text{ }P\left( D+\frac{1}{2}%
\partial ^{2}C\right)  \notag \\
&&+P_{C}^{\prime }\left( \frac{1}{2}SS^{\ast }-\chi \lambda ^{\prime }-%
\overline{\chi }\overline{\lambda ^{\prime }}-\frac{1}{2}A_{\mu }A^{\mu
}\right)  \notag \\
&&+\text{ }\frac{1}{2}P_{C}^{\prime \prime }\left( \frac{i}{2}\overline{\chi 
}\overline{\chi }S-\frac{i}{2}\chi \chi S^{\ast }-\chi \sigma ^{\mu }%
\overline{\chi }A_{\mu }\right) +\frac{1}{8}P_{C}^{\prime \prime \prime
}(\chi \chi \overline{\chi }\overline{\chi })\text{ .}  \label{lag3}
\end{eqnarray}%
where, for more clarity, we still omitted in $L_{SQED}$ matter superfields
reserving them for section 6. As one can see, extra degrees of freedom
related to the $C$ and $\chi $ component fields in a general vector
superfield $V(x,\theta ,\overline{\theta })$ appear through the potential
terms in (\ref{lag3}) rather than from the properly constructed
supersymmetric field strengths, as appear for the vector field $A_{\mu }$
and its gaugino companion $\lambda $.

Note that all terms in the sum in (\ref{slag}) except Fayet-Iliopoulos $D$%
-term\ explicitly break gauge invariance. However, as we will see in the
next section, the special gauge invariance constrained by some gauge
condition will be recovered in the Lagrangian in the broken SUSY phase.
Furthermore, as is seen from (\ref{lag3}), the vector field $A_{\mu }$ may
only appear with bilinear mass term in the polynomially extended superfield
Lagrangian (\ref{slag}) in sharp contrast to the non-SUSY theory case where,
apart from the vector field mass term, some high-linear stabilizing\ terms
necessarily appear in a similar polynomially extended Lagrangian. This means
in turn that physical Lorentz invariance is still preserved in the theory.
Actually, only supersymmetry appears to be spontaneously broken, as
mentioned above.

Indeed, varying the Lagrangian $\mathcal{L}$ with respect to the $D$ field
we come to 
\begin{equation}
D=-P(C)  \label{d}
\end{equation}%
that finally gives the following potential energy for the field system
considered 
\begin{equation}
U(C)=\frac{1}{2}[P(C)]^{2}\text{ .}  \label{pot1}
\end{equation}%
The potential (\ref{pot1}) may lead to spontaneous SUSY breaking in the
visible sector provided that the polynomial $P$ (\ref{pot}) has no real
roots, while its first derivative has, 
\begin{equation}
P\neq 0\text{ , \ }P_{C}^{\prime }=0.\text{\ }  \label{der}
\end{equation}%
This requires $P(C)$ to be an even degree polynomial with properly chosen
coefficients $b_{k}$ in (\ref{pot}) that will force its derivative $%
P_{C}^{\prime }$ to have at least one root, $C=C_{0}$, in which the
potential (\ref{pot1}) is minimized. Therefore, supersymmetry is
spontaneously broken and the $C$ field acquires the VEV 
\begin{equation}
\left\langle C\right\rangle =C_{0}\text{ , \ }P_{C}^{\prime }(C_{0})=0\text{
.}  \label{vvv}
\end{equation}%
As an immediate consequence, that one can readily see from the Lagrangian $%
\mathcal{L}$ (\ref{lag3}), a massless photino $\lambda $ being Goldstone
fermion in the broken SUSY phase make all the other component fields in the
superfield $V(x,\theta ,\overline{\theta })$ including the photon to also
become massless. However, the question then arises whether this masslessness
of the photon will be stable against radiative corrections since gauge
invariance is explicitly broken in the Lagrangian (\ref{lag3}). We show
below that it could be the case if the vector superfield $V(x,\theta ,%
\overline{\theta })$ would appear to be properly constrained.

\subsection{Instability of superfield polynomial potential}

Let us first analyze possible vacuum configurations for the superfield
components in the polynomially extended QED case taken above. In general,
besides the "standard" potential energy expression (\ref{pot1}) determined
solely by the scalar field component $C(x)$ of the vector superfield (\ref%
{par1}), one also has to consider other field component contributions into
the potential energy. A possible extension of the potential energy (\ref%
{pot1}) seems to appear only due to the pure bosonic field contributions,
namely due to couplings of the vector and auxiliary scalar fields, $A_{\mu }$
and $S$, in (\ref{lag3}) 
\begin{equation}
\mathcal{U}=\frac{1}{2}P^{2}+\frac{1}{2}P^{\prime }(A_{\mu }A^{\mu
}-SS^{\ast })\text{ }  \label{pot1a}
\end{equation}%
rather than due to the potential terms containing the superfield fermionic
components\footnote{%
Actually, this restriction is not essential for what follows and is taken
just for simplicity. Generally, the fermion bilinears involved could also
develop VEVs.}. It can be immediately seen that these new couplings in (\ref%
{pot1a}) can make the potential unstable since the vector and scalar fields
mentioned may in general develop any arbitrary VEVs. This happens, as
emphasized above, due the fact that their bilinear term contributions are
not properly compensated by appropriate four-linear field terms which are
generically absent in a SUSY theory context.

For more detail we consider the extremum conditions for the entire potential
(\ref{pot1a}) with respect to all fields involved: $C$, $A_{\mu }$ and $S$.
They are given by the appropriate first partial derivative equations%
\begin{eqnarray}
\mathcal{U}_{C}^{\prime } &=&PP^{\prime }+\frac{1}{2}P^{\prime \prime
}(A_{\mu }A^{\mu }-SS^{\ast })=0,\text{ }  \notag \\
\mathcal{U}_{A_{\mu }}^{\prime } &=&P^{\prime }A^{\mu }=0,\text{ \ \ }%
\mathcal{U}_{S}^{\prime }=-P^{\prime }S^{\ast }=0.  \label{"}
\end{eqnarray}%
where and hereafter all the VEVs are denoted by the corresponding field
symbols (supplied below with the lower index $0$). One can see that there
can occur a local minimum for the potential (\ref{pot1a}) with the unbroken
SUSY solution\footnote{%
Hereafter by $P(C_{0})$ and $P^{\prime }(C_{0})$ are meant the $C$ field
polynomial $P$ (\ref{pot}) and its functional derivative $P^{\prime }$ (\ref%
{dd}) taken in the potential extremum point $C_{0}$.} 
\begin{equation}
C=C_{0},\text{ }P(C_{0})=0\text{, }P^{\prime }(C_{0})\neq 0\text{ };\text{ \ 
}A_{\mu 0}=0,\text{ \ }S_{0}=0  \label{sol}
\end{equation}%
with the vanishing potential energy 
\begin{equation}
\mathcal{U}_{\min }^{s}=0  \label{pe}
\end{equation}%
provided that the polynomial $P$ (\ref{pot}) has some real root $C=C_{0}$.
Otherwise, a local minimum with the broken SUSY solution can occur for some
other $C$ field value (though denoted by the same letter $C_{0}$) 
\begin{equation}
C=C_{0},\text{ }P(C_{0})\neq 0\text{, }P^{\prime }(C_{0})=0\text{ };\text{ \ 
}A_{\mu 0}\neq 0,\text{ \ }S_{0}\neq 0,\text{ }A_{\mu 0}A_{0}^{\mu
}-S_{0}S_{0}^{\ast }=0  \label{sol2}
\end{equation}%
In this case one has the non-zero potential energy 
\begin{equation}
\mathcal{U}_{\min }^{as}=\frac{1}{2}[P(C_{0})]^{2}  \label{pe2}
\end{equation}%
as directly follows from the extremum equations (\ref{"}) and potential
energy expression (\ref{pot1a}).

However, as shows the standard second partial derivative test, the fact is
that the local minima mentioned above are minima with respect to the $C$
field VEV ($C_{0}$) only. Actually, for all three fields VEVs included the
potential (\ref{pot1a}) has indeed saddle points with "coordinates"
indicated in (\ref{sol}) and (\ref{sol2}), respectively. For a testing
convenience this potential can be rewritten in the form 
\begin{equation}
\mathcal{U}=\frac{1}{2}P^{2}+\frac{1}{2}P^{\prime }g^{\Theta \Theta ^{\prime
}}B_{\Theta }B_{\Theta ^{\prime }}\text{ , \ }g^{\Theta \Theta ^{\prime
}}=diag\text{ }(1,-1,-1,-1,-1,-1)\text{\ \ \ }  \label{pot1b}
\end{equation}%
with only two variable fields $C$ and $B_{\Theta }$ where the new field $%
B_{\Theta }$ unifies the $A_{\mu }$ and $S$ field components, $B_{\Theta
}=(A_{\mu },S_{\alpha })$ ($\Theta =\mu ,a;$ $\mu =0,1,2,3$; $\alpha =1,2$)%
\footnote{%
Interestingly, the $B_{\Theta }$\ term in the potential (\ref{pot1b})
possesses the accidental $SO(1,5)$ symmetry. This symmetry, though it is not
shared by kinetic terms, appears in fact to be stable under radiative
corrections since $S$ field is non-dynamical and, therefore, can always be
properly arranged.}. The complex $S$ field is now taken in a real basis, 
\begin{equation}
S_{1}=(S+S^{\ast })/2,\text{ \ \ }S_{2}=(S-S^{\ast })/2i\text{ ,}
\label{bas}
\end{equation}%
so that the "vector" $B_{\Theta }$ field has one time and five space
components. As a result, one finally comes to the following Hessian $7\times
7$ matrix (being in fact the second-order partial derivatives matrix taken
in the extremum point ($C_{0}$, $A_{\mu 0}$, $S_{0}$) (\ref{sol})) 
\begin{equation}
H(\mathcal{U}^{s})=\left[ 
\begin{array}{cc}
\lbrack P^{\prime }(C_{0})]^{2} & 0 \\ 
0 & P^{\prime }(C_{0})g^{\Theta \Theta ^{\prime }}%
\end{array}%
\right] \text{, \ }\left\vert H(\mathcal{U}^{s})\right\vert =-\text{\ }%
[P^{\prime }(C_{0})]^{8}\text{ .}  \label{hess}
\end{equation}%
This matrix clearly has the negative determinant $\left\vert H(\mathcal{U}%
^{s})\right\vert $, as is indicated above, that confirms that the potential
definitely has a saddle point for the solution (\ref{sol}). This means the
VEVs of the $A_{\mu }$ and $S$ fields can take in fact any arbitrary value
making the potential (\ref{pot1a}, \ref{pot1b}) to be unbounded from below
in the unbroken SUSY case that is certainly inaccessible.

One might think that in the broken SUSY case the situation would be better
since due to the conditions (\ref{sol2}) the $B_{\Theta }$ term completely
disappears from the potential $\mathcal{U}$ (\ref{pot1a}, \ref{pot1b}) in
the ground state. Unfortunately, the direct second partial derivative test
in this case is inconclusive since the determinant of the corresponding
Hessian $7\times 7$ matrix appears to vanish 
\begin{equation}
H(\mathcal{U}^{as})=\left[ 
\begin{array}{cc}
P(C_{0})P^{\prime \prime }(C_{0}) & P^{\prime \prime }(C_{0})g^{\Theta
\Theta ^{\prime }}B_{\Theta ^{\prime }} \\ 
P^{\prime \prime }(C_{0})g^{\Theta \Theta ^{\prime }}B_{\Theta ^{\prime }} & 
0%
\end{array}%
\right] \text{, \ \ }\left\vert H(\mathcal{U}^{as})\right\vert =0\text{ .}
\label{hess2}
\end{equation}%
Nevertheless, since in general the $B_{\Theta }$ term can take both positive
and negative values in small neighborhoods around the vacuum point ($C_{0}$, 
$A_{\mu 0}$, $S_{0}$) where the conditions (\ref{sol2}) are satisfied, this
point is also turned out to be a saddle point. Thus, the potential $\mathcal{%
U}$ (\ref{pot1a}, \ref{pot1b}) appears generically unstable both in SUSY
invariant and SUSY broken phase.

\subsection{Stabilization of vacuum by constraining vector superfield}

The only possible way to stabilize the ground states (\ref{sol}) and (\ref%
{sol2}) seems to seek the proper constraints on the superfield component
fields ($C$, $A_{\mu }$, $S$) themselves rather than on their expectation
values. Indeed, if such (potential bounding) constraints are physically
realizable, the vacua (\ref{sol}) and (\ref{sol2}) will be automatically
stabilized. Besides, as we confirm, instead of gauge symmetry broken in the
extended QED Lagrangian (\ref{lag3}) some special gauge invariance is
recovered in (\ref{lag3}) at the SUSY\ breaking minimum of the potential (%
\ref{pot1}).

Let us try to understand first how such constraints may look like. We will
expand the action around the vacuum (\ref{vvv}) by writing 
\begin{equation}
C(x)=C_{0}+c(x)  \label{cc}
\end{equation}%
that gives for the $C$ field polynomial $P(C)$ (\ref{pot}) and its
derivatives (\ref{dd}) to the lowest order in the Higgs-like field $c(x)$ 
\begin{eqnarray}
P(C) &\simeq &P(C_{0})+\frac{1}{2}P_{C}^{\prime \prime }(C_{0})c^{2}\text{ ,
\ }P_{C}^{\prime }(C)\simeq P_{C}^{\prime \prime }(C_{0})c\text{ , \ } 
\notag \\
P_{C}^{\prime \prime }(C) &\simeq &P_{C}^{\prime \prime
}(C_{0})+P_{C}^{\prime \prime \prime }(C_{0})c\text{ , \ }P_{C}^{\prime
\prime \prime }(C)\simeq P_{C}^{\prime \prime \prime }(C_{0})+P_{C}^{\prime
\prime \prime \prime }(C_{0})c\text{\ }  \label{app}
\end{eqnarray}%
with $P_{C}^{\prime }(C_{0})=0$ taken at the minimum point $C_{0}$, as is
determined in (\ref{vvv}). Now, combining the equations of motion for $c(x)$
and for some other component field, say $S(x)$, both derived by varying the
Lagrangian (\ref{lag3}), one has 
\begin{equation}
A_{\mu }A^{\mu }-SS^{\ast }=O(c,c\partial ^{2}c)\text{ , \ \ }\chi \chi =O(c)
\label{sa}
\end{equation}%
where we have used approximate equalities (\ref{app}) with typical nonzero
values of all $P(C_{0})$, $P_{C}^{\prime \prime }(C_{0})$, $P_{C}^{\prime
\prime \prime }(C_{0})$, $P_{C}^{\prime \prime \prime \prime }(C_{0})$ taken
at the minimum point $C_{0}$. For the vanishingly small Higgs-like mode $%
c(x) $ in (\ref{cc}) and (\ref{sa}), one eventually comes to the necessary
constraints which have to be put on the $V$ superfield components to provide
stability of the total potential (\ref{pot1a}).

These pure heuristic arguments can be also realized in a more rigorous way
by properly constraining the vector superfield $V(x,\theta ,\overline{\theta 
})$ from the outset. In a SUSY context a constraint can only be put on an
entire superfield rather than individually on its field components.
Actually, one can constrain our vector superfield (\ref{par1}) by analogy
with the constrained vector field in the nonlinear QED model (\ref{lag}).
This will be done again through some invariant Lagrange multiplier coupling
simply adding its $D$ term to the above Lagrangian (\ref{slag}, \ref{lag3}) 
\begin{equation}
\mathcal{L}^{tot}=\mathcal{L}+\frac{1}{2}\Lambda (V-C_{0})^{2}|_{D}\text{ ,}
\label{ext}
\end{equation}%
where $\Lambda (x,\theta ,\overline{\theta })$ is some auxiliary vector
superfield, while $C_{0}$ is the constant background value of the $C$ field
which minimizes the potential $U$ (\ref{pot1}). Accordingly, the potential
vanishes for the supersymmetric minimum or acquires some positive value
corresponding to the SUSY breaking minimum (\ref{der}) in the visible
sector. We will consider both cases simultaneously using the same notation $%
C_{0}$ for either of the background values of the $C$ field.

Note first of all, the Lagrange multiplier term in (\ref{ext}) has in fact
the simplest possible form that leads to some nontrivial constrained
superfield $V(x,\theta ,\overline{\theta })$. The alternative minimal forms,
such as the bilinear form $\Lambda (V-C_{0})$ or trilinear one $\Lambda
(V^{2}-C_{0}^{2})$, appear too restrictive. One can easily confirm that they
eliminate most component fields in the superfield $V(x,\theta ,\overline{%
\theta })$ including the physical photon and photino fields that is
definitely inadmissible. As to appropriate non-minimal high linear
multiplier forms, they basically lead to the same consequences as follow
from the minimal multiplier term taken in the total Lagrangian (\ref{ext}).
Writing down its invariant $D$ term through the component fields one finds 
\begin{eqnarray}
\Lambda (V-C_{0})^{2}|_{D} &=&C_{\Lambda }\left[ \widetilde{C}D^{\prime
}+\left( \frac{1}{2}SS^{\ast }-\chi \lambda ^{\prime }-\overline{\chi }%
\overline{\lambda ^{\prime }}-\frac{1}{2}A_{\mu }A^{\mu }\right) \right] 
\notag \\
&&+\text{ }\chi _{\Lambda }\left[ 2\widetilde{C}\lambda ^{\prime }+i(\chi
S^{\ast }+i\sigma ^{\mu }\overline{\chi }A_{\mu })\right] +\overline{{\large %
\chi }}_{\Lambda }[2\widetilde{C}\overline{\lambda ^{\prime }}-i(\overline{%
\chi }S-i\chi \sigma ^{\mu }A_{\mu })]  \notag \\
&&+\text{ }\frac{1}{2}S_{\Lambda }\left( \widetilde{C}S^{\ast }+\frac{i}{2}%
\overline{\chi }\overline{\chi }\right) +\frac{1}{2}S_{\Lambda }^{\ast
}\left( \widetilde{C}S-\frac{i}{2}\chi \chi \right)  \notag \\
&&+\text{ }2A_{\Lambda }^{\mu }(\widetilde{C}A_{\mu }-\chi \sigma _{\mu }%
\overline{\chi })+2\lambda _{\Lambda }^{\prime }(\widetilde{C}\chi )+2%
\overline{{\large \lambda }}_{\Lambda }^{\prime }(\widetilde{C}\overline{%
\chi })+\frac{1}{2}D_{\Lambda }^{\prime }\widetilde{C}^{2}  \label{lm1}
\end{eqnarray}%
where 
\begin{equation}
C_{\Lambda },\text{ }\chi _{\Lambda },\text{ }S_{\Lambda },\text{ }%
A_{\Lambda }^{\mu },\text{ }\lambda _{\Lambda }^{\prime }=\lambda _{\Lambda
}+\frac{i}{2}\sigma ^{\mu }\partial _{\mu }\overline{{\large \chi }}%
_{\Lambda },\text{ }D_{\Lambda }^{\prime }=D_{\Lambda }+\frac{1}{2}\partial
^{2}C_{\Lambda }  \label{comp}
\end{equation}%
are the component fields of the Lagrange multiplier superfield $\Lambda
(x,\theta ,\overline{\theta })$ in the standard parametrization ((\ref{par1}%
)) and $\widetilde{C}$ stands for the difference $C(x)-C_{0}$. Varying the
Lagrangian (\ref{ext}) with respect to these fields and properly combining
their equations of motion 
\begin{equation}
\frac{\partial \mathcal{L}^{tot}}{\partial \left( C_{\Lambda },\chi
_{\Lambda },S_{\Lambda },A_{\Lambda }^{\mu },\lambda _{\Lambda },D_{\Lambda
}\right) }=0  \label{var}
\end{equation}%
we find the constraints which appear to put on the $V$ superfield components%
\footnote{%
Indeed, the equations $\partial \mathcal{L}_{tot}/\partial D_{{\large %
\Lambda }}=0$ and $\partial \mathcal{L}_{tot}/\partial S_{{\large \Lambda }%
}=0$ immediately give the constraints $C=C_{0}$ and \ $\chi =0$,
respectively, while the equation $\partial \mathcal{L}_{tot}/\partial C_{%
{\large \Lambda }}=0$ leads to the constraint $A_{\mu }A^{\mu }=SS^{\ast }$
once the previous two constraints are satisfied. They coincide, as expected,
with constraints arisen for the vanishingly small Higgs-like mode $c(x)$ in
the equations (\ref{cc}) and (\ref{sa}).} 
\begin{equation}
C=C_{0},\text{ \ }\chi =0,\text{\ \ }A_{\mu }A^{\mu }=SS^{\ast }\text{.}
\label{const1}
\end{equation}%
They also determine the corresponding $D$-term (\ref{d}), $D=-P(C_{0}),$ for
the spontaneously broken supersymmetry. Again, as in non-SUSY case (\ref{lag}%
), we only take a solution with initial values for all fields (and their
momenta) chosen so as to restrict the phase space to vanishing values of the
multiplier component fields (\ref{comp}). This will provide, as before, a
ghost-free theory with a positive Hamiltonian\footnote{%
As in the non-supersymmetric case discussed above (see also the footnote$%
^{3} $), this solution with all vanishing components of the basic Lagrangian
multiplier superfield ${\large \Lambda }(x,\theta ,\overline{\theta })$ can
be reached by introducing in the total Lagrangian (\ref{ext}) an appropriate
extra Lagrange multiplier term of the type $\Sigma \Lambda ^{2}$, where $%
\Sigma (x)$ is a new multiplier superfield.}.

Remarkably, the constraints (\ref{const1}) does not touch the physical
degrees of freedom of the \ superfield $V(x,\theta ,\overline{\theta })$
related to photon and photino fields. The point is, however, that apart from
the constraints (\ref{const1}), one has the equations of motion for all
fields involved in the basic superfield $V(x,\theta ,\overline{\theta })$.
With vanishing multiplier component fields (\ref{comp}), as was proposed
above, these equations appear in fact as extra constraints on components of
the superfield $V(x,\theta ,\overline{\theta })$. Indeed, equations of
motion for the fields $C$, $S$ and ${\large \chi }$ received by the
corresponding variations of the total Lagrangian $\mathcal{L}^{tot}$ (\ref%
{ext}, \ref{lag3}) are turned out to be, respectively, 
\begin{equation}
P(C_{0})P^{\prime }(C_{0})=0,\text{ \ }S(x)P^{\prime }(C_{0})=0\text{ , \ }%
\lambda (x)P^{\prime }(C_{0})=0  \label{nc}
\end{equation}%
where the basic constraints (\ref{const1}) emerging at the potential
extremum point $C=C_{0}$ have also been used. One can immediately see now
that these equations turn to trivial identities in the broken SUSY case, in
which the factor $P^{\prime }(C_{0})$ in each of them appears to be
identically vanished, $P^{\prime }(C_{0})$ $=0$ (\ref{sol2}). In the
unbroken SUSY case, in which the potential (\ref{pot1}) vanishes instead,
i.e. $P(C_{0})=0$ and $P^{\prime }(C_{0})\neq 0$ (\ref{sol}), the situation
is drastically changed. Indeed, though the first equation in (\ref{nc})
still automatically turns into identity at the extremum point $C(x)=C_{0}$,
other two equations require that the auxiliary field $S$ and the photino
field $\lambda $ have to be identically vanished as well. This causes in
turn that the photon field should also be vanished according to the basic
constraints (\ref{const1}). Besides, the $D$ field component in the vector
superfield is also vanished in the unbroken SUSY case according to the
equation (\ref{d}), $D=-P(C_{0})=0$. Thus, one is ultimately left with a
trivial superfield $V(x,\theta ,\overline{\theta })$ which only contains the
constant $C$ field component $C_{0}$ that is unacceptable. So, we have to
conclude that the unbroken SUSY fails to provide stability of the potential (%
\ref{pot1a}) even by constraining the superfield $V(x,\theta ,\overline{%
\theta })$. In contrast, in the spontaneously broken SUSY case extra
constraints do not appear at all, and one has a physically meaningful theory
that we basically consider in what follows.

Finally, implementing the constraints (\ref{const1}) into the total
Lagrangian $\mathcal{L}^{tot}$ (\ref{ext}, \ref{lag3}) through the Lagrange
multiplier terms for component fields, we come to the emergent SUSY QED
appearing in the broken SUSY phase%
\begin{equation}
\mathcal{L}^{\mathfrak{em}}=\mathcal{L}_{SQED}+P(C)D\text{ }+\frac{%
D_{\Lambda }}{4}(C-C_{0})^{2}-\frac{C_{\Lambda }}{4}\left( A_{\mu }A^{\mu
}-SS^{\ast }\right) \text{ .}  \label{fin}
\end{equation}%
The last two term with the component multiplier functions $C_{\Lambda }$ and 
$D_{\Lambda }$ of the auxiliary superfield $\Lambda $ (\ref{comp}) provide
the vacuum stability condition of the theory. In essence, one does not need
now to postulate from the outset gauge invariance for the physical SUSY QED
Lagrangian $\mathcal{L}_{SQED}$. Rather, one can derive it following the
emergence conjecture specified for Abelian theories in section 3.2. Indeed,
due to the constraints (\ref{const1}), the Lagrangian $\mathcal{L}_{SQED}$
is only allowed to have a conventional gauge invariant form 
\begin{equation}
\mathcal{L}_{SQED}=-\text{ }\frac{1}{4}F^{\mu \nu }F_{\mu \nu }+i\lambda
\sigma ^{\mu }\partial _{\mu }\overline{\lambda }+\frac{1}{2}D^{2}
\label{444}
\end{equation}%
Thus, for the constrained vector superfield involved 
\begin{equation}
\widehat{V}(x,\theta ,\overline{\theta })=C_{0}+\frac{i}{2}\theta \theta S-%
\frac{i}{2}\overline{\theta }\overline{\theta }S^{\ast }-\theta \sigma ^{\mu
}\overline{\theta }A_{\mu }+i\theta \theta \overline{\theta }\overline{%
\lambda }-i\overline{\theta }\overline{\theta }\theta \lambda +\frac{1}{2}%
\theta \theta \overline{\theta }\overline{\theta }D,  \label{sup}
\end{equation}%
we have the almost standard SUSY QED Lagrangian with the same states - a
photon, a photino and an auxiliary scalar $D$ field - in its gauge
supermultiplet, while another auxiliary complex scalar field $S$ gets only
involved in the vector field constraint. The linear (Fayet-Iliopoulos) $D$%
-term with the effective coupling constant $P(C_{0})$ in (\ref{fin}) shows
that supersymmetry in the theory is spontaneously broken due to which the $D$
field acquires the VEV, $D=-P(C_{0})$. Taking the nondynamical $S$ field in
the constraint (\ref{const1}) to be some constant background field (for a
more formal discussion, see below) we come to the SLIV\ constraint (\ref%
{const}) which we discussed above regarding an ordinary non-supersymmetric
QED theory (section 2.4). As is seen from this constraint in (\ref{fin}),
one may only have the time-like SLIV in a SUSY framework but never the
space-like one. There also may be a light-like SLIV, if the $S$ field
vanishes\footnote{%
Indeed, this case, first mentioned in \cite{nambu}, may also mean
spontaneous Lorentz violation with a nonzero VEV $<A_{\mu }>$ $=(\widetilde{M%
},0,0,\widetilde{M})$ and Goldstone modes $A_{1,2}$ and $(A_{0}+A_{3})/2$\ $-%
\widetilde{M}.$ The "effective" Higgs mode $(A_{0}-A_{3})/2$ can be then
expressed through Goldstone modes so as the light-like condition $A_{\mu
}A^{\mu }=0$ to be satisfied.}. So, any possible choice for the $S$ field
corresponds to the particular gauge choice for the vector field $A_{\mu }$
in an otherwise gauge invariant theory. So, the massless photon appearing
first as a companion of a massless photino (being a Goldstone fermion in the
visible broken SUSY phase) remains massless due to this recovering gauge
invariance in the emergent SUSY QED. At the same time, the "built-in"
nonlinear gauge condition in (\ref{fin}) allows to treat the photon as a
vector Goldstone boson induced by an inactive SLIV.

\subsection{Constrained vector superfield: a formal view}

We proceed by showing that our extended Lagrangian $\mathcal{L}^{tot}$ (\ref%
{ext}, \ref{lag3}), underlying the emergent QED model, is SUSY invariant,
and also the constraints (\ref{const1})\ on the field space appearing due to
the Lagrange multiplier term in (\ref{ext}) are consistent with
supersymmetry. The first part of this assertion is somewhat immediate since
the Lagrangian $\mathcal{L}^{tot}$, aside from the standard supersymmetric
QED part $L_{SQED}$ (\ref{slag}), only contains $D$-terms of various vector
superfield products. They are, by definition, invariant under conventional
SUSY transformations \cite{wess} which for the component fields ((\ref{par1}%
)) of a general superfield $V(x,\theta ,\overline{\theta })$ ((\ref{par1}))
are written as 
\begin{eqnarray}
\delta _{\xi }C &=&i\xi \chi -i\overline{\xi }\overline{\chi }\text{ , \ }%
\delta _{\xi }\chi =\xi S+\sigma ^{\mu }\overline{\xi }(\partial _{\mu
}C+iA_{\mu })\text{ , \ }\frac{1}{2}\delta _{\xi }S=\overline{\xi }\overline{%
\lambda }+\overline{\sigma }_{\mu }\partial ^{\mu }\chi \text{ ,}  \notag \\
\delta _{\xi }A_{\mu } &=&\xi \partial _{\mu }\chi +\overline{\xi }\partial
_{\mu }\overline{\chi }+i\xi \sigma _{\mu }\overline{\lambda }-i\lambda
\sigma _{\mu }\overline{\xi }\text{ , \ }\delta _{\xi }\lambda =\frac{1}{2}%
\xi \sigma ^{\mu }\overline{\sigma }^{\nu }F_{\mu \nu }+\xi D\text{ ,} 
\notag \\
\delta _{\xi }D &=&-\xi \sigma ^{\mu }\partial _{\mu }\overline{\lambda }+%
\overline{\xi }\sigma ^{\mu }\partial _{\mu }\lambda \text{ .}  \label{trans}
\end{eqnarray}%
However, there may still be left a question whether supersymmetry remains in
force when the constraints (\ref{const1})\ on the field space are "switched
on" thus leading to the final Lagrangian $\mathcal{L}^{\mathfrak{em}}$(\ref%
{fin}) in the broken SUSY phase with both dynamical fields $C$ and $\chi $
eliminated. This Lagrangian appears similar to the standard supersymmetric
QED taken in the Wess-Zumino gauge, except that supersymmetry is
spontaneously broken in our case. In both cases the photon stress tensor $%
F_{\mu \nu }$, the photino $\lambda $ and the nondynamical scalar $D$ field
form an irreducible representation of the supersymmetry algebra (the last
two lines in (\ref{trans})). Nevertheless, any reduction of component fields
in the vector superfield is not consistent in general with the linear
superspace version of supersymmetry transformations, whether it is the
Wess-Zumino gauge case or our constrained superfield (\ref{sup}). Indeed, a
general SUSY transformation does not preserve the Wess-Zumino gauge: a
vector superfield in this gauge,%
\begin{equation}
V_{WZ}(x,\theta ,\overline{\theta })=\theta \sigma ^{\mu }\overline{\theta }%
A_{\mu }+i\theta \theta \overline{\theta }\overline{\lambda }-i\overline{%
\theta }\overline{\theta }\theta \lambda +\frac{1}{2}\theta \theta \overline{%
\theta }\overline{\theta }D\text{ },  \label{wz}
\end{equation}%
acquires all possible extra terms when being SUSY transformed. The same also
occurs with our constrained superfield $\widehat{V}$ (\ref{sup}). The point,
however, is that in both cases a total supergauge transformation 
\begin{equation}
V\rightarrow V+\frac{i}{2}(\Omega -\Omega ^{\ast })\text{ ,}  \label{spg}
\end{equation}%
where $\Omega $ is an arbitrary chiral superfield transformation parameter 
\cite{wess}%
\begin{equation}
\Omega =\varphi +\sqrt{2}\theta \psi +\theta \theta F+i\theta \sigma ^{\mu }%
\overline{\theta }\partial _{\mu }\varphi -\frac{i}{\sqrt{2}}\theta \theta 
\overline{\theta }\sigma ^{\mu }\partial _{\mu }\psi -\frac{1}{4}\theta
\theta \overline{\theta }\overline{\theta }\partial ^{2}\varphi ,
\label{chir}
\end{equation}%
can always restore the vector superfield initial (restricted) form (\ref{sup}%
) or (\ref{wz}), respectively. In a conventional supersymmetric QED taken in
the Wess-Zumino supergauge an ordinary gauge freedom is left untouched. This
means that the non-trivial part of the $V_{WZ}$ superfield transformation
amounts to 
\begin{equation}
V_{WZ}\rightarrow V_{WZ}-\theta \sigma ^{\mu }\overline{\theta }\partial
_{\mu }\varphi \text{ , \ \ }A_{\mu }\rightarrow A_{\mu }-\partial _{\mu
}\varphi  \label{wzz}
\end{equation}%
where the scalar component $\varphi $ in the SUSY transformation parameter $%
\Omega $ (\ref{chir}) is used. In contrast, in the emergent SUSY QED (\ref%
{fin}) the ordinary gauge is fixed by the vector field constraint (\ref%
{const1}). However, this constraint remains under supergauge transformation (%
\ref{spg}) applied to our superfield $\widehat{V}$ (\ref{sup}). Indeed, the
essential part of this transformation which directly acts on the constraint (%
\ref{const1}) has the form 
\begin{equation}
\widehat{V}\rightarrow \widehat{V}+\frac{i}{2}\theta \theta F-\frac{i}{2}%
\overline{\theta }\overline{\theta }F^{\ast }-\theta \sigma ^{\mu }\overline{%
\theta }\partial _{\mu }\varphi \text{ }  \label{sg}
\end{equation}%
where the real and complex scalar field components, $\varphi $ and $F$, in a
chiral superfield parameter $\Omega $ \ are properly activated. As a result,
the vector and scalar fields, $A_{\mu }$ and $S$, in the supermultiplet $%
\widehat{V}$ (\ref{sup}) transform as 
\begin{equation}
A_{\mu }\rightarrow a_{\mu }=A_{\mu }-\partial _{\mu }\varphi \text{ , \ \ }%
S\rightarrow s=S+F\text{ .}  \label{gra}
\end{equation}%
It can be immediately seen that our basic Lagrangian $\mathcal{L}^{\mathfrak{%
em}}$ (\ref{fin}, \ref{444}) being gauge invariant and containing no the
scalar $S$ field is automatically invariant under either of these two
transformations individually. In contrast, the supplementary vector field
constraint (\ref{const1}), though it is also turned out to be invariant
under supergauge transformations (\ref{gra}), but only if they act jointly.
Indeed, for any choice of the scalar $\varphi $ in (\ref{gra}) there can
always be found such a scalar $F$ (and vice versa) that the constraint
remains invariant. In other words, the vector field constraint is invariant
under supergauge transformations (\ref{gra}) but not invariant under an
ordinary gauge transformation. As a result, in contrast to the Wess-Zumino
case, the supergauge fixing in our case will also lead to an ordinary gauge
fixing. We will use this supergauge freedom to reduce the scalar field
bilinear $SS^{\ast }$ to some constant background value and find the final
equation for the gauge function $\varphi (x)$. It is convenient to come to
real field basis (\ref{bas}) $S_{\alpha }$ and $F_{\alpha }$ ($\alpha ,\beta
,...=1,2$) and choose the parameter fields $F_{\alpha }$ as 
\begin{equation}
F_{\alpha }=r_{\alpha }(\mathrm{M}+f)\text{ , }r_{\alpha }^{2}=1  \label{f}
\end{equation}%
so that the old $S_{\alpha }$ fields in (\ref{gra}) are related to the new
ones $s_{\alpha }$ in the following way 
\begin{equation}
S_{\alpha }=s_{\alpha }-r_{\alpha }(\mathrm{M}+f),\text{ \ }r_{\alpha
}S_{\alpha }=0,\text{ \ }S_{\alpha }S_{\alpha }=s_{\alpha }s_{\alpha }+(%
\mathrm{M}+f)^{2}\text{ \ }  \label{par11}
\end{equation}%
where $\mathrm{M}$ is a mass parameter, $f(x)$ is some Higgs field like
function, while $r_{\alpha }$ is a two-component unit "vector" being
orthogonal to the scalar "doublet" $S_{\alpha }$. Actually, the
parametrization (\ref{par11}) formally looks as if the old fields $S_{\alpha
}$ would develop the VEV, $\left\langle S_{\alpha }\right\rangle =-r_{\alpha
}\mathrm{M}$, due to which some related $SO(2)$ symmetry were spontaneously
violated and corresponding zero modes in terms of the new fields $s_{\alpha
} $ could be consequently produced. Eventually, for the properly chosen
"Higgs field" $f$%
\begin{equation}
f=-\mathrm{M}+\sqrt{\mathrm{M}^{2}-s_{\alpha }s_{\alpha }}
\end{equation}%
we come to 
\begin{equation}
A_{\mu }A^{\mu }=\mathrm{M}^{2}\text{ .}  \label{ert}
\end{equation}%
that is nothing but our old constraint (\ref{const}) taken for the time-like
SLIV. Recall that this constraint, as was thoroughly discussed in section
2.3, does \ not physically breaks gauge invariance. It rather fixes gauge to
which such a gauge function $\varphi (x)$ has to satisfy. Actually,
comparing the relation between the old and new vector fields in (\ref{gra})
with a conventional SLIV parametrization (\ref{gol}) one can find a simple
expression for this gauge function%
\begin{equation}
\varphi =\int^{x}d(n_{\mu }x^{\mu })\sqrt{\mathrm{M}^{2}-n^{2}a^{2}}\text{ }
\label{gf}
\end{equation}%
that explicitly demonstrates that this gauge condition is possible, at least
in the case when the new vector fields in (\ref{gra}) are taken in the terms
of the Lorentz breaking zero modes ($a^{2}=a_{\mu }a^{\mu }$,$\ n_{\mu
}a^{\mu }=0$).

To summarize, it was shown that the constraints on the allowed
configurations of the vector-superfield component fields\ (\ref{const1}),
that provide the potential energy stability in a general polynomially
extended Lagrangian (\ref{ext}),\ are entirely consistent with
supersymmetry. One might think that, unlike the gauge invariant linear
(Fayet-Iliopoulos) superfield term, the quadratic and higher order
superfield terms in the starting Lagrangian (\ref{ext}) would seem to break
gauge invariance. However, the fear proves groundless. In the broken SUSY
phase one eventually comes to the standard SUSY QED type Lagrangian (\ref%
{fin}) being supplemented by the vector field constraint invariant under
supergauge transformations. As a consequence, the gauge noninvariance
mentioned above amounts to the gauge fixing condition with a gauge function\
which can be explicitly calculated (\ref{gf}).

\section{On \textbf{emergent \textbf{SUSY} Standard Models and GUTs}\ }

\subsection{Potential of Abelian and non-Abelian vector superfields}

In this section we extend our discussion to the non-Abelian internal
symmetry case given by some group $G$ with generators $t^{p}$ (\ref{22}).
This case may correspond in general to some Grand Unified Theory which
includes the Standard Model and its possible extensions. For definiteness,
we will be further focused on the $U(1)\times SU(N)$ symmetrical theories,
though any other non-Abelian group in place of $SU(N)$ is also admissible.
Such a split group form is dictated by the fact that in the pure non-Abelian
symmetry case supersymmetry does not get spontaneously broken in a visible
sector that makes it inappropriate for an outgrowth of an emergence process%
\footnote{%
In principle, SUSY may be spontaneously broken in the visible sector even in
the pure non-Abelian symmetry case provided that the vector superfield
potential includes some essential high-dimension couplings.}. So, the theory
now contains the Abelian vector superfield $V$, as is given in (\ref{par1}),
and non-Abelian superfield multiplet $\boldsymbol{V}^{p}$ 
\begin{eqnarray}
\boldsymbol{V}^{p}(x,\theta ,\overline{\theta }) &=&\boldsymbol{C}%
^{p}+i\theta \boldsymbol{\chi }^{p}-i\overline{\theta }\overline{\boldsymbol{%
\chi }}^{p}+\frac{i}{2}\theta \theta \boldsymbol{S}^{p}-\frac{i}{2}\overline{%
\theta }\overline{\theta }\boldsymbol{S}^{p}  \notag \\
&&-\theta \sigma ^{\mu }\overline{\theta }\boldsymbol{A}_{\mu }^{p}+i\theta
\theta \overline{\theta }\overline{\boldsymbol{\lambda }^{\prime }}^{p}-i%
\overline{\theta }\overline{\theta }\theta \boldsymbol{\lambda }^{\prime p}+%
\frac{1}{2}\theta \theta \overline{\theta }\overline{\theta }\boldsymbol{D}%
^{\prime p},  \label{par3}
\end{eqnarray}%
where its vector field components $\boldsymbol{A}_{\mu }^{p}$ are usually
associated with an adjoint gauge field multiplet, $(\boldsymbol{A}_{\mu
})_{j}^{i}\equiv (\boldsymbol{A}_{\mu }^{p}t^{p})_{j}^{i}$ \ ($%
i,j,k=1,2,...,N$ ; $p,q,r=1,2,...,N^{2}-1$). Note that, apart from the
conventional gaugino multiplet $\boldsymbol{\lambda }^{p}$ and the auxiliary
fields $\boldsymbol{D}^{p}$, the superfield $\boldsymbol{V}^{p}$ contains in
general the additional degrees of freedom in terms of the dynamical scalar
and fermion field multiplets $\boldsymbol{C}^{p}$ and $\mathbf{\chi }^{p}$
and nondynamical complex scalar field $\boldsymbol{S}^{p}$. Note that for
the non-Abelian superfield components we use hereafter the bold symbols and
take again the brief notations, $\boldsymbol{\lambda }^{\prime p}=%
\boldsymbol{\lambda }^{p}+\frac{i}{2}\sigma ^{\mu }\partial _{\mu }\overline{%
\boldsymbol{\chi }}^{p}$\ and $\boldsymbol{D}^{\prime p}=\boldsymbol{D}^{p}+%
\frac{1}{2}\partial ^{2}\boldsymbol{C}^{p}$.

Augmenting the SUSY and $U(1)\times SU(N)$ invariant GUT by some polynomial
potential of vector superfields $V$ and $\boldsymbol{V}^{p}$ one comes to 
\begin{equation}
\mathfrak{L}=L_{SGUT}+\frac{1}{2}D^{2}+\frac{1}{2}\boldsymbol{D}^{p}%
\boldsymbol{D}^{p}+[\xi V+b_{1}V^{3}/3+b_{2}V(\boldsymbol{VV})+b_{3}(%
\boldsymbol{VVV})/3]_{D}  \label{np2}
\end{equation}%
where $\xi $ and $b_{1,2,3}$ stand for coupling constants, and the last term
in (\ref{np2}) contains products of the Abelian superfield $V$ and the
adjoint $SU(N)$ superfield multiplet $\boldsymbol{V_{j}^{i}\equiv (%
\boldsymbol{V}^{p}}t\boldsymbol{^{p})_{j}^{i}}$. The round brackets denote
hereafter traces for the superfield $\boldsymbol{V_{j}^{i}}$ 
\begin{equation}
(\boldsymbol{VV...})\equiv Tr(\boldsymbol{VV...})\text{ }  \label{tra}
\end{equation}%
and its field components (see below). For simplicity, we restricted
ourselves to the third degree superfield terms in the Lagrangian $\mathfrak{L%
}$ to eventually have a theory at a renormalizible level. Furthermore, we
have only taken the odd power superfield terms that provides, as we see
below, an additional discrete symmetry of the potential with respect to\ the
scalar field components in the $V$ and $\boldsymbol{V}^{p}$ superfields%
\begin{equation}
C\rightarrow -\text{ }C,\text{ \ \ }\boldsymbol{C}^{p}\rightarrow -\text{ }%
\boldsymbol{C}^{p}.  \label{dis}
\end{equation}%
Finally, eliminating the auxiliary $D$ and $\boldsymbol{D}^{p}$ fields in
the Lagrangian $\mathfrak{L}$ we come to the total potential for all
superfield bosonic field components written in terms of traces mentioned
above (\ref{tra}) 
\begin{eqnarray}
\mathfrak{U}_{B} &=&\mathfrak{U}_{B}(C,\boldsymbol{C})+\frac{1}{2}\text{ }%
b_{1}C(A_{%
{\mu}%
}A^{\mu }-S_{\alpha }S_{\alpha })+\frac{1}{2}\text{ }b_{2}C[(\boldsymbol{A}_{%
{\mu}%
}\boldsymbol{A}^{\mu })-(\boldsymbol{S}_{\alpha }\boldsymbol{S}_{\alpha }) 
\notag \\
&&+\frac{1}{2}b_{2}[A_{%
{\mu}%
}(\boldsymbol{A}^{\mu }\boldsymbol{C})-S_{\alpha }(\boldsymbol{S}_{\alpha }%
\boldsymbol{C})]+\frac{1}{2}b_{3}[(\boldsymbol{A}_{%
{\mu}%
}\boldsymbol{A}^{\mu }\boldsymbol{C})-(\boldsymbol{S}_{\alpha }\boldsymbol{S}%
_{\alpha }\boldsymbol{C})]\text{ }  \label{uuuu}
\end{eqnarray}%
where the potential terms depending only on scalar fields $C$ and $%
\boldsymbol{C_{j}^{i}\equiv (\boldsymbol{C}^{a}}t\boldsymbol{^{a})_{j}^{i}}$
are collected in 
\begin{equation}
\mathfrak{U}_{B}(C,\boldsymbol{C})=\frac{1}{8}[\xi +b_{1}C^{2}+b_{2}(%
\boldsymbol{CC})]^{2}+\frac{1}{2}[b_{2}^{2}C^{2}(\boldsymbol{CC}%
)+b_{2}b_{3}C(\boldsymbol{CCC})+\frac{1}{4}b_{3}^{2}(\boldsymbol{CCCC})]
\label{u'}
\end{equation}%
and complex scalar fields $S_{\alpha }$ and $\boldsymbol{S}_{\alpha }^{p}$
are now taken in the real field basis (\ref{bas}). One can see that all
these terms are invariant under the discrete symmetry (\ref{dis}), whereas
the vector field couplings in $\mathfrak{U}_{B}$ breaks it. However, they
vanish when the $V$ and $\boldsymbol{V}^{p}$ superfields are properly
constrained that we actually confirm in the next section.

As in the SUSY QED case (section 4.2), consider first the\ pure scalar field
potential $\mathfrak{U}_{B}(C,\boldsymbol{C})$. The \ corresponding extremum
conditions for $C$ and\ $\boldsymbol{C}^{a}$ fields are, 
\begin{eqnarray}
\mathfrak{U}_{C}^{\prime } &=&b_{1}(\xi +b_{1}C^{2})C+b_{2}(b_{1}-2b_{2})C(%
\boldsymbol{CC})=0,\text{ }  \notag \\
Tr(\mathfrak{U}_{\boldsymbol{C_{j}^{i}}}^{\prime }) &=&3b_{2}C(\boldsymbol{CC%
})+b_{3}(\boldsymbol{CCC})=0\text{ },  \label{extr}
\end{eqnarray}%
respectively\footnote{%
In more detail, we have first calculated here the variations $\mathfrak{U}%
_{C}^{\prime }=0$ and $\mathfrak{U}_{\boldsymbol{C_{j}^{i}}}^{\prime }=0$,
then took the trace from the second one (thus properly simplifying it due to
the traceless condition for the adjoint $SU(N)$ multiplet $Tr(\boldsymbol{%
C_{j}^{i})}=0$), and finally substituted it into the first one.}. As shows
the second partial derivative test, the simplest solution to the above
equations 
\begin{equation}
C_{0}=0\text{ , \ }\boldsymbol{C_{j}^{i}}=0  \label{sol3}
\end{equation}%
provides, under conditions put on the potential parameters,%
\begin{equation}
\xi ,\text{ }b_{1}>0\text{ ,\ }b_{2}\geq 0\text{ \ or \ }\xi ,\text{ }b_{1}<0%
\text{ ,\ }b_{2}\leq 0  \label{ccc}
\end{equation}%
its global minimum 
\begin{equation}
\mathfrak{U}_{B}(C,\boldsymbol{C})_{\min }^{as}=\frac{1}{8}\xi ^{2}\text{ .}
\label{min2}
\end{equation}%
This minimum corresponds to the broken SUSY phase with the unbroken internal
symmetry $U(1)\times SU(N)$ that is just what one would want to trigger an
emergence process. This minimum appears in fact due to the Fayet-Iliopoulos
linear term in the superfield polynomial in (\ref{np2}). As can easily be
confirmed, in absence of this term, namely, for $\xi =0$ and any arbitrary
values of all other parameters, there is only the SUSY symmetrical solution
with unbroken internal symmetry%
\begin{equation}
\mathfrak{U}_{B}(C,\boldsymbol{C})_{\min }^{sym}=0\text{ .}  \label{min3}
\end{equation}%
Interestingly, the symmetrical solution corresponding to the global minimum (%
\ref{min3}) may appear for the nonzero parameter $\xi $ as well 
\begin{equation}
\text{ }C_{0}^{(\pm )}=\pm \sqrt{-\xi /b_{1}}\text{, \ }\boldsymbol{C_{j}^{i}%
}=0  \label{sol4}
\end{equation}%
provided that 
\begin{equation}
\xi b_{1}<0\text{ .}  \label{cccc}
\end{equation}%
However, as we saw in the QED case, in the unbroken SUSY case one comes to
the trivial constant superfield when all factual constraints are included
into consideration (see equations (\ref{nc}) and the subsequent discussion)
and, therefore, this case is in general of little interest\footnote{%
It is worth noting that for nonzero $b_{1}$ values there are also lots of
local and global SUSY breaking minima with both nonzero scalar field vevs $%
C_{0}$ and $(\boldsymbol{C_{j}^{i})}_{0}$ in some parameter area ($b_{1,3}>0$
$(b_{1,3}<0)$ , $b_{2}<0$\ ($b_{2}>0$). This means that the $SU(N)$ symmetry
is also spontaneously broken in this case that otherwise (when $b_{1}=0$)
would not be happen in itself, as is clearly seen from the extremum
conditions (\ref{extr}).}.

\subsection{Constrained vector supermultiplets}

Let us now take the vector fields $A_{\mu }$ and $\boldsymbol{A}_{\mu }^{p}$
into consideration that immediately reveals that, in contrast to the pure
scalar field part (\ref{u'}), $\mathfrak{U}_{B}(C,\boldsymbol{C})$, the
vector field couplings in the total potential (\ref{uuuu}) make it unstable.
This happens, as was emphasized before, due the fact that bilinear term VEV
contributions of the vector fields $A_{\mu }$ and $\boldsymbol{A}_{\mu }^{p}$%
, as well as the auxiliary scalar fields $S_{\alpha }$ and $\boldsymbol{S}%
_{\alpha }^{p}$, are not properly compensated by appropriate four-linear
field terms which are generically absent in a supersymmetric theory
framework.

Again, as in the supersymmetric QED case considered above, the only possible
way to stabilize the ground state (\ref{sol3}, \ref{ccc}, \ref{min2}) seems
to seek the proper constraints on the superfields component fields ($C$, $%
\boldsymbol{C}^{p}$; $A_{\mu }$, $\boldsymbol{A}^{p}$; $S_{\alpha }$, $%
\boldsymbol{S}^{p}$) themselves rather than on their expectation values.
Provided that such constraints are physically realizable, the required
vacuum will be automatically stabilized. This will be done again through
some invariant Lagrange multiplier couplings simply adding their $D$ terms
to the above Lagrangian (\ref{np2}) 
\begin{equation}
\mathfrak{L}^{tot}=\mathfrak{L}+\frac{1}{2}\Lambda {\large (}%
V-C_{0})^{2}|_{D}+\frac{1}{2}\Pi (\boldsymbol{VV})|_{D}\text{ ,}
\label{ext1}
\end{equation}%
where $\Lambda (x,\theta ,\overline{\theta })$ and $\Pi (x,\theta ,\overline{%
\theta })$ are auxiliary vector superfields. Note that $C_{0}$ presented in
the first multiplier coupling is just the constant background value of the $%
C $ field for which the potential part $\mathfrak{U}_{B}(C,\boldsymbol{C})$
in (\ref{uuuu}) vanishes as appears for the supersymmetric minimum (\ref%
{min3}) or has some nonzero value corresponding to the SUSY breaking minimum
(\ref{min2}) in the visible sector. We will consider both cases
simultaneously using the same notation $C_{0}$ for either of the potential
minimizing values of the $C$ field. The second multiplier coupling in (\ref%
{ext1}) provides, as we will soon see, the vanishing background value for
the non-Abelian scalar field, $\boldsymbol{C}^{a}=0$, due to which the
underlying internal symmetry $U(1)\times SU(N)$ is left intact in both
unbroken and broken SUSY phase. As was emphasized before, the Lagrange
multiplier terms presented in (\ref{ext1}) have in fact the simplest
possible form that leads to some nontrivial constrained superfields $%
V(x,\theta ,\overline{\theta })$ and $\boldsymbol{V}^{p}(x,\theta ,\overline{%
\theta })$. Writing down their invariant $D$ terms through the component
fields one finds the precisely the same expression (\ref{lm1}) as in the
SUSY QED case for the Abelian superfield $V$ and the slightly modified one
for the non-Abelian superfield $\boldsymbol{V}^{a}$ 
\begin{eqnarray}
\Pi (\boldsymbol{VV})|_{D} &=&C_{\Pi }\left[ \boldsymbol{CD}^{\prime }%
\boldsymbol{+}\left( \frac{1}{2}\boldsymbol{SS}^{\ast }\boldsymbol{-\chi
\lambda }^{\prime }\boldsymbol{-}\overline{\boldsymbol{\chi }}\overline{%
\boldsymbol{\lambda }^{\prime }}\boldsymbol{-}\frac{1}{2}\boldsymbol{A}_{\mu
}\boldsymbol{A}^{\mu }\right) \right]  \notag \\
&&+\text{ }\chi _{\Pi }\left[ 2\boldsymbol{C\lambda }^{\prime }\boldsymbol{+}%
i(\boldsymbol{\chi S}^{\ast }\boldsymbol{+}i\sigma ^{\mu }\overline{%
\boldsymbol{\chi }}\boldsymbol{A}_{\mu })\right] +\overline{\chi }_{\Pi }[2%
\boldsymbol{C}\overline{\boldsymbol{\lambda }^{\prime }}-i(\overline{%
\boldsymbol{\chi }}\boldsymbol{S}-i\boldsymbol{\chi \sigma }^{\mu }%
\boldsymbol{A}_{\mu })]  \notag \\
&&+\text{ }\frac{1}{2}S_{\Pi }\left( \boldsymbol{CS}^{\ast }\boldsymbol{+}%
\frac{i}{2}\overline{\boldsymbol{\chi }}\overline{\boldsymbol{\chi }}\right)
+\frac{1}{2}S_{\Pi }^{\ast }\left( \boldsymbol{CS-}\frac{i}{2}\boldsymbol{%
\chi \chi }\right)  \notag \\
&&+\text{ }2A_{\Pi }^{\mu }(\boldsymbol{CA}_{\mu }\boldsymbol{-\chi }\sigma
_{\mu }\overline{\boldsymbol{\chi }})+2{\large \lambda }_{\Pi }^{\prime }(%
\boldsymbol{C\chi })+2\overline{{\large \lambda }}_{\Pi }^{\prime }(%
\boldsymbol{C}\overline{\boldsymbol{\chi }})+\frac{1}{2}D_{\Pi }^{\prime }(%
\boldsymbol{CC})  \label{lm2}
\end{eqnarray}%
where the pairly grouped field bold symbols mean hereafter the $SU(N)$
scalar products of the component field multiplets (for instance, $%
\boldsymbol{CD}^{\prime }=\boldsymbol{C}^{p}\boldsymbol{D}^{\prime p}$, and
so forth) and 
\begin{equation}
C_{\Pi },\text{ }\chi _{\Pi },\text{ }S_{\Pi },\text{ }A_{\Pi }^{\mu },\text{
}\lambda _{\Pi }^{\prime }=\lambda _{\Pi }+\frac{i}{2}\sigma ^{\mu }\partial
_{\mu }\overline{{\large \chi }}_{\Pi },\text{ }D_{\Pi }^{\prime }=D_{\Pi }+%
\frac{1}{2}\partial ^{2}C_{\Pi }  \label{123}
\end{equation}%
are the component fields of the Lagrange multiplier superfield $\Pi
(x,\theta ,\overline{\theta })$ in the standard parametrization (\ref{par3}%
). Varying the total Lagrangian (\ref{ext1}) with respect to the component
fields of both multipliers,\ (\ref{comp}) and (\ref{123}), and properly
combining their equations of motion we find the constraints which appear to
put on the $V$ and $\boldsymbol{V}^{a}$\ superfields components (in the same
way$^{14}$ for both Abelian and non-Abelian superfield case) 
\begin{eqnarray}
C &=&C_{0}\text{,\ \ }\chi =0,\text{\ \ }A_{\mu }A^{\mu }=S_{\alpha
}S_{\alpha },\text{ \ }  \notag \\
\boldsymbol{C}^{p} &=&0,\text{ \ }\boldsymbol{\chi }^{p}=0,\text{ \ }(%
\boldsymbol{A}_{\mu }\boldsymbol{A}^{\mu })=(\boldsymbol{S}_{\alpha }%
\boldsymbol{S}_{\alpha })\text{ , \ }\alpha =1,2\text{ .}  \label{const4}
\end{eqnarray}%
As before in the SUSY QED case, one may only have the time-like SLIV in a
supersymmetric $U(1)\times SU(N)$ framework but never the space-like one
(there also may be a light-like SLIV, if the $S$ and $\boldsymbol{S}$ fields
vanish). Also note that we only take the solution with initial values for
all fields (and their momenta) chosen so as to restrict the phase space to
vanishing values of the multiplier component fields (\ref{comp}) and (\ref%
{123}) that will provide a ghost-free theory with a positive Hamiltonian%
\footnote{%
As in the non-supersymmetric case discussed above (see footnote$^{3}$), this
solution with all vanishing components of the basic Lagrangian multiplier
superfields ${\large \Lambda }(x,\theta ,\overline{\theta })$ and ${\large %
\Pi }(x,\theta ,\overline{\theta })$ can be reached by introducing some
extra Lagrange multipliers.}.

Again, apart from the constraints (\ref{const4}) , one has the equations of
motion for all fields involved in the basic superfields $V(x,\theta ,%
\overline{\theta })$ and $\boldsymbol{V}^{p}(x,\theta ,\overline{\theta })$.
With vanishing multiplier component fields (\ref{comp}) and (\ref{123}), as
was proposed above, these equations appear in fact as extra constraints on
components of the $V$ and $\boldsymbol{V}^{p}$ superfields. Indeed,
equations of motion for the $S_{\alpha }$ , ${\large \chi }$ and $C$ fields,
on the one hand hand, and for the $\boldsymbol{S}_{\alpha }^{p},$ $%
\boldsymbol{\chi }^{p}$ and $\boldsymbol{C}^{p}$ fields, on the other, are
obtained by the corresponding variations of the total Lagrangian $\mathfrak{L%
}^{tot}$ (\ref{ext1}) including the potential (\ref{uuuu}). They are turned
out to be, respectively, 
\begin{eqnarray}
S_{\alpha }C_{0} &=&0\text{ , \ }\lambda C_{0}=0\text{ ,\ \ }(\xi
+b_{1}C_{0}^{2})C_{0}=0\text{ , }  \notag \\
\boldsymbol{S}_{\alpha }^{p}C_{0} &=&0\text{ , }\boldsymbol{\lambda }%
^{p}C_{0}=0,\text{ }b_{2}[A_{%
{\mu}%
}\boldsymbol{A}^{\mu }\boldsymbol{_{j}^{i}}-S_{\alpha }\boldsymbol{S}%
_{\alpha }\boldsymbol{_{j}^{i}}]+b_{3}[(\boldsymbol{A}_{%
{\mu}%
}\boldsymbol{A}^{\mu })\boldsymbol{_{j}^{i}}-(\boldsymbol{S}_{\alpha }%
\boldsymbol{S}_{\alpha })\boldsymbol{_{j}^{i}}]=0  \label{nc6}
\end{eqnarray}%
where the basic constraints (\ref{const4}) emerging at the potential $%
\mathfrak{U}_{B}(C,\boldsymbol{C})$ extremum point ($C_{0}$, $\boldsymbol{C}%
_{0}^{p}=0$) have been also used for both broken and unbroken SUSY case.
Note also that the equations for gauginos $\lambda $ and $\boldsymbol{%
\lambda }^{p}$ in (\ref{nc6}) are received by variation of the potential
terms in (\ref{np2}) containing fermion field couplings%
\begin{eqnarray}
\mathfrak{U}_{F} &=&b_{1}C(\chi \lambda ^{\prime }+\overline{\chi }\overline{%
\lambda ^{\prime }})+\text{ }b_{2}C[(\boldsymbol{\chi \lambda }^{\prime })+(%
\overline{\boldsymbol{\chi }}\overline{\boldsymbol{\lambda }^{\prime }})] 
\notag \\
&&+\frac{1}{2}b_{2}[\chi (\boldsymbol{\lambda }^{\prime }\boldsymbol{C})+%
\overline{\chi }(\overline{\boldsymbol{\lambda }^{\prime }}\boldsymbol{C}%
)+\lambda ^{\prime }(\boldsymbol{\chi C})+\overline{\lambda ^{\prime }}(%
\overline{\boldsymbol{\chi }}\boldsymbol{C})]  \notag \\
&&+b_{3}(\boldsymbol{\chi \lambda }^{\prime }\boldsymbol{C})+(\overline{%
\boldsymbol{\chi }}\overline{\boldsymbol{\lambda }^{\prime }}\boldsymbol{C})]%
\text{ .}  \label{fer}
\end{eqnarray}%
One can immediately see now that all equations in (\ref{nc6}) but the last
equation system\footnote{%
This equation system is not at all dependent on the critical $C$ field
value. It allows, as we will see in the next section, to eliminate the
auxiliary scalar fields $S_{\alpha }$ and $\boldsymbol{S}_{\alpha }^{a}$
from the theory thus properly expressing them through the vector fields $A_{%
{\mu}%
}$ and $\boldsymbol{A}_{\mu }^{a}$.} turn to trivial identities in the
broken SUSY case (\ref{sol3}) in which the corresponding $C$ field value
appears to be identically vanished, $C_{0}=0$. In the unbroken SUSY case (%
\ref{sol4}), this field value \ is definitely nonzero, $C_{0}=\pm \sqrt{-\xi
/b_{1}}$, and the situation is radically changed. Indeed, as follows from
the equations (\ref{nc6}), the auxiliary fields $S(x)$ and $\boldsymbol{S}%
^{p}$, as well as the gaugino fields $\lambda (x)$ and $\boldsymbol{\lambda }%
^{p}(x)$ have to be identically vanished. This causes in turn that the gauge
vector fields field $A_{\mu }$ and $\boldsymbol{A}_{%
{\mu}%
}^{p}$\ should also be vanished according to the basic constraints (\ref%
{const4}). So, we have to conclude, as in the SUSY QED case, that the
unbroken SUSY fails to provide stability of the potential (\ref{pot1a}) even
by constraining the superfields $V$ and $\boldsymbol{V}^{p}$ and, therefore,
only the spontaneously broken SUSY case could in principle lead to a
physically meaningful emergent theory.

\subsection{Broken SUSY phase: an emergent $U(1)\times SU(N)$ theory}

With the constraints (\ref{const4}) providing vacuum stability for the total
Lagrangian $\mathfrak{L}^{tot}$ (\ref{ext1}) we eventually come to the
emergent theory with a local $U(1)\times SU(N)$ symmetry that appears in the
broken SUSY phase (\ref{sol3}). Actually, implementing these constraints
into the Lagrangian through the Lagrange multiplier terms for component
fields one has 
\begin{eqnarray}
\mathfrak{L}^{\mathfrak{em}} &=&\mathfrak{L}_{SGUT}+\frac{1}{2}\xi D\text{ }+%
\frac{D_{\Lambda }}{4}(C-C_{0})^{2}-\frac{C_{\Lambda }}{4}\left( A_{\mu
}A^{\mu }-SS^{\ast }\right)  \notag \\
&&+\frac{D_{\Pi }}{4}(\boldsymbol{CC})-\frac{C_{\Pi }}{4}\left( \boldsymbol{A%
}_{\mu }\boldsymbol{A}^{\mu }\boldsymbol{-SS}^{\ast }\right)  \label{laaag}
\end{eqnarray}%
with the multiplier component functions $C_{\Lambda }$ and $D_{\Lambda }$ of
the auxiliary superfield $\Lambda $ (\ref{comp}) and component functions $%
C_{\Pi }$ and $D_{\Pi }$ of the auxiliary superfield $\Pi $ (\ref{123})
presented in the Lagrangian (\ref{ext1}). Again, with these constraints and
the emergence conjecture specified for non-Abelian theories in section 3.3,
one does not need to postulate gauge invariance for the physical SUSY GUT
Lagrangian $\mathfrak{L}_{SGUT}$ from the outset. Instead, one can derive
it. Indeed, even if the Lagrangian $\mathfrak{L}_{SGUT}$ is initially taken
to only possess the global $U(1)\times SU(N)$ symmetry it will tend to
uniquely acquire a standard gauge invariant form 
\begin{eqnarray}
\mathfrak{L}_{SGUT} &=&-\text{ }\frac{1}{4}F^{\mu \nu }F_{\mu \nu }+i\lambda
\sigma ^{\mu }\partial _{\mu }\overline{\lambda }+\frac{1}{2}D^{2}  \notag \\
&&-\text{ }\frac{1}{4}\boldsymbol{F}^{p\mu \nu }\boldsymbol{F}_{\mu \nu
}^{p}+i\boldsymbol{\lambda }^{p}\sigma ^{\mu }\mathcal{D}_{\mu }\overline{%
\boldsymbol{\lambda }}^{p}+\frac{1}{2}\boldsymbol{D}^{p}\boldsymbol{D}^{p}
\label{555}
\end{eqnarray}%
where the conventional gauge field strengths for both $U(1)$ and $SU(N)$
part and terms with proper covariant derivatives for gaugino fields $%
\boldsymbol{\lambda }^{p}$ necessarily appear. Again as in the pure Abelian
case, for the respectively constrained vector superfields $V$ and $%
\boldsymbol{V}^{p}$ we come in fact to a conventional SUSY GUT Lagrangian
with a standard gauge supermultiplet containing gauge bosons $A_{\mu }$ and $%
\boldsymbol{A}^{p}$, gauginos $\lambda $ and $\boldsymbol{\lambda }^{p}$,
and auxiliary scalar $D$ and $\boldsymbol{D}^{p}$ fields, whereas other
auxiliary scalar fields $S_{\alpha }$ and $\boldsymbol{S}_{\alpha }^{p}$ get
solely involved in the Lagrange multiplier terms (\ref{555}). Actually, the
only remnant of the polynomial potential of vector superfields $V$ and $%
\boldsymbol{V}^{p}$ (\ref{np2}) survived in the emergent theory (\ref{laaag}%
) appears to be the Fayet-Iliopoulos $D$-term which shows that supersymmetry
in the theory is indeed spontaneously broken and the $D$ field acquires the
VEV, $D=-\frac{1}{2}\xi $.

Let us show now that this theory is in essence gauge invariant and the
constraints (\ref{const4})\ on the field space appearing due to the Lagrange
multiplier terms in (\ref{ext1}) are consistent with supersymmetry. Namely,
as was argued above (section 4.4) though \ restricted vector superfields are
not strictly compatible with the linear superspace version of SUSY
transformations, their supermultiplet structure can be restored by
appropriate supergauge transformations. Following the same argumentation,
one can see that these transformations keep invariant the constraints (\ref%
{const4}) put on the vector fields $A_{\mu }$ and $\boldsymbol{A}^{p}$.
Leaving aside the $U(1)$ sector considered above in significant details, we
will now focus on the $SU(N)$ symmetry case with the constrained superfield $%
\boldsymbol{V}^{p}$ transformed as%
\begin{equation}
\boldsymbol{V}^{p}\rightarrow \boldsymbol{V}^{p}+\frac{i}{2}(\boldsymbol{%
\Omega -\Omega }^{\ast })^{p}  \label{ver}
\end{equation}%
The essential part of this transformation which directly acts on the vector
field constraint%
\begin{equation}
\boldsymbol{A}_{\mu }^{p}\boldsymbol{A}^{p\mu }=\boldsymbol{S}^{p}%
\boldsymbol{S}^{\ast p}  \label{vfc}
\end{equation}%
has the form 
\begin{equation}
\boldsymbol{V}^{p}\rightarrow \boldsymbol{V}^{p}+\frac{i}{2}\theta \theta 
\boldsymbol{F}^{p}-\frac{i}{2}\overline{\theta }\overline{\theta }%
\boldsymbol{F}^{\ast p}-\theta \sigma ^{\mu }\overline{\theta }\partial
_{\mu }\boldsymbol{\varphi }^{p}\text{ }  \label{trr}
\end{equation}%
where the real and complex scalar field components, $\boldsymbol{\varphi }%
^{p}$ and $\boldsymbol{F}^{p}$, in a chiral superfield parameter $%
\boldsymbol{\Omega }^{p}$ are properly activated. As a result, the
corresponding vector and scalar component fields, $\boldsymbol{A}_{\mu }^{p}$
and $\boldsymbol{S}_{\alpha }^{p}$, in the constrained supermultiplet $%
\boldsymbol{V}^{p}$ transform as 
\begin{equation}
\boldsymbol{A}_{\mu }^{p}\rightarrow \boldsymbol{a}_{\mu }^{p}=\boldsymbol{A}%
_{\mu }^{p}-\partial _{\mu }\boldsymbol{\varphi }^{p}\boldsymbol{,\ \ S}%
^{p}\rightarrow \boldsymbol{s}^{p}=\boldsymbol{S}^{p}+\boldsymbol{F}^{p}%
\text{ .}  \label{trrr}
\end{equation}%
One can readily see that our basic Lagrangian $\mathfrak{L}^{\mathfrak{em}}$
(\ref{laaag}) being gauge invariant and containing no the auxiliary scalar
fields $\boldsymbol{S}^{p}$ is automatically invariant under either of these
two transformations individually. In contrast, the supplementary vector
field constraint (\ref{vfc}), though it is also turned out to be invariant
under supergauge transformations (\ref{trrr}), but only if they act jointly.
Indeed, for any choice of the scalar $\boldsymbol{\varphi }^{p}$ in (\ref%
{trrr}) there can always be found such a scalar $\boldsymbol{F}^{a}$ (and
vice versa) that the constraint remains invariant. In other words, the
vector field constraint is invariant under supergauge transformations (\ref%
{trrr}) but not invariant under an ordinary gauge transformation. As a
result, in contrast to the Wess-Zumino case, the supergauge fixing in our
case will also lead to the ordinary gauge fixing. We will use this
supergauge freedom to reduce the scalar field bilinear $\boldsymbol{S}^{p}%
\boldsymbol{S}^{\ast p}$ to some constant background value and find a final
equation for the gauge function $\boldsymbol{\varphi }^{p}(x)$. It is
convenient to come to real field basis (\ref{bas}) for scalar fields $%
\boldsymbol{S}_{\alpha }^{p}$ and $\boldsymbol{F}_{\alpha }^{p}$ ($\alpha
=1,2$), and choose the parameter fields $\boldsymbol{F}_{\alpha }^{a}$ as 
\begin{equation}
\boldsymbol{F}_{\alpha }^{p}=r_{\alpha }\boldsymbol{\epsilon }^{p}(\mathbf{M}%
+\boldsymbol{f}),\boldsymbol{\ }r_{\alpha }\boldsymbol{s}_{\alpha }^{p}=0,%
\text{ \ }r_{\alpha }^{2}=1,\text{ }\boldsymbol{\epsilon }^{p}\boldsymbol{%
\epsilon }^{p}=1
\end{equation}%
so that the old $\boldsymbol{S}_{\alpha }^{p}$ fields in (\ref{trrr}) are
related to the new ones $\boldsymbol{s}_{\alpha }^{p}$ in the following way 
\begin{equation}
\boldsymbol{S}_{\alpha }^{p}=\boldsymbol{s}_{\alpha }^{p}-r_{\alpha }%
\boldsymbol{\epsilon }^{p}(\mathbf{M}+\boldsymbol{f}),\text{ }r_{\alpha }%
\boldsymbol{s}_{\alpha }^{p}=0,\text{ }\boldsymbol{S}_{\alpha }^{p}%
\boldsymbol{S}_{\alpha }^{p}=\boldsymbol{s}_{\alpha }^{p}\boldsymbol{s}%
_{\alpha }^{p}+(\mathbf{M}+\boldsymbol{f})^{2}\text{.}  \label{qq}
\end{equation}%
where $\mathbf{M}$ is a new mass parameter, $\boldsymbol{f}(x)$ is some
Higgs field like function, $r_{\alpha }$ is again the two-component unit
"vector" chosen to be orthogonal to the scalar $\boldsymbol{s}_{\alpha }^{p}$%
, while $\boldsymbol{\epsilon }^{p}$ is the unit $SU(N)$ adjoint vector.
Again, this parametrization for the old fields $\boldsymbol{S}_{\alpha }^{p}$
formally looks as if they develop the VEV, $\left\langle \boldsymbol{S}%
_{\alpha }^{p}\right\rangle =$ $-r_{\alpha }\boldsymbol{\epsilon }^{p}%
\mathbf{M}$, due to which the related $SO(2)\times SU(N)$ symmetry would be
spontaneously violated and corresponding zero modes in terms of the new
fields $\boldsymbol{s}_{\alpha }^{p}$ could be consequently produced
(indeed, they they never appear in the theory). Eventually, for an
appropriate choice of \ the Higgs field like function $\boldsymbol{f}(x)$ in
(\ref{qq}) 
\begin{equation}
\boldsymbol{f}=-\mathbf{M}+\sqrt{\mathbf{M}^{2}-\boldsymbol{s}_{\alpha }^{p}%
\boldsymbol{s}_{\alpha }^{p}}  \label{w}
\end{equation}%
we come in (\ref{vfc}) to the condition 
\begin{equation}
\boldsymbol{A}_{\mu }^{p}\boldsymbol{A}^{p\mu }=\mathbf{M}^{2}\text{ .\ }
\label{111}
\end{equation}%
conforming with a general non-Abelian vector field constraint (\ref{constt})
established above in section 3.3. As the vector field constraint (\ref{ert})
for the $U(1)$ symmetry case this constraint also leads exclusively to the
time-like SLIV. Again, one can calculate the gauge function $\boldsymbol{%
\varphi }^{p}(x)$ comparing the relation between the old and new vector
fields in (\ref{trrr}) with a conventional SLIV parametrization for
non-Abelian vector fields (\ref{supp}) 
\begin{equation}
\boldsymbol{\varphi }^{p}=\boldsymbol{\epsilon }^{p}\int^{x}d(n_{\mu }x^{\mu
})\sqrt{\mathbf{M}^{2}-\boldsymbol{n}^{2}\boldsymbol{a}^{2}}\text{ }
\end{equation}%
expressing it through the Goldstone and pseudo-Goldstone modes $\boldsymbol{a%
}_{\mu }^{p}$ involved ($\boldsymbol{a}^{2}\equiv \boldsymbol{a}_{\mu }^{p}%
\boldsymbol{a}^{p\mu }$).

Remarkably, thanks to a generic high symmetry of the constraint (\ref{111})
one can apply the emergence conjecture with dynamically produced massless
gauge modes to any non-Abelian internal symmetry case as well, though SLIV
itself could produce only one zero vector mode. The point is, as was it
presented in significant detail in section 3.3, that although we only
propose Lorentz invariance $SO(1,3)$ and internal symmetry $U(1)\times SU(N)$
of the Lagrangian $\mathfrak{L}^{\mathfrak{em}}$ (\ref{laaag}), the emerged
constraint (\ref{111}) in fact possesses a much higher accidental symmetry $%
SO(\Upsilon ,3\Upsilon )$ determined by the dimension $\Upsilon =N^{2}-1$ of
the $SU(N)$ adjoint representation to which the vector fields $\boldsymbol{A}%
_{\mu }^{p}$ belong\footnote{%
Actually, a total symmetry even higher if one keeps in mind both constraints
(\ref{const}) and (\ref{111}) put on the vector fields $A_{\mu }$ and $%
\boldsymbol{A}_{\mu }^{a}$, respectively. As long as they are independent
the related total symmetry is in fact $SO(1,3)\times SO(\Upsilon ,3\Upsilon
) $ until it starts breaking.}. This symmetry is indeed spontaneously broken
at a scale $\mathbf{M}$ leading exclusively to the time-like SLIV case (\ref%
{ss}), as is determined by the positive sign in the SUSY SLIV constraint (%
\ref{111}). The emerging pseudo-Goldstone vector bosons, as was thoroughly
explained in section 3.3, may be in fact considered as candidates for
non-Abelian gauge fields which together with the true vector Goldstone boson
entirely complete the adjoint multiplet of the internal symmetry group $%
SU(N) $. Remarkably, they remain strictly massless being protected by the
simultaneously generated non-Abelian gauge invariance. When expressed in
these zero modes, the theory look essentially nonlinear and contains many
Lorentz and CPT violating couplings. However, as in the SUSY QED case, they
do not lead to physical SLIV effects which due to simultaneously generated
gauge invariance appear to be strictly cancelled out.

Finally, it is worth noting that with parameterization (\ref{gol}, \ref{supp}%
, \ref{par11}, \ref{qq}, \ref{w}) taken above for Abelian and non-Abelian
vector and scalar field components, one comes to the following relations
between them 
\begin{eqnarray}
&&s_{\alpha }\boldsymbol{s}_{\alpha j}^{\text{ \ }i}+\boldsymbol{\epsilon }%
_{j}^{i}\sqrt{\mathrm{M}^{2}-s^{2}}\sqrt{\mathbf{M}^{2}-\boldsymbol{s}^{2}}+%
\frac{b_{3}}{b_{2}}\left[ (\boldsymbol{s}_{\alpha }\boldsymbol{s}_{\alpha
})_{j}^{i}-(\boldsymbol{\epsilon \epsilon })_{j}^{i}\boldsymbol{s}^{2}\right]
\notag \\
&=&a_{\mu }\boldsymbol{a}_{\mu j}^{\text{ \ }i}+\boldsymbol{\epsilon }%
_{j}^{i}\sqrt{\mathrm{M}^{2}-a^{2}}\sqrt{\mathbf{M}^{2}-\boldsymbol{a}^{2}}+%
\frac{b_{3}}{b_{2}}\left[ (\boldsymbol{a}_{\mu }\boldsymbol{a}^{\mu
})_{j}^{i}-(\boldsymbol{\epsilon \epsilon })_{j}^{i}\boldsymbol{a}^{2})%
\right] \text{ \ }  \label{rell}
\end{eqnarray}%
as is determined by the equation system in (\ref{nc6}) (with a full
contraction of the field indices in $s^{2}\equiv s_{\alpha }s_{\alpha },$ $%
\boldsymbol{s}^{2}\equiv \boldsymbol{s}_{\alpha }^{p}\boldsymbol{s}_{\alpha
}^{p}$, $\ a^{2}\equiv a_{\mu }a^{\mu }$ and $\boldsymbol{a}^{2}\equiv 
\boldsymbol{a}_{\mu }^{p}\boldsymbol{a}^{p\mu }$).\ They allow to express
the auxiliary scalar fields $s_{\alpha }$ and $\boldsymbol{s}_{\alpha j}^{%
\text{ \ }i}$ through the vector zero modes $a_{\mu }$ and $\boldsymbol{a}%
_{\mu }^{p}$, thus completely excluding the formers from the theory.

\subsection{Some immediate outcomes}

Quite remarkably, an obligatory split symmetry form $U(1)\times SU(N)$ (or $%
U(1)\times G$, in general) of plausible emergent theories which could exist
beyond the prototype QED case, leads us to the standard electroweak theory
with the $U(1)\times SU(2)$ symmetry as the simplest possibility. The
potential of type (\ref{np2}) written for the corresponding superfields
requires spontaneous SUSY breaking in the visible sector to avoid the vacuum
instability in the theory. Eventually, this requires the SLIV type
constraints to be put on the hypercharge and weak isospin vector fields,
respectively,%
\begin{equation}
B_{\mu }B^{\mu }=\mathrm{M}^{2}\text{ , \ }\boldsymbol{W}_{\mu }^{p}%
\boldsymbol{W}^{p\mu }=\mathbf{M}^{2}\text{ \ }(p=1,2,3).  \label{ew}
\end{equation}%
These constraints are independent from each other and possess, as was
generally argued above, the total symmetry $SO(1,3)\times SO(3,9)$ which is
much higher than the actual Lorentz invariance and electroweak $U(1)\times
SU(2)$ symmetry in the theory. Thanks to this fact, one Goldstone and three
pseudo-Goldstone zero vector modes $b_{\mu }$ and $\boldsymbol{w}_{\mu }^{p}$
are generated to eventually complete the gauge multiplet of the Standard
Model%
\begin{eqnarray}
B_{\mu } &=&b_{\mu }+n_{\mu }\sqrt{\mathrm{M}^{2}-b_{\mu }b^{\mu }}\text{ ,
\ }n_{\mu }b_{\mu }=0\text{ ,}  \notag \\
\boldsymbol{W}_{\mu }^{p} &=&\boldsymbol{w}_{\mu }^{p}+n_{\mu }\boldsymbol{%
\epsilon }^{p}\sqrt{\mathbf{M}^{2}-\boldsymbol{w}_{\mu }^{q}\boldsymbol{w}%
^{q\mu }}\text{ , }n_{\mu }\boldsymbol{w}^{p\mu }=0\text{ }  \label{ew1}
\end{eqnarray}%
where the unit vectors $n_{\mu }$ and $\boldsymbol{\epsilon }^{p}$ are
defined in accordance with a rectangular unit matrix $\boldsymbol{n}_{\mu
}^{p}$ taken in the two-vector form (\ref{vec}). The true vector Goldstone
boson appear to be some superposition of the zero modes $b_{\mu }$ and $%
\boldsymbol{w}_{\mu }^{3}$. This superposition is in fact determined by the
conventional Higgs doublet in the model since just through the Higgs field
couplings these modes are only mixed. Thus, when the electroweak symmetry
gets spontaneously broken an accidental degeneracy related to the total
symmetry of constraints mentioned above is lifted. As a consequence, the
vector pseudo-Goldstones acquire masses and only photon, being the true
vector Goldstone boson in the model, is left massless\footnote{%
More details on how the zero vector modes can acqure masses both in emergent
QED and SM can be found in \cite{kep, par}.}. In this sense, there is not
much difference for photon in emergent QED and SM: it emerges as a true
vector Goldstone boson in both frameworks.

Going beyond the Standard Model we unavoidably come to the flipped $SU(5)$
GUT \cite{fl} as a minimal and in fact distinguished possibility. Indeed,
the $U(1)$ symmetry part being mandatory for emergent theories now naturally
appears as a linear combination of a conventional electroweak hypercharge
and another hypercharge belonging to the standard $SU(5).$ The flipped $%
SU(5) $ GUT has several advantages over the standard $SU(5)$ one - the
doublet-triplet splitting problem is resolved with use of only minimal Higgs
representations and protons are naturally long lived, neutrinos are
necessarily massive, and supersymmetric hybrid inflation can easily be
implemented successfully. Also in string theory, the flipped $SU(5)$ model
are of significant interest for a variety of reasons. In essence, the
above-mentioned natural solution to the doublet-triplet splitting problem
without using large GUT representations is in the remarkable conformity with
string theories where such representations are typically unavailable. Also,
in weakly coupled heterotic models, the flipped $SU(5)$ allows to achieve
gauge coupling unification at the string scale $10^{17}$ \textrm{GeV }if
some extra vector-like particles are added. They are normally taken to
transform in the $10$ and $\overline{10}$ representations, that is easy to
engineer in string theory.

So, supersymmetric emergent theories look attractive both theoretically and
phenomenologically whether they are considered at low energies in terms of
the Standard Model or at very high energies as the flipped $SU(5)$ GUTs
being inspired by superstrings. However, their most generic manifestations
seem to be related to a spontaneous SUSY violation in the visible sector
that we discuss in the next section.

\section{Phenomenological implications: photino as pseudo-goldstino}

Let us now turn to the matter sector being described by chiral matter
superfields which have not yet been included both in QED and the Standard
Model. In their presence the SUSY breaking in the tree approximation we have
used above is in fact phenomenologically ruled out by the well-known
supertrace sum rule \cite{wess}. In a supersymmetric QED it looks especially
simple%
\begin{equation}
STr\mathcal{M}^{2}\equiv \sum_{\mathrm{J}}(-1)^{2\mathrm{J}}(2\mathrm{J}%
+1)Tr(\mathrm{m}_{\mathrm{J}}^{2})=2Tr\mathrm{Q}\left\langle D\right\rangle
\label{sr}
\end{equation}%
where $\mathrm{m}_{\mathrm{J}}$ is the mass matrix for spin $\mathrm{J}$
fields, $\mathrm{Q}$ is the electric charge matrix of the chiral superfields
under consideration, and $\left\langle D\right\rangle $ is the VEV of the
gauge superfield $D$ component. One can easily confirm that for all
realistic cases requiring $Tr\mathrm{Q}=0$ to cancel the anomalies related
to $U(1)_{em}$ this sum rule leads to some unacceptably light superpartners
in a theory\footnote{%
Even worse, because particles with different electric charges cannot mix,
the supertrace (\ref{sr}) vanishes separately in each charge sector, thus
leading to light sparticles for all types of charges individually.}.

Usually, solution to this problem is related to a softly broken SUSY \cite%
{wess} that in our case would be inaccessible. Indeed, inclusion of direct
soft mass terms for superpartners in the model would mean that the visible
SUSY is explicitly, rather than spontaneously, broken that would immediately
invalidate the whole idea of an emergence nature of QED and SM. Therefore,
we need models where SUSY spontaneously breaks, at least partially, in the
visible sector as well. Actually, in the presence of a hidden sector, an
additional visible SUSY breaking is not forbidden phenomenologically. Below,
we will also consider a class of the pure visible SUSY breaking models,
where supersymmetry is solely broken at tree level. Since this section is
largely concerned with the phenomenological aspects of emergent SUSY
theories, it is reasonable to consider them in a context of the entire $%
SU(3)_{\mathrm{C}}\times SU(2)_{\mathrm{L}}\times U(1)_{\mathrm{Y}}$
Standard Model, rather than in the pure QED framework.

\subsection{Two-sector SUSY breaking {}}

According to a conventional two-sector paradigm, supersymmetry breaking
entirely occurs in a hidden sector and then this breaking is mediated to the
visible sector by some indirect interactions whose nature depends on a
particular mediation scenario \cite{wess}. An emergent approach for QED and
SM advocated here requires some modification of this idea. While a hidden
sector is largely responsible for supersymmetry breaking providing a
reliable solution to the problem of superpartner masses in a theory,
supersymmetry itself can also be spontaneously broken in the visible sector
that ultimately leads to a double spontaneous SUSY breaking pattern. As a
result, the simplified picture discussed above in the SUSY QED case (section
4) is properly changed: a strictly massless fermion eigenstate, a true
goldstino $\zeta _{g}$, should now be some mix of the visible sector photino 
$\lambda $ and the hidden sector\ goldstino $\kappa ^{\prime }$,

\begin{equation}
\zeta _{g}=\frac{\left\langle D\right\rangle \lambda +\left\langle F^{\prime
}\right\rangle \kappa ^{\prime }}{\sqrt{\left\langle D\right\rangle
^{2}+\left\langle F^{\prime }\right\rangle ^{2}}}\text{ ,}  \label{tr}
\end{equation}%
where $\left\langle D\right\rangle $ and $\left\langle F^{\prime
}\right\rangle $ are the corresponding $D$- and $F$-term VEVs in the visible
and hidden sectors, respectively (we use the primed letters for the hidden
sector entities)\footnote{%
Note that what we call photino in QED is the linear combination of bino and
neutral wino in the SM faramework. Thus the term photino means hereafter the
"photino content" of the neutralino states involved rather than the pure
photino state.}. We have also proposed that spontaneous SUSY breaking in the
hidden sector goes basically through the $F$-terms VEVs and, in addition, we
neglected possible mixing in (\ref{tr}) with other neutralinos both in
visible and hidden sectors. So, the orthogonal combination of these states,
which may be referred to as a pseudo-goldstino, is 
\begin{equation}
\zeta _{pg}=\frac{\left\langle F^{\prime }\right\rangle \lambda
-\left\langle D\right\rangle \kappa ^{\prime }}{\sqrt{\left\langle
D\right\rangle ^{2}+\left\langle F^{\prime }\right\rangle ^{2}}}\text{ .}
\label{ps}
\end{equation}%
In the supergravity context, a true goldstino $\zeta _{g}$ is eaten through
the super-Higgs mechanism to form the longitudinal component of a massive
gravitino $\zeta _{G}$, while a pseudo-goldstino $\zeta _{pg}$ gets some
mass whose value depends on a particular mediation scenario taken. However,
in any case, due to large soft masses required to be mediated, one may
generally expect that SUSY is much stronger broken in the hidden sector than
in the visible one, $\left\langle F^{\prime }\right\rangle >>$ $\left\langle
D\right\rangle $. This means in turn the pseudo-goldstino (\ref{ps}) is
largely given by the pure photino state,%
\begin{equation}
\zeta _{pg}\simeq \lambda \text{ .}  \label{ph}
\end{equation}%
These pseudo-Goldstone photinos seem to be of special observational interest
in the model that, apart from some indication of the SM emergence nature,
may shed light on SUSY breaking physics. The possibility that the
supersymmetric SM visible sector might also spontaneously break SUSY thus
giving rise to some pseudo-goldstino state was also considered, though in a
different context, in \cite{vis, tha}. Though this idea may be implemented
in supersymmetric QED or SM with practically any hidden sector SUSY breaking
scenario, we choose gauge-mediated scheme. This scenario allows for a
natural suppression of flavour violations in the supersymmetric sector \cite%
{wess} and have very distinctive phenomenological features.

Let us note first of all that our polynomially extended QED and SM
Lagrangians (\ref{slag}) and (\ref{np2}) are not only SUSY invariant but
also generically possesses continuous $R$-symmetry $U(1)_{R}$ \cite{wess}.
Indeed, vector superfields always have zero $R$-charge, since they are real.
Accordingly, it follows that the physical components in the constrained
vector superfield $\widehat{V}$ (\ref{sup}) transform as%
\begin{equation}
A_{\mu }\rightarrow A_{\mu }\text{ , \ }\lambda \rightarrow e^{i\alpha
}\lambda \text{ , \ }D\rightarrow D  \label{tran}
\end{equation}%
and, therefore, they have $R$ charges $0$, $1$ and $0$, respectively. Along
with that, we assume a suitable $R$-symmetric matter superfield setup as
well making a proper $R$-charge assignment for basic fermions and scalars
(and messenger fields) involved. This will lead to the light
pseudo-Goldstone matter in the gauge-mediated scenario \cite{vis, tha}.
Normally, if the visible sector possesses the $R$-symmetry which is
preserved in the course of the mediation, then the masslessness of a photino
(a gaugino, in general) is protected up to the supergravity effects which
violate $R$-symmetry\footnote{%
Note that Majorana masses for gauginos always break a continuous $R$%
-symmetry, as is clearly seen from transformations (\ref{tran}). For $R$%
-invariance one might properly extend a field content in the theory so as to
achieve Dirac gaugino masses (that is not yet assumed in our case).
Remarkably, the properly arranged $R$-symmetry in the theory supplemented by
additional matter and Higgs chiral supermultiplets may lead to a very
efficient suppression of flavor-changing effects \cite{kr}.}. As a result,
our pseudo-goldstino will acquire the mass being proportional to the
gravitino mass. The latter can be typically estimated as 
\begin{equation}
\mathrm{m}_{3/2}\simeq \left\langle F^{\prime }\right\rangle /\mathrm{M}_{P}
\label{3/2}
\end{equation}%
(where we omitted the negligible $D$-term VEV contribution from the visible
sector) that simply follows from dimensional analysis, since this mass must
vanish in the limits when supersymmetry is restored ($\left\langle F^{\prime
}\right\rangle \rightarrow 0$) and when gravity is turned off ($\mathrm{M}%
_{P}\rightarrow \infty $). Once the gravitino mass is fixed by the properly
chosen scale $\left\langle F^{\prime }\right\rangle $ of the hidden sector
SUSY breaking, it is straightforward to calculate the supergravity
contribution to the pseudo-goldstino mass (see \cite{tha} and references
therein). It appears that in theories with both $F$-term and $D$-term
visible sector breakings, pseudo-goldstino acquires a mass which is always
lighter (much lighter in the most parameter space) than twice the gravitino
mass, $\mathrm{m}_{pg}<2\mathrm{m}_{3/2}$. This means that the
pseudo-goldstino $\zeta _{pg}$, being practically the visible sector photino 
$\lambda $ (\ref{ph}), is in fact the lightest supersymmetric particle (LSP)
in the model considered. Taking the mass $\mathrm{m}_{3/2}$ to be much
smaller than the weak scale, say being of the \textrm{k}$\mathrm{eV}$ order
or less, one naturally comes to a possible solution for both gravitino and
pseudo-goldstino overproduction problems in the early universe \cite{tha}.

Apart from cosmological problems, many other sides of new physics related to
pseudo-goldstinos appearing through the multiple SUSY breaking were also
studied recently (see \cite{vis, tha, gol} and references therein). The
point is, however, that there have been exclusively used non-vanishing $F$%
-terms as the only mechanism of visible sector SUSY breaking\footnote{%
We briefly consider below this case to make clear a significant difference
between the $F$-term visible sector SUSY breaking with our $D$-term breaking
(see below). In the framework of supersymmetric SM, some minimal setup \cite%
{vis} of the visible sector $F$-term SUSY breaking includes, together with
ordinary Yukawa interactions for quarks and leptons, a simple O'Raifeartaigh
type superpotential. So, the total superpotential is 
\begin{equation*}
W=W_{Yuk}+fX(H_{u}H_{d}-\eta )+\mu _{u}H_{u}R_{u}+\mu _{d}H_{d}R_{d}
\end{equation*}%
where, apart from the standard Higgs doublets $H_{u,d}$, the new Higgs
doublets $R_{u,d}$ \ appear and also, like the next-to-minimal \
supersymmetric SM, there is a gauge singlet field $X$ ($f,\eta ,\mu _{u,d}$
stand for some coupling constants and mass parameters). This superpotential
possesses $R$-symmetry with $R$ charges $0$, $1$, and $2$ for standard Higgs
doublets $H_{u,d}$, quarks and leptons ($Q,U^{c},D^{c},L,E^{c}$) and extra
superfields ($R_{u},R_{d},X$), respectively. Remarkably, in the absence of
gauge interactions, this\ superpotential on its own is an example of a
Wess-Zumino model having, as argued in \cite{vr}, the persistent zero mode
which remains for arbitrary scalar field configurations emerged. In the
entire framework of supersymmetric SM with a hidden sector included this
mode appears as a massless (at tree level) pseudo-goldstino mode being
cosmologically safe or dangerous depending whether $R$-symmetry is exact or
appreciably broken.}. In this connection, our pseudo-Goldstone photinos
caused by non-vanishing $D$-term in the visible sector SUSY may lead to
somewhat different observational consequences.

One interesting difference concerns the $R$-symmetry role in these
approaches, though they both may typically start with $R$-invariant setup,
as we discussed above. However, for an appreciable $R$-symmetry violation
due to the SUSY breaking mediation one would come to dramatic consequences
in the $F$-term visible sector SUSY breaking case being basically determined
by the superpotential mentioned above$^{27}$. The reason is that even after
coupling of the visible sector to a hidden source of SUSY breaking a light
pseudo-goldstino persists as a remnant of the original visible SUSY breaking
dynamics \cite{tha}. Its tree-level mass is suppressed because it is only
induced by small mixings with gauginos, while at one loop its mass is\ still
protected by the visible sector $R$-symmetry. Actually, though $R$-violating
mediation causes in general some rise of the pseudo-goldstino mass, it is
always one loop factor suppressed relative to the weak scale and typically
located in the cosmologically dangerous range $O(10$ $MeV-1$ $\mathrm{GeV})$%
. As to interactions, the pseudo-goldstino inherits rather small couplings
to supersymmetric SM fields through the mixing with gauginos and higgsinos
that determines its lifetime being typically longer than a second, the time
at which Big Bang Nucleosynthesis begins. As a result, one is unavoidably
led to the conclusion that the visible sector pseudo-goldstino is
generically overproduced in the early universe, unless $R$-symmetry remains.
In contrast, in $D$-term visible sector SUSY breaking case nothing dramatic
would happen if $R$-symmetry were really violated in the course of the
mediation. Depending on the particular type of this violation the
pseudo-goldstino which now is essentially the visible sector photino $%
\lambda $ (being properly mixed with other neutralinos) could in principle
become the next-to-lightest supersymmetric particle (NLSP) which then decays
into gravitino and photon (see the next section).

Another, and more touchable, difference belongs to the Higgs boson decays in
the supersymmetric SM framework. For light pseudo-goldstino and gravitino
these decays are appreciably modified. Actually, for the $F$-term visible
sector SUSY breaking$^{27}$ the dominant channel becomes \cite{vis,tha} a
conversion of the Higgs boson (say, the lighter $CP$-even Higgs boson $h^{0}$%
) into a conjugated pair of corresponding pseudo-sgoldstinos $\phi _{pg}$
and $\overline{\phi }_{pg}$ 
\begin{equation}
h^{0}\rightarrow \phi _{pg}+\overline{\phi }_{pg}
\end{equation}%
being superpartners of\ pseudo-goldstinos $\zeta _{pg}$ and $\overline{\zeta 
}_{pg}$, respectively. If this decay is kinematically allowed, one may
conclude that the Higgs boson could dominantly decay invisibly. By contrast,
for $D$-term SUSY\ \ breaking case considered here the roles of a
pseudo-goldstino and a pseudo-sgoldstino are just played by a photino and a
photon, respectively, that could make the standard two-photon decay channel
of Higgs boson to be even somewhat enhanced. In the light of recent
discovery of Higgs-like state \cite{h} just through its visible decay modes,
the $F$-term SUSY breaking in the visible sector seems to be disfavored by
data, while $D$-term SUSY breaking is not yet in trouble with them. \ 

\subsection{Pure visible sector SUSY breaking scenario}

Let us consider now the pure visible sector SUSY breaking models which, on
contrary to conventional lore, can also be constructed (see \cite{lyk} and
references therein). They appear to include some relatively low-scale extra
hypercharge $U(1)_{\mathrm{Y}^{\prime }}$ gauge symmetry which, when being
properly assigned to quarks and leptons and their superpartners, allows to
construct some phenomenologically viable supersymmetric SM extensions. So,
for the tree level supertrace equation (\ref{sr}) one has on its right-hand
side 
\begin{equation}
STr\mathcal{M}^{2}=2\left[ \mathrm{g}_{\mathrm{Y}}Tr(\mathrm{Y})\left\langle
D\right\rangle +\mathrm{g}_{\mathrm{Y}^{\prime }}Tr(\mathrm{Y}^{\prime
})\left\langle D^{\prime }\right\rangle \right] \text{ .}  \label{sr1}
\end{equation}%
where $\mathrm{g}_{\mathrm{Y}}$ and $\mathrm{g}_{\mathrm{Y}^{\prime }}$ are
the corresponding gauge coupling constants. The first term in the bracket
related to the standard $U(1)$ hypercharge symmetry will vanish since the
quark and lepton representations are chosen to be anomaly free that leads to
the traceless condition $Tr(\mathrm{Y})=0$. However, if in the second term
in (\ref{sr1}) the $D$-term VEV $\left\langle D^{\prime }\right\rangle $ is
nonvanishing and the trace $Tr(\mathrm{Y}^{\prime })$ over quarks and
leptons is separately nonzero, as is the case when all quark and lepton
superfields (as well as Higgs superfields) are given $\mathrm{Y}^{\prime }$%
-hypercharges of the same sign\footnote{%
The simplest choice would be to assign positive $Y^{\prime }$-hypercharges ($%
Y^{\prime }=+1$) to all quark and lepton superfields and negative $Y^{\prime
}$-hypercharges ($Y^{\prime }=-2$) to the Higgs superfields $H_{u,d}$ (for
some earlier discussions, see the first paper in \cite{wess} and references
therein).}, then all the sparticles can receive large masses. Normally, the
extra $U(1)_{\mathrm{Y}^{\prime }}$ hypercharge gauge symmetry is broken at
tree level and the corresponding gauge boson $Z^{\prime }$ acquires a mass.
Its lower bound has been recently pushed up to $\mathrm{M}_{Z^{\prime }}$ $%
>2.33$ \textrm{T}$\mathrm{eV}$ at LHC \cite{ber}. In general, the $Z^{\prime
}$ boson is mixed with an ordinary $Z$ boson of the SM. As of now, for the $%
\mathrm{M}_{Z^{\prime }}$ bound value mentioned, their mixing angle\ appears
well below its experimental upper limit \cite{ber}.

Generally, such models \cite{lyk} are indeed rather complicated. They, apart
from gauge and matter superfields of the conventional MSSM (minimal
supersymmetric Standard Model), contain several exotic chiral superfields
with SM quantum numbers: an $SU(3)_{\mathrm{C}}$ octet superfield, an $%
SU(2)_{\mathrm{L}}$ triplet superfield, two vectorlike pairs of the $U(1)_{%
\mathrm{Y}}$ hypercharged superfields, and several MSSM singlet fields being
only charged under $U(1)_{\mathrm{Y}^{\prime }}$. These fields are
introduced to cancel all the anomalies related to $SU(3)_{\mathrm{C}%
}^{2}U(1)_{\mathrm{Y}^{\prime }}$, $SU(2)_{\mathrm{L}}^{2}U(1)_{\mathrm{Y}%
^{\prime }}$, $U(1)_{\mathrm{Y}}^{2}U(1)_{\mathrm{Y}^{\prime }},$ and\
others. Supersymmetry is spontaneously broken at tree level by
Fayet-Iliopoulos terms for both $U(1)$ and $U(1)_{\mathrm{Y}^{\prime }}$
hypercharges leading to $D$- and $D^{\prime }$- term VEVs shown above in the
supertrace equation (\ref{sr1}). Apart from that, a special O'Raifeartaigh
type superpotential is introduced to break SUSY and the $U(1)_{\mathrm{Y}%
^{\prime }}$ spontaneously at tree level by generating the proper $F$-term
VEVs (referred to as the $F^{\prime }$-term VEVs for what follows). Due to
this $F^{\prime }$-term breaking, all of the MSSM matter superpartners
(squarks and sleptons) and gauginos receive soft-breaking diagonal masses%
\begin{equation}
\mathrm{m}_{sq/sl}^{2}\simeq \mathrm{g}_{\mathrm{Y}^{\prime
}}^{2}\left\langle D^{\prime }\right\rangle ^{2}+(\Delta \mathrm{m}%
)_{1-loop}^{2}\text{ , \ }\mathrm{M}_{gaugino}\simeq (\Delta \mathrm{M}%
)_{1-loop}  \label{mm}
\end{equation}%
at tree level and one-loop, respectively. Remarkably, not only the universal
tree-level SUSY breaking contribution related to the extra $U(1)_{\mathrm{Y}%
^{\prime }}$ symmetry but also all radiative corrections implied in (\ref{mm}%
) are turned out to be "flavor-blind". Actually, spontaneous SUSY breaking
caused by a generic $U(1)_{\mathrm{Y}^{\prime }}$ symmetry mechanism is
transmitted to superparticles according to some gauge-mediated like scenario
with the $\mathrm{SM}\times U(1)_{\mathrm{Y}^{\prime }}$ gauge bosons
playing a role of messenger fields.

In order to generate one-loop gaugino masses in (\ref{mm})\ which are large
enough to satisfy current experimental bounds (e.g. $\mathrm{m}_{\widetilde{g%
}}>800$ $\mathrm{GeV}$ \cite{ber} for the gluino mass), the heavy sector $%
F^{\prime }$- and $D^{\prime }$-term VEVs must be of order $(30$ \textrm{T}$%
\mathrm{eV})^{2}$. This is in fact a single input scale in the theory. Note
that due to the same sign $\mathrm{Y}^{\prime }$-hypercharges assigned to
all the quarks and leptons, the bare $\mathrm{\mu }$ term is forbidden in
the theory but an effective $\mathrm{\mu }$ term is generated once the $%
U(1)_{\mathrm{Y}^{\prime }}$ symmetry is spontaneously broken at an input
scale mentioned. To obtain the proper electroweak scale, one has to require
a single tree level tuning of the Fayet-Iliopoulos parameters in $U(1)_{%
\mathrm{Y}}$ and $U(1)_{\mathrm{Y}^{\prime }}$ sectors. However, it is not a
fine tuning in the ordinary sense, since radiative corrections to the Higgs
boson masses are appreciably suppressed in the theory. Thus these masses
naturally remain the tree level order values which are chosen to be of the
electroweak scale order. As some immediate outcome, the theory predicts
relatively light gauginos and quite heavy squarks and sleptons with masses
around $7$-$8$ \textrm{T}$\mathrm{eV}$ for the input scale indicated. Such
heavy squarks and sleptons may not be easily observable at the LHC in the
foreseen future. One of the most attractive features of the theory is, as
mentioned above, that flavor changing processes are naturally suppressed,
similar to that in gauge mediated SUSY theories. For more details on this
class of models, we refer the reader to the original paper \cite{lyk} and
only consider here some of their generic predictions concerning the
goldstino phenomenology.

Indeed, all models of low energy supersymmetry breaking predict that the
gravitino may be the LSP, as is determined in the entire supergravity
framework where the gravitino acquires a mass by eating the goldstino
through the super-Higgs mechanism. This goldstino in the model considered is
mostly made of heavy sector fields. This is in fact a combination of the
respective $U(1)_{\mathrm{Y}^{\prime }}$ gaugino and chiral fermions
underlying the above-mentioned O'Raifeartaigh type superpotential which
breaks SUSY and the $U(1)_{\mathrm{Y}^{\prime }}$ at tree level. In
addition, it also may have some small higgsino content, which might be
relevant for a subsequent gravitino phenomenology. The mass of the gravitino
can be estimated this time as (the standard $F$- and $D$-term VEV
contributions are neglected) 
\begin{equation}
\mathrm{m}_{3/2}\simeq \frac{\sqrt{\left\langle F^{\prime }\right\rangle
^{2}+\left\langle D^{\prime }\right\rangle ^{2}}}{\mathrm{M}_{P}}
\label{3/2'}
\end{equation}%
where the relatively low $F^{\prime }$ and $D^{\prime }$ term VEVs mentioned
above\ give for its value $\mathrm{m}_{3/2}\sim 0.07$ $\mathrm{eV}$ that is
definitely safe for cosmology \cite{wess}. The gravitino, by absorbing the
goldstino, inherits its non-gravitational interactions and so can play an
important role in collider physics.

The generic interactions of the goldstino $\zeta _{g}$\ (being the
longitudinal part of a massive gravitino $\zeta _{G}$) follow, as usual \cite%
{wess}, from the total supercurrent conservation that determines its
effective low-energy Lagrangian as 
\begin{equation}
L_{eff}=-i\overline{\zeta }_{g}\overline{\sigma }^{\mu }\partial _{\mu
}\zeta _{g}-\frac{1}{\sqrt{\left\langle F^{\prime }\right\rangle
^{2}+\left\langle D^{\prime }\right\rangle ^{2}}}\left( \zeta _{g}\partial
_{\mu }j^{\mu }+h.c.\right) \text{ .}  \label{eff}
\end{equation}%
where $j^{\mu }$ is the supercurrent which includes contributions from all
matter and gauge supermultiplets involved. As a consequence, one has the
basic goldstino-scalar-chiral fermion vertex%
\begin{equation}
\zeta _{g}^{w}\partial _{\mu }(\sigma ^{\nu }\overline{\sigma }^{\mu }\psi
_{i})_{w}\partial _{\nu }\varphi ^{\ast i}  \label{3}
\end{equation}%
and goldstino-gaugino-gauge boson vertex%
\begin{equation}
-\zeta _{g}^{w}\partial _{\mu }[(\sigma ^{\nu }\overline{\sigma }^{\rho
}\sigma ^{\mu }\overline{\lambda }^{p})_{w}F_{\nu \rho }^{p}]/2\sqrt{2}\text{
.}  \label{4}
\end{equation}%
in the theory (here $w$ stands for a spinor index, while indices $i$ and $p$
belong to the SM group representations for matter and gauge supermultiplets,
respectively). Since this derivation depends only on the total supercurrent
conservation, the Lagrangian (\ref{eff}) holds independently of the details
of supersymmetry breaking. It universally determines the decay rate of any
sparticle $\widetilde{X}$ into its superpartner $X$ plus the
goldstino/gravitino ($\zeta _{g}/\zeta _{G}$) whether ($X$, $\widetilde{X}$)
is a chiral superfield pair ($\varphi $, $\psi $) or a vector superfield
pair ($A_{\mu }$, $\lambda $), respectively.

Remarkably, an orthogonal combination to the goldstino $\zeta _{g}$, namely
the pseudo-goldstino $\zeta _{pg}$, happens to be mostly a bino\footnote{%
For a typical range of parameters in the model considered in \cite{lyk} this
pseudo-goldstino has a content%
\begin{equation*}
\zeta _{pg}=-0.9999\widetilde{B}-0.003\widetilde{W}^{0}-0.002\widetilde{H}%
_{u}^{0}+0.004\widetilde{H}_{d}^{0}\text{ }
\end{equation*}%
including, apart from the bino, the vanishingly small admixtures of the wino 
$\widetilde{W}^{0}$ and the higgsinos $\widetilde{H}_{u,d}^{0}$ .}, or a
photino (\ref{ph}) if we turn to a pure QED framework. In the SM context,
this bino is a NLSP having the electroweak scale order mass. As a
consequence, the photino being the linear combination of a bino and a
neutral wino will dominantly decay into the photon and the gravitino with a
decay rate being entirely determined by the interaction vertex (\ref{4})

\begin{equation}
\Gamma (\widetilde{\gamma }\rightarrow \gamma +\zeta _{G})\simeq \frac{%
\mathrm{m}_{\widetilde{\gamma }}^{5}k_{\widetilde{\gamma }}}{16\pi \left(
\left\langle F^{\prime }\right\rangle ^{2}+\left\langle D^{\prime
}\right\rangle ^{2}\right) }  \label{gam}
\end{equation}%
where $k_{\widetilde{\gamma }}$ is the pure photino content of the
pseudo-goldstino $\zeta _{pg}$ in the supersymmetric SM. For typical values $%
k_{\widetilde{\gamma }}\sim 0.15,$ $m_{\widetilde{\gamma }}\sim 100$ $%
\mathrm{GeV}$ in the model and heavy sector VEVs $\left\langle F^{\prime
}\right\rangle \sim \left\langle D^{\prime }\right\rangle \sim 30$ \textrm{T}%
$\mathrm{eV}$ taken above one has for the photino lifetime $\tau _{%
\widetilde{\gamma }}$ $\sim 2\cdot 10^{-15}\sec $ that could make its mean
decay length to reach up to $0.5$ $\mathrm{\mu m}$ under LHC energies.

To summarize, the emergent Standard Models with spontaneous SUSY breaking
which is only occurred in the visible sector seem not to violate any current
phenomenological constraint. In general, these models predict light gauginos
and quite heavy squarks and sleptons which may not be observable at the LHC.
The LSP is a stable very light gravitino with a significant higgsino
admixture, while the NLSP is mostly a bino. Apart from that, it is worth
noticing some other advantages of these low-scale models thoroughly
described in \cite{lyk}. Proton decay is sufficiently and naturally
suppressed, even for a rather low cutoff scale about $10^{8}$ $\mathrm{GeV}$%
. The strong $CP$ problem is naturally solved through the Nelson-Barr
mechanism \cite{nb}. In addition, an introduction of the extra $U(1)_{%
\mathrm{Y}^{\prime }}$ helps to sufficiently suppress the $\mathrm{B}$ and $%
\mathrm{L}$ violating interactions. An interesting generic cold dark matter
candidate is also found. This is the lightest particle among several SM
singlet fields introduced in the theory heavy sector to cancel all possible
anomalies related to the $U(1)_{\mathrm{Y}^{\prime }}$ symmetry. Although it
typically has the \textrm{T}$\mathrm{eV}$ scale order mass, it appears
absolutely stable due to some surviving discrete symmetry of the appropriate
O'Raifeartaigh superpotential taken.

\section{Summary and conclusions}

As we argued above, spontaneous Lorentz violation in a vector field theory
framework may be active as in the composite and potential-based models
leading to physical Lorentz violation, or inactive as in the
constraint-based models resulting in the nonlinear gauge choice in an
otherwise Lorentz invariant theory. Remarkably, between these two basic SLIV
versions SUSY unambiguously chooses the inactive SLIV case. Indeed, SUSY
theories only admit the bilinear mass term in the vector field potential
energy. As a result, without a stabilizing quartic vector field terms, the
physical spontaneous Lorentz violation never occurs in SUSY theories. Hence
it follows that the composite and potential-based SLIV models can in no way
be realized in the SUSY context. This may have far-going consequences in
that supergravity and superstring theories could also disfavor such models
in general.

Nevertheless, even in the case when SLIV is not physical it inevitably leads
to the generation of massless photons as vector NG bosons provided that SUSY
itself is spontaneously broken. In this sense, a generic trigger for
massless photons to dynamically emerge happens to be spontaneously broken
supersymmetry rather than physically manifested Lorentz noninvariance. To
see how this idea might work we considered supersymmetric QED model extended
by an arbitrary polynomial potential of a general vector superfield that
induces spontaneous SUSY violation in the visible sector, and gauge
invariance gets broken as well. \ Notably, massless photons at this point
are related to spontaneously broken supersymmetry (SBS) itself rather than
gauge invariance. Actually, SBS only provides the tree-level masslessness of
a photon (as a photino companion) but cannot protect it against radiative
corrections since its generic massless mode is only a photino rather than a
whole gauge supermultiplet. Nevertheless, though gauge invariance is
explicitly broken by the superfield potential, the special gauge invariance
is in fact recovered in the broken SUSY phase that universally protects the
photon masslessness. This invariance is only restricted by the nonlinear
gauge condition (\ref{const1}) put on the vector field. The point, however,
is that this length-fixing gauge condition happens at the same time to be
the SLIV type constraint which treats in turn the physical photon as the
Lorentzian NG mode. So, figuratively speaking, the photon passes through
three evolution stages being initially the massive vector field component of
a general vector superfield (\ref{lag3}), then the three-level massless
companion of the Goldstone photino in the broken SUSY stage (\ref{vvv}) and
finally the generically massless state as the emergent Lorentzian NG mode in
the inactive SLIV stage (\ref{const1}).

All basic arguments developed in SUSY QED were then generalized to Standard
Model and Grand Unified Theories. Remarkably, thanks to a generic high
symmetry of the length-fixing SLIV constraint put on the vector fields the
emergence conjecture with dynamically produced massless gauge modes can be
applied to any non-Abelian internal symmetry case. Specifically, one can
argue that in a theory with an internal symmetry group $G$ not only the pure
Lorentz symmetry $SO(1,3)$, but the larger accidental symmetry $SO(\Upsilon
,3\Upsilon )$ of the SLIV constraint (\ref{111}) in itself appears to be
spontaneously broken as well ($\Upsilon $ is a dimension of the group $G$).
As a result, although the pure Lorentz violation on its own still generates
only one genuine Goldstone vector boson, the accompanying pseudo-Goldstone
vector bosons related to the $SO(\Upsilon ,3\Upsilon )$ breaking also come
into play properly completing the whole gauge multiplet of the internal
symmetry group $G$ taken. Remarkably, they appear to be strictly massless as
well, being protected by the simultaneously generated non-Abelian gauge
invariance. For definiteness, we focused on the $U(1)\times SU(N)$
symmetrical theories. Such a split group form is dictated by the fact that
in the pure non-Abelian symmetry case one only has the SUSY invariant phase
in the theory that would make it inappropriate for an outgrowth of an
emergence process. As briefly discussed, supersymmetric emergent theories
look attractive both theoretically and phenomenologically whether they are
considered at low energies in terms of the Standard Model or at very high
energies as the flipped $SU(5)$ GUTs inspired by superstrings.

However, their most generic manifestations seem to be related to a
spontaneous SUSY violation in the visible sector that we finally considered.
The photino emerging due to this violation will be then mixed with another
goldstino which stems from a spontaneous SUSY violation in the hidden
sector. Eventually, it largely turns into light pseudo-goldstino whose
physics seems to be of special interest. Such a pseudo-Goldstone photino
appears typically as the $\mathrm{eV}$ scale stable LSP or the electroweak
scale long-lived NLSP, being accompanied by a very light gravitino in both
cases, that can be considered as some observational signature of the class
of models where SUSY breaks, at least partially, in the visible sector as
well. This is the only class of models where emergent supersymmetric QED or
Standard Model can be successfully realized. So, in contrast to non-SUSY
analogs, the emergent SUSY theories even with the Lorentz-preserving
inactive SLIV could naturally have some clear observational signal. Its
validation, apart from some indication of an emergence nature of gauge
symmetries, could shed considerable light on the SUSY breaking physics that
is actively studied in recent years.

We conclude by a general remark that supersymmetry with its well known
advantages, such as naturalness, grand unification and dark matter candidate
seems to possess one more attractive feature: it may trigger, through its
own spontaneous violation, a dynamical generation of massless gauge fields
as massless NG modes during which physical Lorentz invariance itself is
generically preserved. An extension of this idea to the local supersymmetry
case, which could presumably underlie an emergent supergravity theory
unifying all elementary forces, seems to be especially interesting and worth
pursuing.

\section*{ Acknowledgments}

I thank Pierre Fayet, Colin Froggatt, Alan Kostelecky, Rabi Mohapatra and
Holger Nielsen for interesting discussions and correspondence, and Pierre
Fayet also for drawing my attention to his papers \cite{fayet}. This work is
partially supported by Georgian National Science Foundation (Contracts No.
31/89 and No. DI/12/6-200/13).

\end{document}